\pdfoutput=1
\documentclass[aps,prb,superscriptaddress,twocolumn,10pt
]{revtex4-2}
\usepackage{graphicx}
\usepackage[svgnames,psnames]{xcolor}
\usepackage[T1]{fontenc}
\usepackage{textcomp}
\usepackage{bm}
\usepackage{amsmath,amssymb,amsthm}
\usepackage{mathtools}
\usepackage{mathrsfs}
\usepackage{comment}
\usepackage[normalem]{ulem}
\usepackage{braket}
\usepackage[separate-uncertainty]{siunitx}
\newcommand{\abs}[1]{\lvert #1\rvert}
\newcommand{\Abs}[1]{\left|#1\right|}
\newcommand{\rmd}{{\mathrm{d}}}
\newcommand{\iu}{{\mathrm{i}}}
\renewcommand{\dfrac}[2]{\frac{\displaystyle{#1}}{\displaystyle{#2}}}
\usepackage[
colorlinks=true,
citecolor=DarkGreen,
linkcolor=FireBrick,
unicode
]{hyperref}

\allowdisplaybreaks[0]
\begin{document}

\title{
Boundary condition for phonon distribution functions
at a smooth crystal interface
and interfacial angular momentum transfer}

\author{Yuta Suzuki}
\email{suzuki.y.8cc2@m.isct.ac.jp}
\affiliation{Department of Physics, Institute of Science Tokyo, 2-12-1 Ookayama, Tokyo 152-8551, Japan {(JSPS Research Fellow)}}
\affiliation{Department of Applied Physics, The University of Tokyo, 7-3-1 Hongo, Tokyo 113-0033, Japan}
\author{Shuntaro Sumita}
\affiliation{Department of Basic Science, The University of Tokyo, 3-8-1 Komaba, Tokyo 153-8902, Japan}
\affiliation{Komaba Institute for Science, The University of Tokyo, 3-8-1 Komaba, Tokyo 153-8902, Japan}
\affiliation{Condensed Matter Theory Laboratory, RIKEN CPR, Wako, Saitama 351-0198, Japan}
\author{Yusuke Kato}
\affiliation{Department of Basic Science, The University of Tokyo, 3-8-1 Komaba, Tokyo 153-8902, Japan}
\affiliation{Quantum Research Center for Chirality, Institute for Molecular Science, Okazaki, Aichi 444-8585, Japan}
\affiliation{Department of Physics, Graduate School of Science, The University of Tokyo, 7-3-1 Hongo, Tokyo 113-0033, Japan}

\date{\today}

\begin{abstract}%
We theoretically elucidate the boundary conditions for phonon distribution functions of long-wavelength acoustic phonons at smooth crystal interfaces. We first derive boundary conditions that fully incorporate reflection, transmission, and mode conversion. We obtain these conditions for phonons from those for classical lattice vibrations, using the correspondence between the quantum and classical descriptions. This formulation provides a theoretical foundation for the acoustic mismatch model, widely used to analyze Kapitza resistance. We then refine the boundary conditions to include spatial dependence parallel to the interface. The refined form captures transverse shifts of elastic wave packets, analogous to the optical Imbert--Fedorov shift, and ensures conservation of total angular momentum. Consequently, circularly polarized phonons carrying spin angular momentum (SAM) generate phonon orbital angular momentum (OAM) at the interface. We analytically determine the spatial profile of this OAM and demonstrate that SAM and OAM are both involved in the interfacial diffusion of chiral phonons. Our theory provides concise boundary conditions for phonons, with applications ranging from heat transport to phonon angular momentum transport.
\end{abstract}
\maketitle
\section{Introduction}

Phonons can carry angular momentum (AM), providing a channel for transmitting rotational motion through solids~\cite{Zhang2014,Zhang2015,
Chen2015,Zhu2018,Kishine2020,Hamada2018,Hamada2020a,Park2020,Zhang2021,Chen2022,Yao2022,Zhang2022,Romao2023,AKato2023,Ishito2023a,Ishito2023b,Oishi2024,Tateishi2025,Ishizuka2025}. 
These excitations---known as chiral phonons~\cite{Wang2024,TZhang2025} or more precisely axial phonons~\cite{Juraschek2025,Kusunose2024,Ishito2023a,Ishito2023b,Oishi2024}---have attracted growing attention for
their impact on AM transport across crystal interfaces. 
A central motivation comes from their coupling to magnetism~\cite{Anastassakis1972,Rebane1983,
Mentink2019,AKato2022,Weissenhofer2023,Shokeen2024,Juraschek2017,
Juraschek2019,Hamada2020,Ren2021,
Juraschek2022,Tauchert2022,Fransson2023,Luo2023,Basini2024,Chaudhary2024,Funato2024,Sano2024,Yao2024,Li2024,Yokoyama2024,Yao2025,Yao2026}.  
Phonons have been shown experimentally to transfer AM both to conduction electrons and to localized spins in junction systems~\cite{Korenev2016,
Nova2017,Holanda2018,Sasaki2021,Jeong2022,Kim2023,Ueda2023,Ohe2024,Davies2024,Choi2024,Nabei2026}.  
Because acoustic phonons propagate well in the media, their AM transport cannot be explained solely by 
\emph{interfacial exchange} with electrons~\cite{Funato2024,Nishimura2025}; 
the \emph{direct transmission} of phonons carrying AM must also be considered~\cite{SuzukiSumitaKato2024a}.

Meanwhile, the microscopic mechanism of phonon AM transport across an interface remains much less understood than that of heat transport.  
The contrast originates from the assumptions of standard models of interfacial heat transfer.  
The Landauer and Caroli formulae~\cite{Caroli1971} treat the interface as that couples two nearly equilibrium reservoirs at slightly different temperatures.  
Such models justify the use of equilibrium phonon distributions.  
{In equilibrium of nonmagnetic systems,
however, spin angular momentum (SAM) carried by circularly polarized phonons~\cite{
Vonsovskii1962,Levine1962,Portigal1968,Pine1970,Ishii1975,Mclellan1988} cancels because right- and left-circular modes are equally populated. }
As a result, no net AM transport occurs.  
Therefore, methods developed for heat transport cannot be applied directly to AM transport.

\begin{figure*}
\centering
\includegraphics[pagebox=artbox,width=0.99\textwidth]{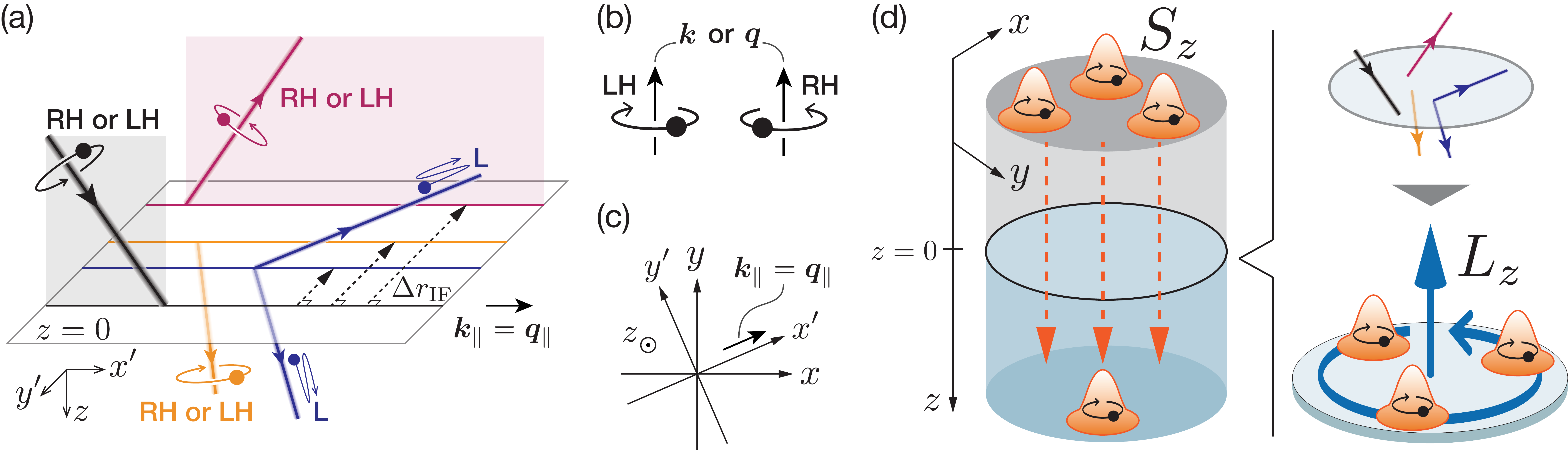}
\caption{
Schematic illustration of the reflection and transmission of a phonon or elastic wave packet at the interface $z = 0$ between two crystals, 
and the associated generation of orbital angular momentum~(OAM).
(a)~When the incident wave packet (black line) is circularly polarized, the reflected and transmitted wave packets 
(colored lines) emerge from positions laterally displaced 
from the incident plane by $\Delta r_{\text{IF}}$. 
The wavevector in the region $z < 0$ is denoted by $\bm{k}$, and in $z > 0$ by $\bm{q}$, with equal in-plane components: $\bm{k}_{\parallel} = \bm{q}_{\parallel}$.
(b)~Definition of right-handed (RH) and left-handed (LH) circular polarizations as used in this study. 
They are defined from the point of view of the source.
(c)~Orthogonal coordinates $x'y'$ introduced along the in-plane wavevector component $\bm{k}_{\parallel} = \bm{q}_{\parallel}$.
(d)~Spin angular momentum (SAM) $S_z$ associated with circular polarization accumulates in the region $z < 0$ and diffuses across the interface, 
which in turn induces a finite OAM $L_z$ at the interface. 
The phonon wave packet is depicted as an orange bell shape with a black ring indicating circular polarization. 
The OAM $L_z$ arises from the transverse shifts $\Delta r_{\text{IF}}$, 
which yield an extrinsic angular momenta (AM) of the form $\Delta r_{\text{IF}} \times \hbar \Abs{\bm{k}_{\parallel}}$.
}
\label{fig: intro schematics}
\end{figure*}

Phonon AM transport instead requires non-equilibrium conditions where SAM is spatially biased or localized.  
One realization is a junction of two crystals where a temperature gradient is applied only to one side, 
as demonstrated experimentally~\cite{Kim2023,Ohe2024}. 
In this case, the phonon Edelstein effect~\cite{Hamada2018,Zhang2025} generates SAM inside the heated crystal and creates a step in the SAM distribution at the interface.  
Even under such simple conditions, there is still no established formalism for calculating the SAM flux across an interface.

To fill this gap, in this paper, we formulate boundary conditions for the phonon distribution function at 
the interface of a two-crystal junction. 
We focus on long-wavelength acoustic phonons, which dominate thermal excitations at low temperatures.  
Assuming a smooth interface, we consider specular reflection and refraction of phonons governed by Snell's law. 
Using these assumptions, we first construct the \emph{coarse-grained} boundary conditions, 
which can describe the SAM transport across the interface (see our companion {Letter} for a demonstration of this transport~\cite{SuzukiSumitaKato2024a}). 
These conditions take the form of linear relations between incident and scattered distribution functions
with the real coordinates parallel to the interface being coarse-grained~[Eq.~\eqref{eq: boundary condition assump}]. 
The coefficients in our relation follow the acoustic mismatch model~\cite{Little1959,Khalatnikov1965,Swartz1989},  
a standard approach to interfacial thermal resistance, and are given by the power reflectance and transmittance from elasticity theory. 
A central result is that these boundary conditions explicitly incorporate phonon mode and polarization, and
are derived directly from elastic wave and lattice vibration theory.

Furthermore, we formulate the \emph{detailed} boundary conditions by refining the coarse-grained ones to describe the interfacial transport of phonon orbital angular momentum (OAM) as well as SAM. 
Although OAM has been overlooked in most previous studies on phonon AM transport, the concept of phonon OAM is naturally introduced through analogy with elasticity theory~\cite{Garanin2015,Nakane2018,Bliokh2006}.
In that context, OAM is defined as the cross product of the center-of-mass position and momentum of a wave packet. 
For elastic waves, when a circularly polarized wave packet [Fig.~\ref{fig: intro schematics}(b)] encounters a discontinuity in the medium, part of its SAM converts into OAM~\cite{Bliokh2006}. Consequently, both reflected and transmitted wave packets experience spatial displacements of their centers along the direction perpendicular to the incident plane; this effect is also known as the Imbert--Fedorov shift in optics~\cite{Fedorov1955,Schilling1965,Imbert1972,Onoda2004}. 
Figure~\ref{fig: intro schematics}(a) shows these displacements $\Delta r_{\mathrm{IF}}$, which we refer to as \emph{transverse shifts} from here on. In our detailed boundary conditions for phonon distributions, we thus include the spatial dependence of the phonon distribution function before and after interfacial scattering 
[Eqs.~\eqref{163324_9Jun25} and \eqref{163328_9Jun25}] by relating the transverse shift of a classical elastic wave packet to the position expectation value of the scattered phonon distribution. These conditions ensure the conservation of total angular momentum (TAM), i.e., the sum of phonon SAM and OAM, at the interface.

Finally, based on the detailed boundary conditions and Boltzmann theory, 
we predict the generation of phonon OAM at an interface when a finite flux of SAM crosses it. 
Figure~\ref{fig: intro schematics}(d) schematically shows that the SAM flux exhibits a discontinuity at $z=0$; we attribute the resulting OAM to the compensation of this discontinuity. The generated OAM corresponds to a circulating flow along the edge of the interface, as illustrated in Fig.~\ref{fig: intro schematics}(d).
Unlike previous studies~\cite{Hefner1999,Thomas2003,Bliokh2006,Ayub2011,Wang2021}, which examined individual wave packets or beams, 
we consider a statistical average over a phonon ensemble. 
As a specific example, we examine the application of temperature gradient to one side of the two-crystal junction.
We analytically derive the density and flux of the OAM generated in this junction, along with its subsequent diffusion into the bulk.

\begin{figure}[tbp]
\centering
\includegraphics[pagebox=artbox,width=\columnwidth]{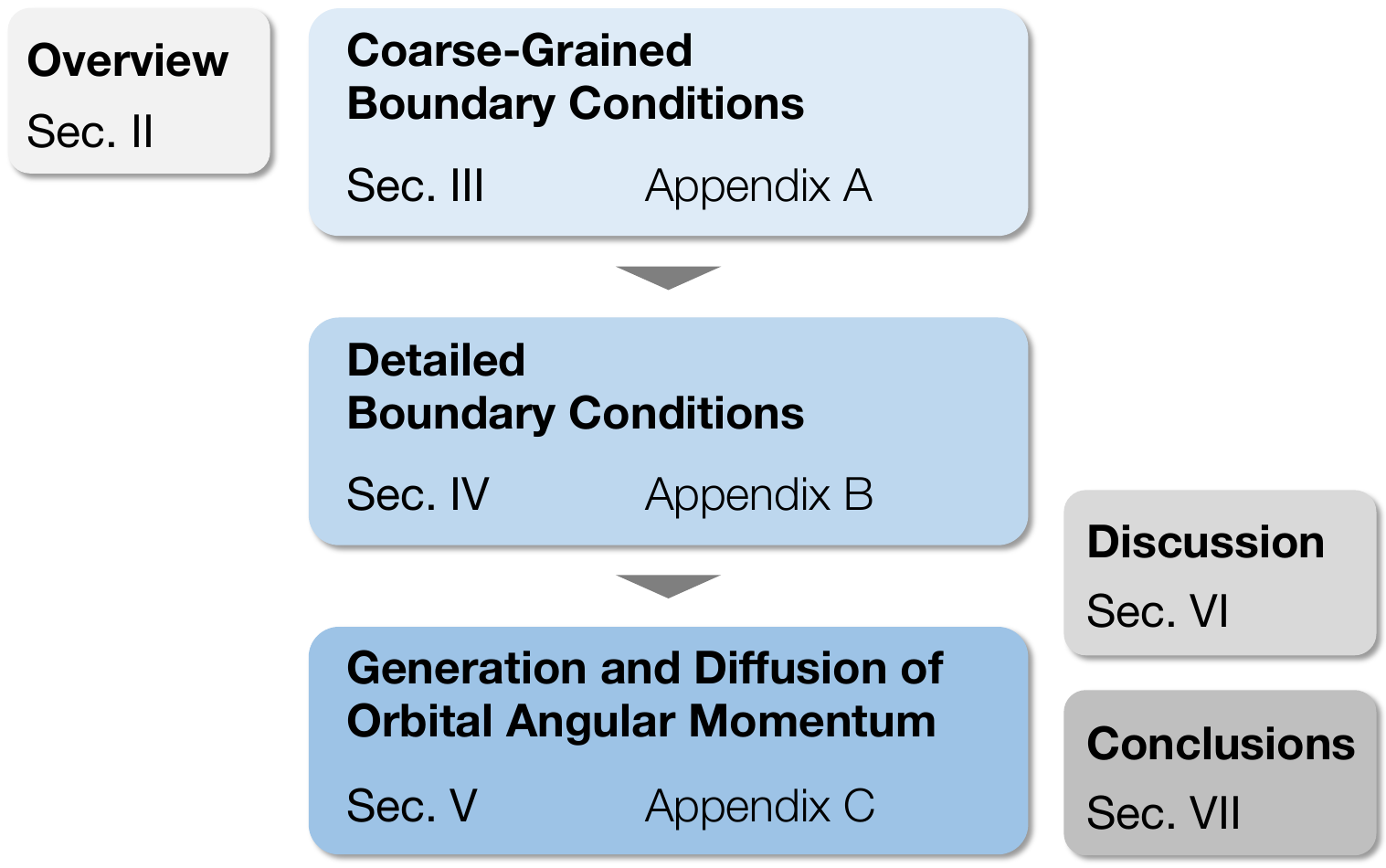}
\caption{Flow of the discussion in this paper.}
\label{fig: flowchart of sections}
\end{figure}

The remainder of this paper is organized as follows (see also Fig.~\ref{fig: flowchart of sections}). 
In Sec.~\ref{sec: assump and b.c.s}, we summarize the assumptions and 
overview the key results associated with the two boundary conditions introduced in this study. 
In Sec.~\ref{sec: lattice vibration ph bc}, we derive the coarse-grained boundary conditions for phonon distribution functions at the interface. 
In Sec.~\ref{sec: detailed b.c. formula}, we formulate the detailed boundary conditions that explicitly depend on the lateral coordinates $(x,y)$. 
In Sec.~\ref{sec: Generation of orbital angular momentum at the interface}, we apply the detailed boundary conditions to demonstrate the generation of OAM at a crystal interface. 
{In Sec.~\ref{sec: Discussion}, we discuss unresolved issues regarding wave-packet confinement, 
the dependence of phonon OAM on system details, 
experimental signatures of phonon AM transport, 
the effect of discrete rotational symmetry on the boundary conditions, 
mechanisms of TAM relaxation, and applications of the formulated boundary conditions to phonon and boson transport.}
In Sec.~\ref{sec: summary}, we conclude the paper.

Appendix~\ref{sec: elastic plane wave} provides an overview of boundary scattering of elastic plane waves. 
In Appendix~\ref{sec: IF shift of elastic beam}, we derive the classical transverse shift of reflected and transmitted beams under circularly polarized incidence. 
Appendix~\ref{sec: appendix validity of xy derivatives in BTE} explains the approximation 
used to neglect parallel drift terms in the OAM calculations of Sec.~\ref{subsec: Boltzmann theory}.

\section{Overview}
\label{sec: assump and b.c.s}

\subsection{Assumptions}
\label{subsec: our assump}

{We consider a model system where two crystals, both treated as isotropic elastic media}~\footnote{
A crystal is regarded as an isotropic elastic medium in either of the following two cases~\cite{LandauLifshitzTextbookVol7}: 
(i)~it is polycrystalline, with grain size
smaller than the wavelength we focus on; or 
(ii)~its multiple elastic constants happen to reduce to the two parameters of isotropic elasticity.
}{, are joined at a smooth interface at $z = 0$.
This assumption is widely employed in the acoustic mismatch model for interfacial thermal transport~\cite{Little1959,Swartz1989}. 
Here, we aim to establish the general mechanism of phonon AM transmission; 
quantitative modeling of specific anisotropic crystals is left for future work. 
We focus on nonmagnetic crystals and assume the absence of external magnetic fields.}
We assume the system is in a steady state~\footnote{
In particular, the entire structure is not undergoing rigid-body rotation relative to the laboratory frame. 
} and at low temperatures.

\begin{figure}[tbp]
\centering
\includegraphics[pagebox=artbox,width=0.65\columnwidth]{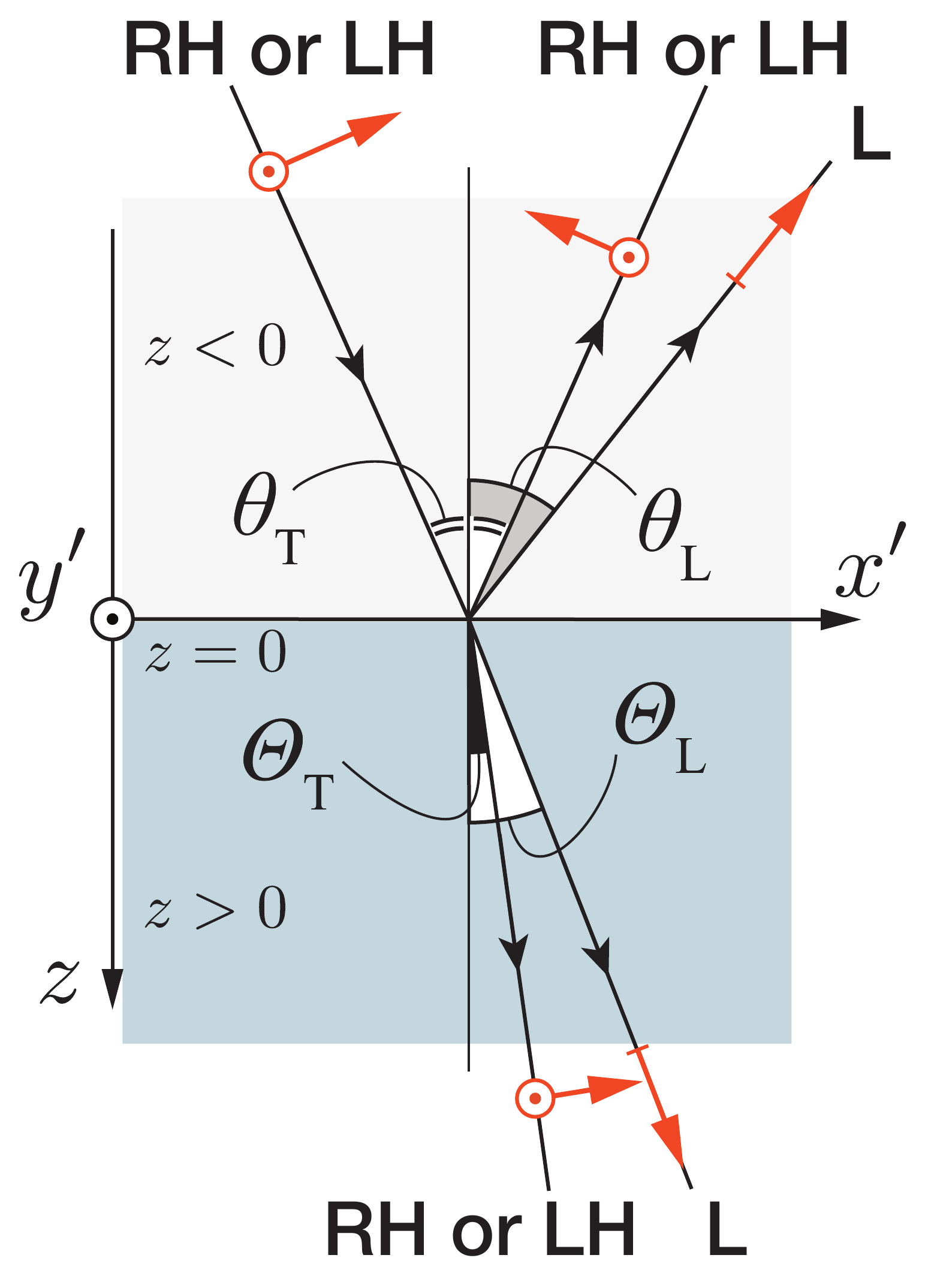}
\caption{
Schematics of the adjacent two crystals, and the reflection and transmission of plane waves at a smooth interface $z = 0$.
The polarization vectors of each wave mode are indicated by orange arrows.
The interface corresponds to the $x'y'$ plane, and the incident plane is the $zx'$ plane.
Note that $x'$ and $y'$ axes are in the direction of $\bm{k}_{\parallel}$ and $\hat{\bm{z}}\times \bm{k}_{\parallel}$, respectively.
We also use a laboratory coordinate $(x, y, z)$ without primes that is independent of the direction of wavevectors, 
as shown in Fig.~\ref{fig: intro schematics}(c).}
\label{fig: incidentplane}
\end{figure}

The characteristic length scales of the system follow this order: 
lattice constant $\ll$ interface roughness $\ll$ phonon wavelength $\ll$ phonon mean free path $\ll$ crystal size. 
We treat the $z$-direction length of the system, normal to the interface, as infinite.
That is, the two crystals occupy the semi-infinite regions $z < 0$ and $z > 0$, respectively (see Fig.~\ref{fig: incidentplane}).

We now introduce the physical quantities used in both crystals. Table~\ref{tab: symbols} lists the notation. 
\begin{table*}
\centering
\begin{minipage}{0.85\textwidth}
\caption{Notation used throughout this paper.}
\label{tab: symbols}
 \begin{ruledtabular}  
\begin{tabular}{lll}
                                     & $z <0$     &  $z >0$    \\ \hline
Mode                                 & $n = \mathrm{L}$, $\mathrm{RH}$, $\mathrm{LH}$ &  $n = \mathrm{L}$, $\mathrm{RH}$, $\mathrm{LH}$\\
Wavevector                           & $\bm{k} = (\bm{k}_{\parallel}, k_z)$  &     $\bm{q} = (\bm{q}_{\parallel}, q_z)$       \\
Position                             & $\bm{r} = (\bm{R}_{\parallel}, z)$   & $\bm{r} = (\bm{r}_{\parallel}, z)$           \\
Dispersion relation                  & $\omega = \Omega_{\bm{k} n} = v_n\Abs{\bm{k}}$  & $\omega = \omega_{\bm{q}n} = c_n \Abs{\bm{q}}$            \\
Distribution function                & $F_{\bm{k}n}(\bm{r}) = F_{s, n}(\omega, \bm{k}_{\parallel}, \bm{r})$   & $f_{\bm{q}n}(\bm{r}) = f_{s, n}(\omega, \bm{q}_{\parallel}, \bm{r})$           \\
Coarse-grained distribution function &  $F_{s, n}(\omega, \bm{k}_{\parallel}, z)$  &   $f_{s, n}(\omega, \bm{q}_{\parallel}, z)$          \\
SAM                                  & $\bm{S}_{\bm{k}n} = \bm{S}_{s, n}(\omega, \bm{k}_{\parallel})$ & $\widetilde{\bm{S}}_{\bm{q}n} = \widetilde{\bm{S}}_{s, n}(\omega, \bm{q}_{\parallel})$           \\
OAM ($z$ component)                               &  $L^z(\bm{k}_{\parallel}, \bm{R}_{\parallel})$  &  $\widetilde{L}^z(\bm{q}_{\parallel}, \bm{r}_{\parallel})$ \\
TAM ($z$ component)                                 & $J^z_{s, n}(\omega, \bm{k}_{\parallel}, \bm{R}_{\parallel})$ & $\widetilde{J}^z_{s, n}(\omega, \bm{q}_{\parallel}, \bm{r}_{\parallel})$ 
\end{tabular}
\end{ruledtabular}  
\end{minipage}
\end{table*}
We first specify the eigenstates of acoustic phonons in the regions $z < 0$ and $z > 0$ 
by pairs of wavevector and mode, denoted by $(\bm{k}, n)$ and $(\bm{q}, n)$, respectively.
The mode index $n$ includes one longitudinal compression mode ($n = \text{L}$) and two-degenerate transverse shear modes.
Since our focus is on the phonon AM, 
we approximate the states of transverse phonons as being characterized by right- and left-handed circular polarizations, 
labeled by $n = \text{RH}$ and $n = \text{LH}$, respectively~\footnote{
Strictly speaking, fully specifying the number and polarization state of transverse phonons at each wavevector
requires four parameters, analogous to the optical Stokes parameters.
We adopt an approximation where only the total intensity and the circular component are retained.
See also Appendix~\ref{151957_6Feb25}.
}.
We illustrate the definition of these circular polarization in Fig.~\ref{fig: intro schematics}(b).
We then introduce phonon distribution functions $F_{\bm{k}n}(x, y, z < 0)$ for the phonon state 
$(\bm{k}, n = \text{L}, \text{RH}, \text{LH})$, or for wave packets centered around these states.
Similarly, for the states $(\bm{q}, n)$ in the region $z > 0$, we define distribution functions $f_{\bm{q}n}(x, y, z)$.

In scattering at a steady and smooth interface, both the energy $\hbar\omega$ and the momentum component $\hbar\bm{k}_{\parallel}$ 
parallel to the interface are conserved. 
Once the pair $(\omega, \bm{k}_{\parallel})$ is specified, the state of a phonon with linear dispersion is uniquely determined 
by specifying just two additional quantities: 
the sign of the group velocity in the direction normal to the interface, denoted by $s = k_z/\abs{k_z}$ or $s = q_z/\abs{q_z}$, 
and the mode index $n$.
We thus label phonon states in the regions $z < 0$ and $z > 0$ using the 4-tuple $(\omega, \bm{k}_{\parallel}, s, n)$. 
The phonon distribution functions are then expressed as
$F_{\bm{k}n} (\bm{r}) = F_{s, n} (\omega, \bm{k}_{\parallel}, \bm{r})$ and
$f_{\bm{q}n} (\bm{r}) = f_{s, n} (\omega, \bm{q}_{\parallel}, \bm{r})$, with the spatial coordinate $\bm{r} = (x, y, z)$.
For example, $F_{+, n}(x, y, z < 0)$ and $f_{-, n}(x, y, z > 0)$ represent distribution functions with group velocities directed toward the interface, 
i.e., they correspond to phonons incident on the interface. 
In contrast, $F_{-, n}(x, y, z < 0)$ and $f_{+, n}(x, y, z > 0)$ describe distribution functions with group velocities directed away from the interface, 
representing phonons after scattering.
As in this example, we will often omit the arguments $(\omega, \bm{k}_{\parallel})$ for simplicity.

We emphasize that the AM carried by phonons has two components: SAM and OAM.  
The SAM can be defined analogously to the circular polarization of light.  
As illustrated in Fig.~\ref{fig: intro schematics}(b), in the region $z < 0$ we assign to each mode  
\begin{subequations}  
\begin{equation}  
\bm{S}_{\bm{k}, \text{RH}} \equiv +\hbar \bm{k} / \abs{\bm{k}}, \qquad  
\bm{S}_{\bm{k}, \text{LH}} \equiv -\hbar \bm{k} / \abs{\bm{k}},  
\label{eq: SAM for T mode def z<0}  
\end{equation}  
In the region $z > 0$, we define in the same way
 \begin{equation}
 \widetilde{\bm{S}}_{\bm{q}, \text{RH}} \equiv + \hbar \bm{q} / \abs{\bm{q}},\qquad
 \widetilde{\bm{S}}_{\bm{q}, \text{LH}} \equiv - \hbar \bm{q} / \abs{\bm{q}}.
 \end{equation}
 Longitudinal modes ($n = \text{L}$) do not carry SAM, i.e.,
 \begin{equation}
 \bm{S}_{\bm{k}, \text{L}} = \widetilde{\bm{S}}_{\bm{q}, \text{L}} = 0. \label{eq: SAM for L mode def z<0}
\end{equation}
\end{subequations}

Next, we define the OAM relevant to our analysis. 
Let the component of the position vectors parallel to the interface
in the regions $z < 0$ and $z > 0$ be denoted by $\bm{R}_{\parallel}$ and $\bm{r}_{\parallel}$, respectively. 
For the distribution function 
$F_{s, n}(\omega, \bm{k}_{\parallel}, x, y, z) =  F_{s, n}(\omega, \bm{k}_{\parallel}, \bm{R}_{\parallel}, z)$ in the region $z < 0$, 
the $z$ component of the OAM it carries is defined as the cross product of position and momentum:
\begin{subequations}
 \begin{equation}
 L^z(\bm{k}_{\parallel}, \bm{R}_{\parallel})\equiv  \left(\bm{R}_{\parallel}\times \hbar \bm{k}_{\parallel}\right)_z.
 \label{eq: OAM definition Lz}
 \end{equation}
 Similarly, for the distribution function $f_{s, n}(\omega, \bm{q}_{\parallel}, \bm{r}_{\parallel}, z)$ in the region $z > 0$, 
the $z$ component of the OAM is defined as
 \begin{equation}
 \widetilde{L}^z(\bm{q}_{\parallel}, \bm{r}_{\parallel}) \equiv (\bm{r}_{\parallel}\times\hbar \bm{q}_{\parallel})_z.
 \label{eq: OAM definition tildeL z}
\end{equation}
\end{subequations}
{Note that the OAM defined above is the extrinsic OAM associated with the trajectory of a phonon wave packet.
It is distinct from pseudo-orbital AM associated with discrete rotational symmetry~\cite{Zhang2015} and from intrinsic OAM associated with 
phonon vortex beams~\cite{ONeil2002,Bliokh2013,Wang2021}.
The present formulation is restricted to this extrinsic contribution.}

We then define the TAM as the sum of the spin and orbital components. 
For the distribution function $F_{s, n}(\omega, \bm{k}_{\parallel}, \bm{R}_{\parallel}, z)$, 
the $z$ component of the TAM is
\begin{subequations}
 \begin{equation}
 J^z_{s,n} (\omega, \bm{k}_{\parallel}, \bm{R}_{\parallel})\equiv S^z_{s, n}(\omega, \bm{k}_{\parallel})
 + L^z(\bm{k}_{\parallel}, \bm{R}_{\parallel}).
 \end{equation}
 Here $S^z_{s, n}(\omega, \bm{k}_{\parallel})$ is equivalent to $S^z_{\bm{k}n}$.
 Likewise, the TAM carried by the distribution function $f_{s,n}(\omega, \bm{q}_{\parallel}, \bm{r}_{\parallel}, z)$ is given by
 \begin{equation}
 \widetilde{J}^z_{s,n} (\omega, \bm{q}_{\parallel}, \bm{r}_{\parallel})
 \equiv \widetilde{S}^z_{s, n}(\omega, \bm{q}_{\parallel}) + \widetilde{L}^z(\bm{q}_{\parallel}, \bm{r}_{\parallel}).
 \end{equation}
\end{subequations}

As the final assumption, we suppose that in the directions parallel to the interface, namely the $x$ and $y$ directions, 
the nonequilibrium state of phonons is spatially uniform when viewed at a resolution of the order of the phonon mean free path.
Accordingly, we often perform coarse-graining in the $x$ and $y$ directions for the distribution functions $F_{s, n}(x, y, z < 0)$ and $f_{s, n}(x, y, z > 0)$
over a length scale comparable to the mean free path,
and refer to the resulting functions as ${F}_{s, n}(z < 0)$ and ${f}_{s, n}(z > 0)$.

\subsection{Boundary conditions for phonon distribution functions}
\label{subsec: b.c. summa}

We impose two types of boundary conditions and use one or the other depending on the physical quantity of interest.

\subsubsection{Coarse-grained boundary conditions}

First, we select a physical quantity whose definition is independent of spatial coordinates, e.g., SAM. 
Such a quantity can be coarse-grained over a length scale comparable to the mean free path.
Most macroscopic quantities of practical interest satisfy this condition.
To analyze interfacial transport of such a quantity, 
we apply the following boundary condition to the phonon distribution functions, coarse-grained in the $x$ and $y$ directions 
($n = \text{L}, \text{RH}, \text{LH}$): 
\begin{widetext}
\begin{equation}
\begin{bmatrix}
 {F}_{-, n}(\omega, \bm{k}_{\parallel}, z = 0) \\[8pt] 
{f}_{+, n}(\omega, \bm{k}_{\parallel}, z = 0)
\end{bmatrix} 
= \sum_{m = \text{L}, \text{RH}, \text{LH}}
\begin{bmatrix}
 \mathcal{R}_{nm}(\omega, \bm{k}_{\parallel}) & \mathcal{T}'_{nm}(\omega, \bm{k}_{\parallel}) \\[8pt]
 \mathcal{T}_{nm}(\omega, \bm{k}_{\parallel}) & \mathcal{R}'_{nm}(\omega, \bm{k}_{\parallel})
\end{bmatrix}
\begin{bmatrix}
 {F}_{+, m}(\omega, \bm{k}_{\parallel}, z = 0) \\[8pt] {f}_{-, m}(\omega, \bm{k}_{\parallel}, z = 0)
\end{bmatrix}.
\label{eq: boundary condition assump}
\end{equation} 
\end{widetext}
Here $\mathcal{R}_{nm}$, $\mathcal{T}_{nm}$, $\mathcal{R}'_{nm}$, and $\mathcal{T}'_{nm}$ represent the classical power reflectance and transmittance.
They are defined in Appendix~\ref{sec: elastic plane wave}, 
and their dependence on the angle of incidence for several parameter sets is presented in Supplemental Material~\cite{SupplementalMaterialFullPaper}.
Specifically, the energy of a wave in mode $m$ incident from the region $z < 0$ is partially transferred into a reflected wave in mode $n$ 
in the same region 
with fraction $\mathcal{R}_{nm}$, and into a transmitted wave in mode $n$ in the region $z > 0$ with fraction $\mathcal{T}_{nm}$.
Similarly, for a wave incident from the region $z > 0$ in mode $m$, a fraction $\mathcal{R}'_{nm}$ of its energy is reflected into mode $n$
and a fraction $\mathcal{T}'_{nm}$ is transmitted into mode $n$.
Equation~\eqref{eq: boundary condition assump} thus linearly relates the phonon numbers incident on the interface, 
${F}_{+, m}$ and ${f}_{-, m}$ (right-hand side), 
to the numbers after scattering at the interface, ${F}_{-, n}$ and ${f}_{+, n}$ (left-hand side).

\subsubsection{Detailed boundary conditions}

For a system consisting of many particles or many wave packets, some physical quantities of interest are defined explicitly in terms of spatial coordinates. 
A representative example is the mechanical AM, 
$\sum_i (\bm{r}_i \times \bm{p}_i)$, defined for particles with positions and momenta $(\bm{r}_i, \bm{p}_i)$. 
If we interpret the particles as phonon wave packets, and the positions and momenta as those of their centers, 
this mechanical AM corresponds directly to the phonon OAM, given in Eqs.~\eqref{eq: OAM definition Lz} and \eqref{eq: OAM definition tildeL z}.

When spatial coarse-graining is applied, the mechanical AM is typically approximated as 
$\sum_i \braket{\bm{r}} \times\bm{p}_i$, where $\braket{\bm{r}}$ is the average of the positions $\{\bm{r}_i\}$. 
However, this procedure generally alters the AM itself.

To describe the transport of phonon OAM, we must retain the $x$ and $y$ dependence of the distribution function 
with spatial resolution of the order of the wavelength; coarse-graining would obscure essential information. 
Accordingly, more detailed boundary conditions are required than those given in Eq.~\eqref{eq: boundary condition assump}. 
These refined boundary conditions are given as follows (for $n = \text{L}, \text{RH}, \text{LH}$):
\begin{subequations}
\begin{widetext}
\begin{equation}
 \begin{bmatrix}
 F_{-, n}(\omega, \bm{k}_{\parallel}, \bm{R}_{\parallel, -, n}, z = 0)\\[8pt]  
f_{+, n}(\omega, \bm{k}_{\parallel}, \bm{r}_{\parallel, +, n}, z = 0)
  \end{bmatrix}
= \sum_{m = \text{L}, \text{RH}, \text{LH}}
\begin{bmatrix}
 \mathcal{R}_{nm}(\omega, \bm{k}_{\parallel}) & \mathcal{T}'_{nm}(\omega, \bm{k}_{\parallel}) \\[8pt]
 \mathcal{T}_{nm}(\omega, \bm{k}_{\parallel}) & \mathcal{R}'_{nm}(\omega, \bm{k}_{\parallel})
\end{bmatrix}
\begin{bmatrix}
 F_{+, m}(\omega, \bm{k}_{\parallel}, \bm{R}_{\parallel, +, m}, z= 0)\\[8pt]
 f_{-, m} (\omega, \bm{k}_{\parallel}, \bm{r}_{\parallel, -, m}, z = 0)
\end{bmatrix}.\label{163324_9Jun25}
\end{equation}
This equation holds only for phonon wave packets with equal TAM.
In other words, the following additional constraint on
the positions $\bm{R}_{\parallel, s, n}$ and $\bm{r}_{\parallel, s, n}$ must be satisfied in Eq.~\eqref{163324_9Jun25}~\footnote{
In addition to the transverse shift, 
a shift along the incident plane, known as the longitudinal or Schoch shift~\cite{Schoch1950}, can also occur in elastic wave packet scattering. 
We neglect this in the present study for simplicity.}:
\begin{equation}
J^z_{-,n} (\omega, \bm{k}_{\parallel}, \bm{R}_{\parallel, -, n})
= \widetilde{J}^z_{+, n} (\omega, \bm{k}_{\parallel}, \bm{r}_{\parallel, +, n}) 
= J^z_{+,m} (\omega, \bm{k}_{\parallel}, \bm{R}_{\parallel, +, m})
= \widetilde{J}^z_{-, m} (\omega, \bm{k}_{\parallel}, \bm{r}_{\parallel, -, m}). \label{163328_9Jun25}%
\end{equation}
\end{widetext}
\end{subequations}
Equations~\eqref{163324_9Jun25} and \eqref{163328_9Jun25} represent the main result of this paper. 
Note that the boundary condition given in Eq.~\eqref{163324_9Jun25} reduces to the original boundary condition~\eqref{eq: boundary condition assump} 
when coarse-grained on the scale larger than the wavelength.

\subsection{Generation of OAM}

Using the detailed boundary conditions in Eqs.~\eqref{163324_9Jun25} and \eqref{163328_9Jun25}, 
we can describe the OAM generation and its interfacial transport. 
As an illustrative setup, we consider a crystal in the region $z<0$ under a temperature gradient and a crystal in the region $z>0$ without external bias.  

\begin{figure}[tbp]
\centering
\includegraphics[pagebox=artbox,width=0.99\columnwidth]{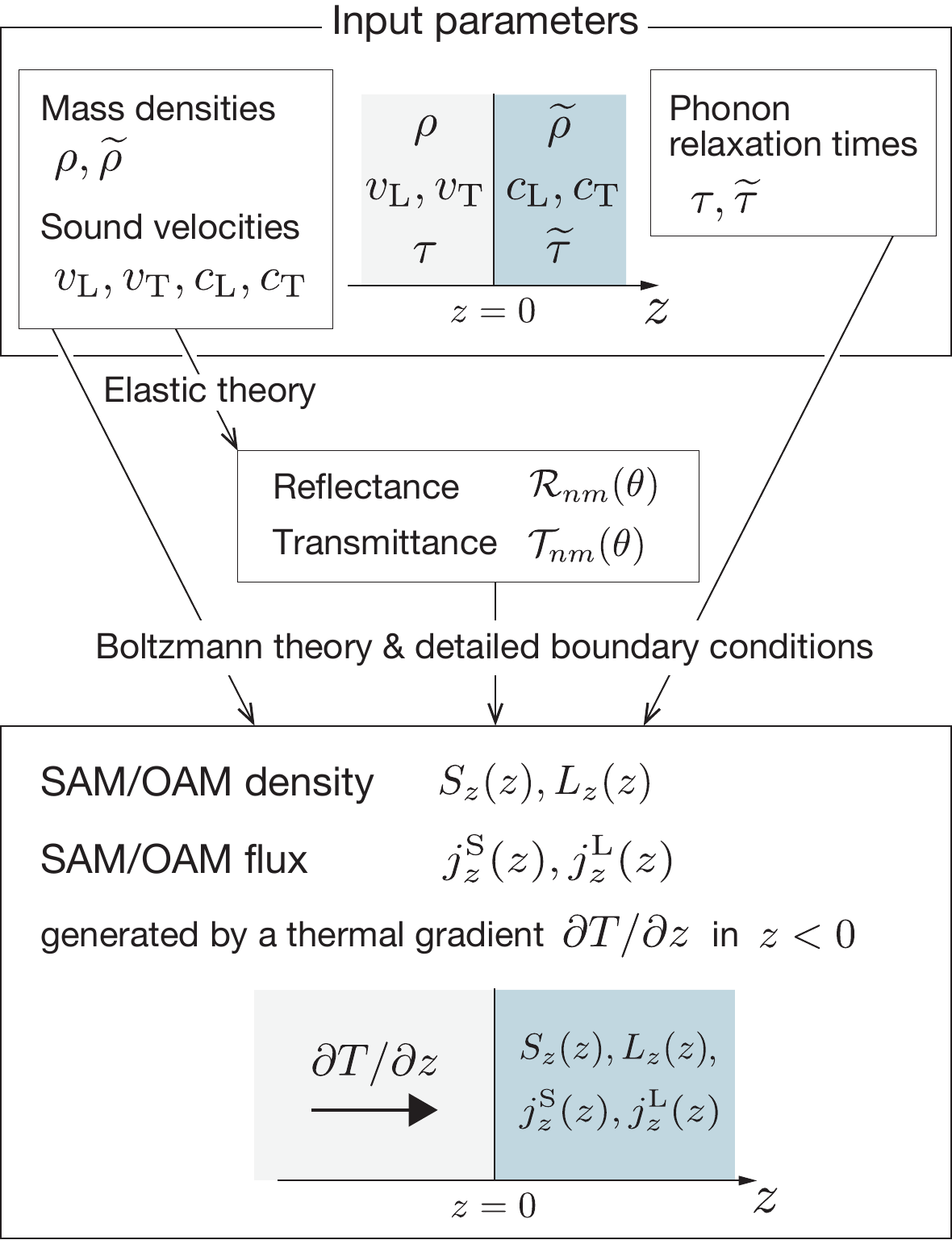}
\caption{{Workflow for calculating interfacial phonon SAM and OAM from material parameters.}}
\label{002240_29Jun26}
\end{figure}
\begin{figure*}
 \centering
\includegraphics[pagebox=artbox,width=0.99\textwidth]{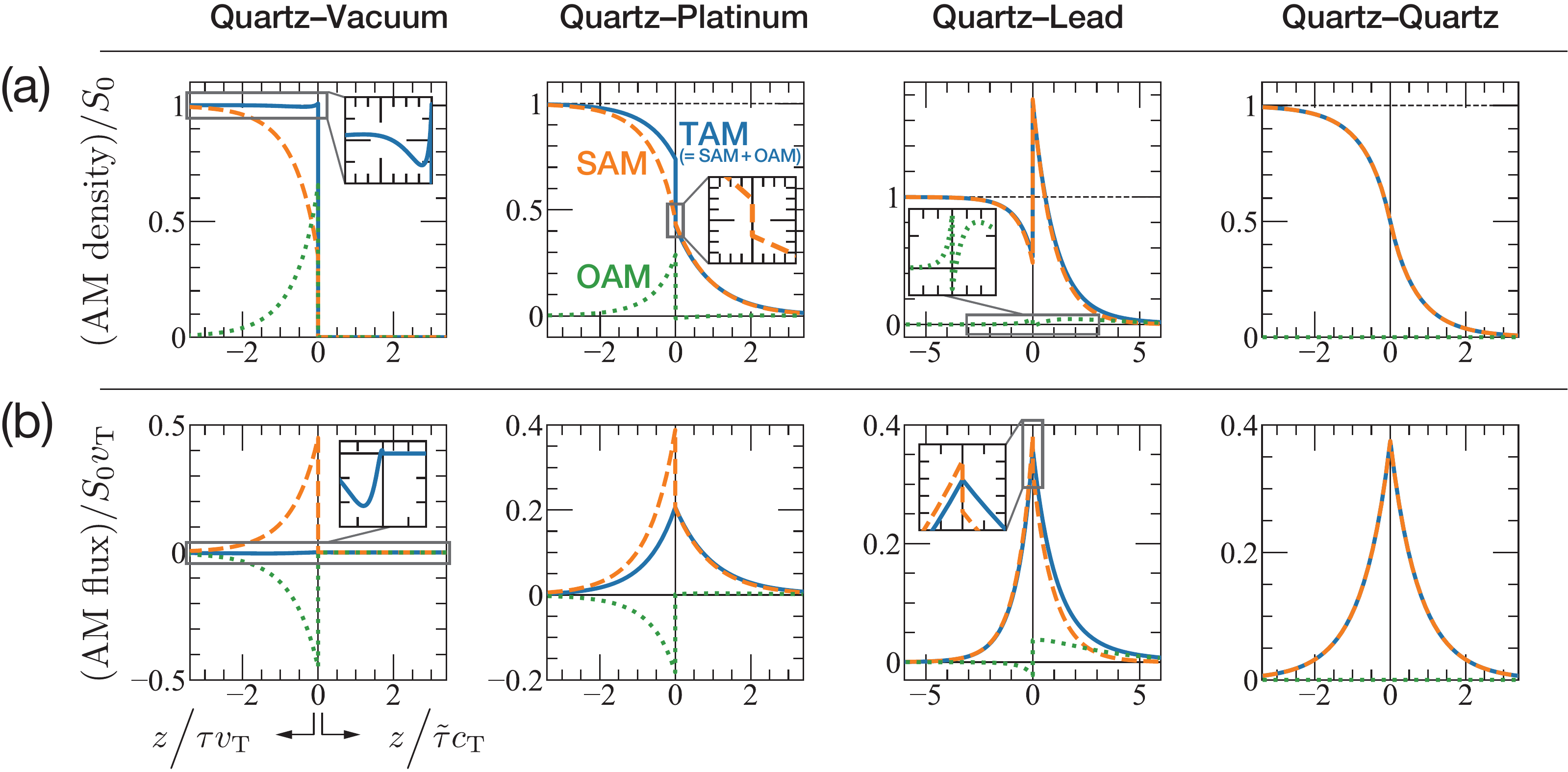}
\caption{
Spatial profile of phonon angular momentum (AM), including spin (SAM), orbital (OAM), and total (TAM) components, at four different interfaces.  
The interface separates a chiral quartz crystal at $z < 0$ under a thermal gradient $\partial T/\partial z$ from an achiral crystal at $z > 0$ without external driving. 
{For reference, the rightmost panels 
show the idealized Quartz--Quartz case, in which a thermal gradient is applied only to the chiral quartz at $z<0$, 
while the achiral medium at $z>0 $ has the same isotropic elastic parameters as quartz.}
(a)~AM densities: $ S_z $ (SAM), $ L_z $ (OAM), and $ J_z = S_z + L_z $ (TAM).  
(b)~AM fluxes: $ j^{\text{S}}_z $ (SAM), $ j^{\text{L}}_z $ (OAM), and $ j^{\text{J}}_z = j^{\text{S}}_z + j^{\text{L}}_z $ (TAM).  
{The insets enlarge the following regions: (a) Quartz--Vacuum, $[-3.5,0]\times[0.99,1.01]$; Quartz--Platinum, $[-0.02,0.02]\times[0.42,0.46]$; Quartz--Lead, $[-3,3]\times[-0.03,0.06]$. (b) Quartz--Vacuum, $[-3.5,3.5]\times[-0.006,0.001]$; Quartz--Lead, $[-0.2,0.2]\times[0.3,0.4]$.}
AM densities and fluxes are normalized by the bulk SAM density $ S_0 $ and transverse-wave velocity $ v_{\text{T}} $ in the quartz.  
We set parameters as $v_{\text{L}}/v_{\text{T}} = 1.59$ and 
[Quartz--Vacuum]~$\widetilde{\rho} c_{\text{T}}/\rho v_{\text{T}} = 0$,
[Quartz--Platinum]~$c_{\text{L}}/c_{\text{T}} = 2.25$,
$c_{\text{T}}/v_{\text{T}} = 0.493$, and $\widetilde{\rho} c_{\text{T}}/\rho v_{\text{T}} = 3.99$,
[Quartz--Lead]~$c_{\text{L}}/c_{\text{T}} = 2.84$,
$c_{\text{T}}/v_{\text{T}} = 0.183$, and $\widetilde{\rho} c_{\text{T}}/\rho v_{\text{T}} = 0.947$,
[Quartz--Quartz]~$c_{\text{L}}/v_{\text{T}} = 1.59$ and 
$c_{\text{T}}/v_{\text{T}} = \widetilde{\rho} /\rho = 1$.
{The material parameters are taken from Ref.~\cite{Rikanempyo2024} within the isotropic elastic approximation.}
The results of the SAM in (a) and (b) are identical to Fig.~2(a) in our companion {Letter}~\cite{SuzukiSumitaKato2024a}. 
{All profiles are representative results obtained within 
the population-imbalance approximation described 
in our companion Letter~\cite{SuzukiSumitaKato2024a}.}
}
\label{fig: AM spatial distribution}
\end{figure*}

{Figure~\ref{002240_29Jun26} summarizes the overall procedure for calculating the interfacial SAM and OAM generated by the temperature gradient $\partial T/\partial z$. 
The required material parameters are the mass densities $\rho$ and $\widetilde{\rho}$, the longitudinal and transverse sound velocities 
$(v_{\mathrm{L}},v_{\mathrm{T}})$ and $(c_{\mathrm{L}},c_{\mathrm{T}})$, and the phonon relaxation times $\tau$ and $\widetilde{\tau}$ in the two crystals.
From these inputs, the elastic boundary conditions determine the interfacial power reflectance $\mathcal{R}_{nm}$ and transmittance $\mathcal{T}_{nm}$
(see Appendix~\ref{sec: elastic plane wave}). 
These coefficients, together with the Boltzmann theory and the detailed boundary conditions, 
yield analytical expressions for the SAM density and flux [Eqs.~\eqref{201937_28Jan25}--\eqref{110137_30Sep25}], 
and for the OAM density and flux [Eqs.~\eqref{201219_7Oct25}--\eqref{201139_7Oct25}]. 
{The explicit profiles are evaluated using the population-imbalance approximation of our companion Letter~\cite{SuzukiSumitaKato2024a}, which retains the chirality-induced RH–LH population imbalance but does not include all first-order effects of the mode splitting.}
Figures~\ref{fig: AM spatial distribution}(a) and \ref{fig: AM spatial distribution}(b) present representative results for four interfaces derived using this procedure.}
These panels summarize the results derived in Sec.~\ref{sec: Generation of orbital angular momentum at the interface}.

The temperature gradient induces a net SAM in the crystal at $z<0$, particularly when it is chiral~\cite{Hamada2018,Zhang2025}.
We find that this SAM diffuses toward the adjacent crystal, and its flux exhibits a discontinuity at $z=0$ (see dashed lines in Fig.~\ref{fig: AM spatial distribution}(b)).
The detailed boundary conditions ensure that this jump in the SAM flux is compensated by an opposite jump in the OAM flux. This compensation generates finite OAM at the interface; the OAM then diffuses into the bulk, as shown by the dotted lines in Figs.~\ref{fig: AM spatial distribution}(a) and \ref{fig: AM spatial distribution}(b). 
The solid lines in Fig.~\ref{fig: AM spatial distribution}(b) demonstrate the continuity of the TAM flux across the interface.

\section{Derivation of the coarse-grained boundary conditions}
\label{sec: lattice vibration ph bc}

We derive the boundary conditions presented in Eq.~\eqref{eq: boundary condition assump}. 
Let us start with the comparison between elastic plane waves and lattice vibrations; 
the former propagate in an infinite medium, while the latter are confined to finite crystals in contact.
From this comparison, we formulate boundary conditions for lattice vibrations at the interface.  
The key idea for the formulation is that 
a wave packet composed of an acoustic mode with nearly identical wavevectors and frequencies can be approximated by a monochromatic plane wave in an elastic medium.  
We then reinterpret the boundary conditions for lattice vibrations as constraints on phonon excitations in the crystal junction.  
These constraints determine the boundary conditions for phonon distribution functions at the interface.

\subsection{Elastic plane waves}

\subsubsection{Setup}

We consider two elastic media in contact, as illustrated in Fig.~\ref{fig: incidentplane}.
The entire system oscillates at a single frequency $\omega$.
For the wavevectors $\bm{k}$ and $\bm{q}$, the equality $\bm{k}_{\parallel} = \bm{q}_{\parallel}$ holds due to translational symmetry of the interface.

We focus first on the region $z < 0$ and write the mass density as $\rho$.
To simplify the analysis, we assume the medium is isotropic.
Then, the three acoustic modes exhibit the following linear dispersion relations:
\begin{subequations}
 \begin{equation}
 \Omega_{\bm{k}, \text{L}} = v_{\text{L}}|\bm{k}|,
 \end{equation}
and
 \begin{equation}
 \Omega_{\bm{k}, \text{RH}}, \Omega_{\bm{k}, \text{LH}} \simeq \Omega_{\bm{k}, \text{T}} = v_{\text{T}}|\bm{k}|,
 \end{equation}
\end{subequations}
corresponding to the longitudinal (L) mode and two degenerate transverse modes (RH and LH), respectively.
The sound velocities for the longitudinal and transverse waves are denoted by $v_{\text{L}}$ and $v_{\text{T}}$.
We introduce a rotational basis for the transverse modes (see also Eq.~\eqref{eq: L SH SV polarization vec} and the subsequent discussion).  
The polarization vectors are defined as
\begin{equation}
\bm{e}_{\bm{k}, \text{L}} \equiv \frac{\bm{k}}{|\bm{k}|}, \quad
\bm{e}_{\bm{k}, \text{RH}} = \bm{e}^{*}_{\bm{k}, \text{LH}} \equiv 
\frac{
(\hat{\bm{z}}\times \bm{k})\times \frac{\bm{k}}{\Abs{\bm{k}}} + \iu \hat{\bm{z}}\times \bm{k}
}{\sqrt{2 (|\bm{k}|^2 - k_z^2)}}.
\label{eq: basis L RH LH}
\end{equation}
The unit vector $\hat{\bm{z}}$ points in the positive $z$ direction.

For a given frequency $\omega$, in-plane wavevector $\bm{k}_{\parallel}$, and mode $n = \text{L},~\text{RH},~\text{LH}$, 
the dispersion relation $\omega = \Omega_{\bm{k}, n} = v_n \sqrt{|\bm{k}_{\parallel}|^2 + k_z^2}$
yields two possible values of $k_z$, denoted as
\begin{align}
&k^z_{s,n}(\omega, \bm{k}_{\parallel}) \equiv s \sqrt{\frac{\omega^2}{v_n^2} - \abs{\bm{k}_{\parallel}}^2} \nonumber\\
&= 
\begin{cases}
\displaystyle\frac{s\omega\cos\theta_{\text{L}}}{v_{\text{L}}} & n=\text{L}\\[10pt]
\displaystyle \frac{s\omega\cos\theta_{\text{T}}}{v_{\text{T}}} & n=\text{RH},\,\text{LH}
\end{cases} 
\label{142550_29Jun26}
\end{align}
with $s = \pm$. 
As illustrated in Fig.~\ref{fig: incidentplane}, we define the angles of incidence, reflection, and refraction as
$\sin\theta_{\text{L}} = v_{\text{L}}|\bm{k}_{\parallel}|/\omega$
and $\sin\theta_{\text{T}} = v_{\text{T}}|\bm{k}_{\parallel}|/\omega$, 
corresponding to the longitudinal and transverse modes, respectively.
We also rewrite the polarization vectors $\bm{e}_{\bm{k}, n}$ as 
$\bm{e}_{\bm{k}, n} = \bm{e}_{s, n}(\omega, \bm{k}_{\parallel})$, labeled by the index pair $(s, n)$.

We now describe the plane wave of mode $n$ propagating along the $s \, \hat{\bm{z}}$ direction ($s = \pm$) 
with amplitude $u_{s, n}$ in the region $z < 0$. 
The displacement field corresponding to this wave is expressed as
\begin{equation}
\bm{u}^{\text{ela}}_{s, n} (\bm{R}_{\parallel}, z, t)
=  u_{s, n} \bm{e}_{s, n} e^{ \iu\left(\bm{k}_{\parallel} \cdot \bm{R}_{\parallel} + k^z_{s,n} z - \omega t\right)} + \mathrm{c.c.}
\label{eq: plane wave s n}
\end{equation}
Here ``c.c.'' denotes the complex conjugate of all preceding terms. 
We omit the arguments $(\omega, \bm{k}_{\parallel})$ on both sides for simplicity.

Similarly, we characterize the elastic plane wave in the region $z > 0$ by the propagation direction $s = \pm$ and the mode 
$n = \text{L},~\text{RH},~\text{LH}$.
The corresponding parameters are summarized as follows:
mass density $\widetilde{\rho}$, dispersion relation $\omega = \omega_{\bm{q}, n} = c_n |\bm{q}|$,
longitudinal and transverse sound velocities $c_{\text{L}}$ and $c_{\text{T}}$,
angles $\Theta_{\text{L}}$ and $\Theta_{\text{T}}$ (see Fig.~\ref{fig: incidentplane}),
polarization vector $\widetilde{\bm{e}}_{s, n}$, 
and amplitude $\widetilde{u}_{s, n}$ of each wave component.

\subsubsection{
\texorpdfstring{Energy flux and $S$-matrix}{Energy flux and S-matrix}
}

We now evaluate the energy flux normal to the interface, per unit area, 
associated with the plane wave~\eqref{eq: plane wave s n}.
This flux is obtained by taking the long-time average of the mechanical power transmitted by the wave across the interface.
Following Ref.~\cite{Synge1956}, the result reads
\begin{equation}
j^{\text{ela}}_{\mathfrak{u}; s, n} \equiv 2s \omega^2 \rho v_n \cos\theta_n \Abs{u_{s, n}}^2, 
\label{eq: ene fl}
\end{equation}
where the subscript $\mathfrak{u}$ indicates that the quantity corresponds to energy transport.
The energy flux in the region $z > 0$ takes the same form, with parameters replaced by those of the medium in $z >0$: 
$2s \omega^2 \widetilde{\rho} c_n \cos\Theta_n |\widetilde{u}_{s, n}|^2$.

In the case of perfect contact,
the wave amplitudes $u_{s,n}$ and $\widetilde{u}_{s,n}$ on either side of the interface are related through a scattering matrix ($S$-matrix) as
\begin{equation}
Z^{1/2}
 \left[
\renewcommand\arraystretch{1.3}
\begin{array}{c}
u_{-,\text{L}}\\
u_{-,\text{RH}}\\
u_{-,\text{LH}}\\ \hline
\widetilde{u}_{+,\text{L}}\\
\widetilde{u}_{+,\text{RH}}\\
\widetilde{u}_{+,\text{LH}}
\end{array} 
\right]
= S (\omega, \bm{k}_{\parallel})\, Z^{1/2}
 \left[
\renewcommand\arraystretch{1.3}
\begin{array}{c}
u_{+,\text{L}}\\
u_{+,\text{RH}}\\
u_{+,\text{LH}}\\ \hline
\widetilde{u}_{-,\text{L}}\\
\widetilde{u}_{-,\text{RH}}\\
\widetilde{u}_{-,\text{LH}}
\end{array} 
\right].
\label{eq: S matrix in circular pol 1}
\end{equation}
The derivation is provided in Appendix~\ref{sec: elastic plane wave}. 
Here $Z$ is a diagonal matrix containing acoustic impedances: 
\begin{align}
 Z &= \mathrm{diag}\left[ 
\rho v_{\text{L}}\cos\theta_{\text{L}},~
\rho v_{\text{T}}\cos\theta_{\text{T}},~
\rho v_{\text{T}}\cos\theta_{\text{T}},\right.\nonumber\\
&\quad\qquad \left. \widetilde{\rho} c_{\text{L}}\cos\Theta_{\text{L}},~
\widetilde{\rho} c_{\text{T}}\cos\Theta_{\text{T}},~
\widetilde{\rho} c_{\text{T}}\cos\Theta_{\text{T}}
\right].
\end{align}
The power reflectance and transmittance of elastic waves are obtained by squaring the elements of $S(\omega, \bm{k}_{\parallel})$.

\subsection{Lattice vibrations in the elastic limit}

We next consider lattice vibrations in a junction system composed of two crystals, as illustrated in Fig.~\ref{fig: incidentplane}, 
in the low-energy, long-wavelength limit. 
In this regime, the lattice deformation $\bm{u}(\bm{r}, t)$ in the bulk region $z < 0$ can be expressed as a superposition of plane waves:
\begin{align}
\bm{u}(\bm{r}, t) & = \sum_{\bm{k}} \sum_{n = \text{L}, \text{RH}, \text{LH}} 
e^{\iu(\bm{k} \cdot \bm{r} - \Omega_{\bm{k}n} t)} 
\sqrt{\frac{\hbar}{2 \rho V \Omega_{\bm{k}n}}} \, a_{\bm{k}n} \,
\bm{e}_{\bm{k}n} \nonumber\\
& + \text{c.c.} \label{172803_2Aug25}
\end{align}
Here $a_{\bm{k}n}$ denotes the dimensionless amplitude of the vibrational state labeled by $(\bm{k}, n)$. 
Microscopic details, such as the lattice constant and sublattice structure, are averaged into macroscopic quantities: 
the mass density $\rho$, the volume $V$ of the crystal in $z < 0$, and the phonon dispersions $\Omega_{\bm{k} n}$. 
We treat $a_{\bm{k}n}$ as a classical number until Sec.~\ref{subsec: b.c. for phonon distrib}.

For notational convenience, we introduce an auxiliary amplitude
\begin{equation}
A^{\text{lat}}_{\bm{k}n} \equiv \sqrt{\frac{\hbar}{2 \rho \Omega_{\bm{k}n}}} \, a_{\bm{k}n},
\label{eq: correspondence between u and a}
\end{equation}
which allows us to rewrite the lattice deformation as
\begin{equation}
\bm{u}(\bm{r}, t) = \sum_{\bm{k}', n} 
\frac{e^{\iu(\bm{k}' \cdot \bm{r} - \Omega_{\bm{k}'n} t)}}{\sqrt{V}} \,
A^{\text{lat}}_{\bm{k}'n} \, \bm{e}_{\bm{k}'n} + \text{c.c.}
\label{eq: sum of vibrational states}
\end{equation}
A similar expansion holds for the lattice deformation in the bulk region $z > 0$. 
We write the dimensionless amplitude of the state $(\bm{q}, n)$ as $\widetilde{a}_{\bm{q}n}$ and define
\begin{equation}
\widetilde{A}^{\text{lat}}_{\bm{q}n}
\equiv \sqrt{\frac{\hbar}{2\widetilde{\rho} \omega_{\bm{q}n}}} \, \widetilde{a}_{\bm{q}n}.
\label{eq: correspondence between tilde u and tilde a}
\end{equation}

\subsection{Wave packet reflection and transmission}  

{\subsubsection{Boundary conditions based on wave packets}}

To describe the transfer of lattice vibrations across the interface, we formulate the boundary conditions using \emph{wave packets}. 
We consider a wave packet centered at a frequency $\omega$ and an in-plane wavevector $\bm{k}_{\parallel}$, constructed from a finite set of vibrational states with discrete frequencies and wavevectors. 
Let $\Delta \omega$ denote the frequency width, and let $\Delta^2 \bm{k}_{\parallel} = \Delta k_x \Delta k_y$ denote the spread in the in-plane wavevector components.
In the limit where $\Delta\omega \to 0$ 
and $\Delta k_x, \Delta k_y \to 0$, the wave packet becomes indistinguishable from a plane wave with frequency $\omega$ and in-plane wavevector $\bm{k}_{\parallel}$; the corresponding energy flux and reflectance/transmittance at the interface coincide with those of that plane wave. 
We use this correspondence as the basis for deriving the boundary conditions. 

The wave-packet formulation offers a clear way to handle the mismatched discrete spectra of lattice vibrations in \emph{finite-sized} crystals.
In contrast to elastic waves in infinite media, 
each crystal has discrete vibrational spectra, given by $\bm{k}$ and $\bm{q}$ and by $\Omega_{\bm{k}n}$ and $\omega_{\bm{q}n}$, and these bulk spectra generally do not match between the two crystals.  
Because of this mismatch, we cannot identify a set of bulk vibrational states on which to impose boundary conditions.  
A wave packet resolves this difficulty by spanning a finite range of frequencies and wavevectors, $\Delta\omega\,\Delta^2\bm{k}_{\parallel}$;  
within this range, we can consistently relate the vibrational states on both sides of the interface.  
This construction provides a practical description of the interfacial transfer of energy and momentum.

An alternative approach is to study the coupled vibrations at the microscopic level near the interface.
When two crystals are brought into contact, the normal modes in each crystal are modified so that a common frequency $\omega$ and in-plane wavevector $\bm{k}_{\parallel}$ appear. 
Although this method is conceptually appealing, it requires detailed modeling of the interface and substantial computational effort; 
furthermore, the reflection and transmission of long-wavelength modes are largely insensitive to such microscopic details~\cite{Zhang2011}.  
In contrast, the wave-packet formulation provides a clear and practical framework to study interfacial transfer, without relying on detailed microscopic modeling, as shown
below.

{\subsubsection{Wave-packet construction}}

We divide the summation over all vibrational states~\eqref{eq: sum of vibrational states} into contributions from individual wave packets.
Each wave packet is composed of a mode $n$ propagating in the $s\, \hat{\bm{z}}$ direction ($s = \pm$) in the region $z<0$.
It consists of a superposition of plane waves centered around a frequency $\omega$ and an in-plane wavevector $\bm{k}_{\parallel}$.
We assume that the amplitude of the lattice vibrations within the wave packet is uniformly distributed over the frequency range 
$[\omega -\Delta \omega/2 ,~\omega+\Delta \omega/2]$ and over an area $\Delta^2 \bm{k}_{\parallel}$ centered at $\bm{k}_{\parallel}$.
This uniformity assumption holds when the total width $\Delta \omega \Delta^2 \bm{k}_{\parallel}$ is sufficiently small.
{The particular choice of a rectangular frequency window is made only for convenience and does not restrict the generality of the formulation.}

In summary, the displacement field $\bm{w}_{s,n}$ of the wave packet and its amplitude are expressed as
\begin{subequations}
 \begin{gather}
 \bm{w}_{s, n}(\omega, \bm{k}_\parallel; \bm{r}, t)
 = \sum_{\bm{k}'}^{ s\, k'_z > 0}
\frac{e^{\iu(\bm{k}'\cdot\bm{r}-\Omega_{\bm{k}'n}t)}}{\sqrt{V}} A^{\text{w.p.}}_{\bm{k}'n} 
 \bm{e}_{\bm{k}'n} + \text{c.c.},\label{eq: uniformly distributed wave packet a}\\
 A^{\text{w.p.}}_{\bm{k}'n} = \sqrt{\mathcal{G} (\Omega_{\bm{k}'n}-\omega, \bm{k}'_{\parallel} - \bm{k}_{\parallel})}\, u_{\bm{k}n}, \label{eq: uniformly distributed wave packet b} 
 \end{gather}
where the condition $s\, k'_z > 0$ restricts the range of the $\bm{k}'$ summation.
We define the amplitude
\begin{equation}
 u_{\bm{k}n} = u_{s,n}(\omega, \bm{k}_{\parallel})\label{eq: ukn usn}
\end{equation}
as that of the elastic plane wave in Eq.~\eqref{eq: plane wave s n}. 
The wave packet approaches this plane wave asymptotically in the limit where 
$\Delta \omega \to 0$ and $\Delta k_x, \Delta k_y \to 0$.
The envelope function $\mathcal{G}$ for each wave packet is defined as
 \begin{equation}
 \mathcal{G} (\delta\omega, \delta\bm{k}_{\parallel}) \equiv 
 \begin{cases}
 g(\omega, \bm{k}_{\parallel}) 
 & \max\left\{ 
 \frac{\Abs{\delta\omega}}{\Delta \omega },
 \frac{\Abs{\delta {k}_{x}}}{\Delta {k}_{x}},
 \frac{\Abs{\delta {k}_{y}}}{\Delta {k}_{y}}
 \right\} \leq \frac{1}{2} \\
 0 & \text{otherwise}
 \end{cases}, \label{eq: uniformly distributed wave packet d}
 \end{equation}
\label{eq: uniformly distributed wave packet}%
\end{subequations}
where $g(\omega, \bm{k}_{\parallel})$ is a normalization factor introduced for convenience and 
is determined by matching the energy flux of the wave packet to that of the plane wave.

{\subsubsection{Wave-packet energy flux}}

We now evaluate the energy flux per unit area across the interface, carried by the wave packet. 
Let $j^{\mathrm{w.p.}}_{\mathfrak{u}; s, n}$ denote the energy flux. 
{According to Ref.~\cite{Synge1956}, 
$j^{\mathrm{w.p.}}_{\mathfrak{u}; s, n}$ is given by the mechanical power exerted by the wave packet on the interface, namely, 
the negative of the product of the displacement velocity
$\frac{\partial }{\partial t} w_{j} = \frac{\partial}{\partial t} \left[\bm{w}_{s,n}(\omega, \bm{k}_{\parallel}; \bm{r}, t)\right]_j$
and the stress acting on the wave packet at the plane $z=0$,
$\sigma_{zj} = \left[\sigma_{s,n}(\omega, \bm{k}_{\parallel}; \bm{r}, t)\right]_{zj}$: 
\begin{equation}
j^{\mathrm{w.p.}}_{\mathfrak{u}; s, n} =  - \int\frac{\rmd x \rmd y}{\varSigma} \, \sum_{j = x, y, z} \left. \overline{
\sigma_{zj}\frac{\partial w_{j}}{\partial t}}\right|_{z=0}.
\label{093633_5Jun26}
\end{equation}
Here, $\varSigma$ is the area of the interface. 
The overline denotes the long-time average, defined by 
\begin{equation}
 {\overline{O(t)}} \equiv \lim_{T \to \infty} \frac{1}{T} \int^T_0 O(t)\, \rmd t.
\label{eq: l time average}
\end{equation}}

{Next, we substitute the constitutive relation for an isotropic elastic medium~\cite{LandauLifshitzTextbookVol7,Bliokh2006}, 
\begin{equation}
\sigma_{ij} = \rho (v_{\text{L}}^2 - 2 v_{\text{T}}^2) \delta_{ij} \sum_{l = x, y, z} \frac{\partial w_l}{\partial r_l}
+ \rho v_{\text{T}}^2\left(\frac{\partial w_j}{\partial r_i} + \frac{\partial w_i}{\partial r_j}\right),
\label{062712_5Jun26}
\end{equation}
into Eq.~\eqref{093633_5Jun26}. 
For arbitrary quantities $P_{\bm{k}}$ and $Q_{\bm{k}}$ that depend only on $\bm{k}$, we use the identity
\begin{widetext}
\begin{align}
&\int\frac{\rmd x \rmd y}{\varSigma} \, \left.\overline{
\left[\sum_{\bm{k}'}^{ s\, k'_z > 0} \frac{e^{\iu(\bm{k}'\cdot\bm{r}-\Omega_{\bm{k}'n}t)}}{\sqrt{V}} P_{\bm{k}'} + \text{c.c.}\right]
\left[\sum_{\bm{k}''}^{ s\, k''_z > 0} \frac{e^{\iu(\bm{k}''\cdot\bm{r}-\Omega_{\bm{k}''n}t)}}{\sqrt{V}} Q_{\bm{k}''} + \text{c.c.}\right]} \right|_{z=0} \nonumber\\
&= \sum_{\bm{k}'}^{ s\, k'_z > 0}\sum_{\bm{k}''}^{ s\, k''_z > 0} 
\frac{P_{\bm{k}'} Q^*_{\bm{k}''}}{V}\int\frac{\rmd x \rmd y}{\varSigma} \, \left. e^{\iu(\bm{k}'-\bm{k}'')\cdot\bm{r}}\right|_{z=0}
\lim_{T\to\infty} \int^T_0\frac{\rmd t}{T} e^{- \iu(\Omega_{\bm{k}'n} - \Omega_{\bm{k}''n})t} + \text{c.c.}\nonumber\\
&= \sum_{\bm{k}'}^{ s\, k'_z > 0}\sum_{\bm{k}''}^{ s\, k''_z > 0} 
\frac{P_{\bm{k}'} Q^*_{\bm{k}''}}{V} \delta_{\bm{k}'_{\parallel}, \bm{k}''_{\parallel}} \delta_{\Omega_{\bm{k}'n}, \Omega_{\bm{k}''n} } + \text{c.c.} \nonumber\\
&= \frac{1}{V} \sum_{\bm{k}'}^{ s\, k'_z > 0} P_{\bm{k}'} Q^*_{\bm{k}'} + \text{c.c.}
\end{align}
\end{widetext}
As a result, the energy flux becomes
\begin{align}
& j^{\mathrm{w.p.}}_{\mathfrak{u}; s, n} = 
\frac{1}{V}\sum_{\bm{k}'}^{ s\, k'_z > 0} \rho\Omega_{\bm{k}'n} \Abs{A^{\mathrm{w.p.}}_{\bm{k}'n}}^2 \nonumber\\
&\times \left\{
(v_{\text{L}}^2 - 2v_{\text{T}}^2) e^z_{\bm{k}'n} \left(\bm{k}'\cdot \bm{e}^*_{\bm{k}'n}\right) \right.\nonumber\\
&\qquad \left. + v_{\text{T}}^2 \left[
k'_z \left(\bm{e}_{\bm{k}'n}\cdot \bm{e}^*_{\bm{k}'n}\right) + \left(\bm{k}'\cdot \bm{e}^*_{\bm{k}'n}\right) e^z_{\bm{k}'n} \right]\right\} + \text{c.c.} \nonumber\\
&= \frac{1}{V}\sum_{\bm{k}'}^{ s\, k'_z > 0} \rho \Omega_{\bm{k}'n} \Abs{A^{\mathrm{w.p.}}_{\bm{k}'n}}^2 k'_z \nonumber\\
& \times \bm{e}^{\dagger}_{\bm{k}'n}\left[
v_{\text{L}}^2 \frac{\bm{k}'\hat{\bm{z}}^{\top} }{k'_z}
+ v_{\text{T}}^2 \left(1- \frac{\bm{k}'\hat{\bm{z}}^{\top} }{k'_z}\right)\right] \bm{e}_{\bm{k}'n} + \text{c.c.}
\end{align}}

{Here, the matrices $\bm{k}'\hat{\bm{z}}^{\top}/k'_z$ and 
$1-\bm{k}'\hat{\bm{z}}^{\top}/k'_z$, constructed from the direct product of the vectors $\bm{k}'$ and $\hat{\bm{z}}$, are projection operators. 
Their images correspond to the one-dimensional subspace spanned by the longitudinal polarization vector 
$\bm{e}_{\bm{k}', \mathrm{L}} = \bm{k}'/\Abs{\bm{k}'}$ 
and the two-dimensional subspace spanned by the transverse polarizations orthogonal to it, respectively. 
Therefore,
\begin{equation}
 \bm{e}^{\dagger}_{\bm{k}'n}\left[v_{\text{L}}^2 \frac{\bm{k}'\hat{\bm{z}}^{\top} }{k'_z}
+ v_{\text{T}}^2 \left(1- \frac{\bm{k}'\hat{\bm{z}}^{\top} }{k'_z}\right)\right] \bm{e}_{\bm{k}'n}
= \bm{e}^{\dagger}_{\bm{k}'n} v_n^2 \bm{e}_{\bm{k}'n}.
\end{equation}
Using this relation together with $\bm{e}^{\dagger}_{\bm{k}'n} \bm{e}_{\bm{k}'n} = 1$, we obtain
\begin{align}
 & j^{\mathrm{w.p.}}_{\mathfrak{u}; s, n} = 
\frac{\rho v_n^2}{V}\sum_{\bm{k}'}^{ s\, k'_z > 0} \Omega_{\bm{k}'n} \Abs{A^{\mathrm{w.p.}}_{\bm{k}'n}}^2 k'_z + \text{c.c.} \nonumber\\
&= \frac{2\rho v_n^2}{V}\sum_{\bm{k}'}^{ s\, k'_z > 0} \Omega_{\bm{k}'n} k'_z 
\mathcal{G} (\Omega_{\bm{k}'n}-\omega, \bm{k}'_{\parallel} - \bm{k}_{\parallel}) \Abs{u_{\bm{k}n}}^2 \nonumber\\
&\simeq \frac{2s \omega^2 \rho v_n \cos\theta_n \Abs{u_{\bm{k}n}}^2}{V}
\sum_{\bm{k}'}^{s \, k'_z > 0} \mathcal{G}(\Omega_{\bm{k}'n}-\omega, \bm{k}'_{\parallel} - \bm{k}_{\parallel}).
\label{eq: energy flux by w.p. in finite vol}
\end{align}
Here, the summation range is restricted by $\mathcal{G}$; this restriction justifies the approximations 
$\Omega_{\bm{k}'n}\simeq \omega$ and $k'_z\simeq s (\omega /v_n)\cos\theta_n$ [see Eq.~\eqref{142550_29Jun26}].}

The summation in the final line of Eq.~\eqref{eq: energy flux by w.p. in finite vol} represents the number of vibrational modes contributing 
to the wave packet.  
We write this number as $N_n(\omega, \bm{k}_{\parallel}) \Delta \omega \Delta^2\bm{k}_{\parallel}$ in the limit of infinitesimal 
$\Delta\omega$ and $\Delta^2\bm{k}_{\parallel}$. Then we obtain
\begin{equation}
\sum_{\bm{k}'}^{s \, k'_z > 0} \mathcal{G}(\Omega_{\bm{k}'n}-\omega, \bm{k}'_{\parallel} - \bm{k}_{\parallel}) 
= g(\omega, \bm{k}_{\parallel}) \, N_n(\omega, \bm{k}_{\parallel}) \Delta \omega \Delta^2\bm{k}_{\parallel}.
\label{eq: sum G num of vib states}
\end{equation}
The quantity $N_n$ is evaluated as
\begin{align}
& N_n(\omega, \bm{k}_{\parallel}) = \sum_{\bm{k}'}^{s \, k'_z > 0}
\delta(\Omega_{\bm{k}'n} - \omega) \, \delta(\bm{k}'_{\parallel} - \bm{k}_{\parallel}) \nonumber\\
&= V \int \frac{\rmd k'_{z} \rmd^2 \bm{k}'_{\parallel}}{(2\pi)^3}
\frac{\delta\left(k'_z - s \sqrt{\frac{\omega^2}{v_n^2} - \abs{\bm{k}_{\parallel}}^2}\right)}
{\Abs{\frac{\partial \Omega_{\bm{k},n}}{\partial k_z}}}
\delta(\bm{k}'_{\parallel} - \bm{k}_{\parallel}) \nonumber \\
&= \frac{V}{(2\pi)^3} \, \frac{1}{v_n \cos\theta_n}, 
\label{eq: number of plane wave components}
\end{align}
where we have converted the discrete sum over $\bm{k}'$ into an integral by imposing periodic boundary conditions for calculational convenience. 
This procedure is justified because the boundary condition does not influence the bulk density of states~\cite{Ledermann1944}.

Substituting Eq.~\eqref{eq: sum G num of vib states} into Eq.~\eqref{eq: energy flux by w.p. in finite vol}, 
we obtain the final expression for the energy flux density:
\begin{align}
j^{\text{w.p.}}_{\mathfrak{u}; s, n} &= 
2s \omega^2 \rho v_n \cos\theta_n \Abs{u_{\bm{k}n}}^2 \nonumber\\
&\quad \times g(\omega, \bm{k}_{\parallel}) \, 
\frac{N_n(\omega, \bm{k}_{\parallel}) \Delta\omega \Delta^2\bm{k}_{\parallel}}{V}.
\label{eq: energy flux by w.p. rearr}
\end{align}

{\subsubsection{Correspondence between wave packets and plane waves}}

Let us examine the correspondence between two frameworks:
(i)~the continuum description based on elasticity theory, and
(ii)~a discrete lattice model for a finite-sized crystal. 
Both frameworks describe the same physical wave propagation in the long-wavelength limit.
In particular, as $\Delta\omega \Delta^2\bm{k}_{\parallel} \to 0$, 
the wave packet becomes indistinguishable from a single plane wave. 
The associated energy flux density must be uniquely determined, regardless of the theoretical framework.

Therefore, the energy flux density $j^{\text{w.p.}}_{\mathfrak{u}; s, n}$ of the wave packet, given in Eq.~\eqref{eq: energy flux by w.p. rearr}, 
coincides with the energy flux density $j^{\text{ela}}_{\mathfrak{u}; s, n}$ of the plane wave in Eq.~\eqref{eq: ene fl}.  
The equality 
\begin{equation}
u_{s,n} = \sqrt{g(\omega, \bm{k}_{\parallel}) \, \frac{N_n \Delta\omega \Delta^2\bm{k}_{\parallel}}{V}} \, u_{\bm{k}n}
\label{eq: amplitude equality between plane and packet}
\end{equation}
then follows, where $u_{s,n}$ and $u_{\bm{k}n}$ are the amplitudes of the plane wave and the wave packet, respectively.  
By equating $u_{\bm{k}n}$ with $u_{s,n}$ as in Eq.~\eqref{eq: ukn usn}, we determine the normalization factor $g(\omega, \bm{k}_{\parallel})$ as  
\begin{equation}
g(\omega, \bm{k}_{\parallel}) = \frac{V}{N_n \Delta\omega \Delta^2\bm{k}_{\parallel}}. 
\label{154426_3Aug25}
\end{equation}

We then reformulate the $S$-matrix relation~\eqref{eq: S matrix in circular pol 1}, originally written for plane-wave amplitudes $u_{s,n}$, 
so that it applies to wave packets.  
Using the correspondence established in Eq.~\eqref{eq: amplitude equality between plane and packet},  
each term $\sqrt{\rho v_n \cos\theta_n}\, u_{s,n}$ in Eq.~\eqref{eq: S matrix in circular pol 1} is replaced by  
\begin{equation}
\sqrt{\rho v_n \cos\theta_n} \, \sqrt{\frac{N_n \Delta\omega \Delta^2\bm{k}_{\parallel}}{V}}\, A^{\text{w.p.}}_{\bm{k}'n}
= \sqrt{\frac{\Delta\omega \Delta^2\bm{k}_{\parallel}}{(2\pi)^3}}\, \sqrt{\rho}\, A^{\text{w.p.}}_{s,n},
\end{equation}
in the region $z<0$.  
Here we used Eqs.~\eqref{eq: uniformly distributed wave packet b}, \eqref{eq: uniformly distributed wave packet d}, and \eqref{154426_3Aug25}, 
and wrote $A^{\text{w.p.}}_{\bm{k}'n}$ as $A^{\text{w.p.}}_{s,n}$.  
Similarly, in the region $z>0$, the expression $\sqrt{\widetilde{\rho} c_n \cos\Theta_n}\,\widetilde{u}_{s,n}$ becomes  
$\sqrt{\frac{\Delta\omega \Delta^2\bm{k}_{\parallel}}{(2\pi)^3}}\, \sqrt{\widetilde{\rho}}\, \widetilde{A}^{\text{w.p.}}_{s,n}$. 
The amplitude $\widetilde{A}^{\text{w.p.}}_{s,n}$ refers to a wave packet in the region $z>0$, 
specified by $(\omega,\bm{k}_{\parallel},s,n)$ and the width $\Delta\omega \Delta^2\bm{k}_{\parallel}$.  

Thus, the $S$-matrix relation~\eqref{eq: S matrix in circular pol 1} reduces to
\begin{equation}
\left[
\renewcommand\arraystretch{1.3}
\begin{array}{c}
\sqrt{\rho}\, A^{\text{w.p.}}_{-,\text{L}}\\
\sqrt{\rho}\, A^{\text{w.p.}}_{-,\text{RH}}\\
\sqrt{\rho}\, A^{\text{w.p.}}_{-,\text{LH}}\\ \hline \rule{0pt}{2.8ex} 
\sqrt{\widetilde{\rho}}\, \widetilde{A}^{\text{w.p.}}_{+,\text{L}}\\
\sqrt{\widetilde{\rho}}\, \widetilde{A}^{\text{w.p.}}_{+,\text{RH}}\\
\sqrt{\widetilde{\rho}}\, \widetilde{A}^{\text{w.p.}}_{+,\text{LH}}
\end{array}
\right]
=
S(\omega, \bm{k}_{\parallel})
\left[
\renewcommand\arraystretch{1.3}
\begin{array}{c}
\sqrt{\rho}\, A^{\text{w.p.}}_{+,\text{L}}\\
\sqrt{\rho}\, A^{\text{w.p.}}_{+,\text{RH}}\\
\sqrt{\rho}\, A^{\text{w.p.}}_{+,\text{LH}}\\ \hline\rule{0pt}{2.8ex}
\sqrt{\widetilde{\rho}}\, \widetilde{A}^{\text{w.p.}}_{-,\text{L}}\\
\sqrt{\widetilde{\rho}}\, \widetilde{A}^{\text{w.p.}}_{-,\text{RH}}\\
\sqrt{\widetilde{\rho}}\, \widetilde{A}^{\text{w.p.}}_{-,\text{LH}}
\end{array}
\right].
\label{eq: S matrix for u wp}
\end{equation}
This relation holds for any choice of infinitesimal $\Delta\omega$ and $\Delta^2 \bm{k}_{\parallel}$, 
and for every value of $(\omega, \bm{k}_{\parallel})$.

We finally reconstruct the full lattice vibration of Eq.~\eqref{eq: sum of vibrational states} from the wave packets in Eq.~\eqref{eq: uniformly distributed wave packet a}. 
Specifically, we identify $A^{\text{w.p.}}_{s,n}$ and $\widetilde{A}^{\text{w.p.}}_{s,n}$ in Eq.~\eqref{eq: S matrix for u wp} 
with the amplitudes $A^{\text{lat}}_{\bm{k}n}$ and $\widetilde{A}^{\text{lat}}_{\bm{q}n}$ of the lattice deformation 
under the conditions $\Omega_{\bm{k}n}, \omega_{\bm{q}n} \simeq \omega$ and $\bm{q}_{\parallel} \simeq \bm{k}_{\parallel}$. 
Using Eqs.~\eqref{eq: correspondence between u and a} and \eqref{eq: correspondence between tilde u and tilde a}, 
we then convert these lattice amplitudes into the dimensionless forms 
$a_{\bm{k}n} = a_{s,n}$ and $\widetilde{a}_{\bm{k}n} = \widetilde{a}_{s,n}$. With these substitutions, Eq.~\eqref{eq: S matrix for u wp} becomes
\begin{equation}
 \left[
\renewcommand\arraystretch{1.3}
\begin{array}{c}
a_{-,\text{L}}\\
a_{-,\text{RH}}\\
a_{-,\text{LH}}\\ \hline
\widetilde{a}_{+,\text{L}}\\
\widetilde{a}_{+,\text{RH}}\\
\widetilde{a}_{+,\text{LH}}
\end{array} 
\right]
= S (\omega, \bm{k}_{\parallel})
 \left[
\renewcommand\arraystretch{1.3}
\begin{array}{c}
a_{+,\text{L}}\\
a_{+,\text{RH}}\\
a_{+,\text{LH}}\\ \hline
\widetilde{a}_{-,\text{L}}\\
\widetilde{a}_{-,\text{RH}}\\
\widetilde{a}_{-,\text{LH}}
\end{array} 
\right].
\label{eq: Smatrix for classical a }
\end{equation}

\subsection{Phonon reflection and transmission}
\label{subsec: b.c. for phonon distrib}

We now quantize lattice vibrations and formulate the boundary conditions for the phonon distribution functions~\footnote{Reference~\cite{Amrit2025Poster} implicitly noted that the interface scattering of phonons, treated as \emph{particles}, and that of elastic waves, 
treated as \emph{wave phenomena}, behave identically in multilayer structures whose dimensions exceed the wavelength.}.  
We replace the classical amplitudes $a_{\bm{k}n}$ and $a^{*}_{\bm{k}n}$ with the annihilation and creation operators 
$\hat{a}_{\bm{k}n}$ and $\hat{a}^{\dagger}_{\bm{k}n}$, which obey  
\begin{equation}
[\hat{a}_{\bm{k}n},~\hat{a}^{\dagger}_{\bm{k}'n'}] = \delta_{\bm{k},\bm{k}'} \delta_{nn'} .
\label{eq: commu rel}
\end{equation}

The classical boundary conditions in Eq.~\eqref{eq: Smatrix for classical a } hold for arbitrary amplitudes.  
Therefore, an analogous set of conditions must hold in the quantum description.  
Equation~\eqref{eq: Smatrix for classical a } imposes constraints on the phonon Fock space at the interface; 
it is not consistent to assume that the $S$-matrix acts directly on the operators, because such an assumption violates the commutation relation in Eq.~\eqref{eq: commu rel}.  
Instead, Eq.~\eqref{eq: Smatrix for classical a } requires the relation
\begin{subequations}
 \begin{equation}
 \left[
 \renewcommand\arraystretch{1.3}
 \begin{array}{c}
 \hat{a}_{-,\text{L}}\ket{\Psi}\\
 \hat{a}_{-,\text{RH}}\ket{\Psi} \\
 \hat{a}_{-,\text{LH}}\ket{\Psi} \\ \hline\rule{0pt}{2.8ex}
 \hat{\widetilde{a}}_{+,\text{L}}\ket{\Psi} \\
 \hat{\widetilde{a}}_{+,\text{RH}}\ket{\Psi} \\
 \hat{\widetilde{a}}_{+,\text{LH}}\ket{\Psi}
 \end{array} 
 \right]
 = S (\omega, \bm{k}_{\parallel})
 \left[
 \renewcommand\arraystretch{1.3}
 \begin{array}{c}
 \hat{a}_{+,\text{L}}\ket{\Psi} \\
 \hat{a}_{+,\text{RH}}\ket{\Psi} \\
 \hat{a}_{+,\text{LH}}\ket{\Psi} \\ \hline\rule{0pt}{2.8ex}
 \hat{\widetilde{a}}_{-,\text{L}}\ket{\Psi} \\
 \hat{\widetilde{a}}_{-,\text{RH}}\ket{\Psi} \\
 \hat{\widetilde{a}}_{-,\text{LH}}\ket{\Psi}
 \end{array} 
 \right],
\label{eq: restriction on the phonon Fock space a}
 \end{equation}
 or more concisely,
 \begin{equation}
 \hat{a}^{\text{out}}_{\mu}\ket{\Psi}
 = \sum_{\nu} S_{\mu\nu}(\omega, \bm{k}_{\parallel}) \hat{a}^{\text{in}}_{\nu}\ket{\Psi},
\label{eq: restriction on the phonon Fock space b}
 \end{equation}
 \label{eq: restriction on the phonon Fock space}%
\end{subequations}
for any physical phonon state $\ket{\Psi}$ in the junction system.
Here the annihilation operators labeled ``in'' and ``out'' represent the incident and scattered wave components at the interface, respectively.
The constraints~\eqref{eq: restriction on the phonon Fock space a} or \eqref{eq: restriction on the phonon Fock space b} 
eliminate unphysical excitations that are classically forbidden near the interface~\footnote{
As a remark, we assume that the crystals on both sides of the interface remain undeformed. 
This assumption also constrains how the classical $S$-matrix relation can be reformulated in the quantum case. 
In principle, one might either adopt Eq.~\eqref{eq: restriction on the phonon Fock space a} as written or replace all annihilation operators by creation operators. 
The undeformed condition favors the former, since the vacuum state $\ket{\Psi}=\ket{0}$ before contact remains within the Fock space after the systems are joined.
}.

We introduce the phonon density matrix at the interface, denoted by $\hat{\rho}$.
To describe the statistical properties of phonons near the interface, we define reduced density matrices for incident and scattered waves as follows:
\begin{equation}
\rho^{\text{in}}_{\nu_1\nu_2} 
\equiv \frac{\mathrm{Tr}\,\left[
\hat{a}^{\text{in}}_{\nu_1}\hat{\rho} \hat{a}^{\text{in}\, \dagger}_{\nu_2}
\right]}{\mathrm{Tr}\,[\hat{\rho}]}, \quad
\rho^{\text{out}}_{\mu_1\mu_2} 
\equiv \frac{\mathrm{Tr}\,\left[
\hat{a}^{\text{out}}_{\mu_1}\hat{\rho}\hat{a}^{\text{out}\, \dagger}_{\mu_2}
\right]}{\mathrm{Tr}\,[\hat{\rho}]},
\end{equation}
where the trace is taken over the restricted Fock space specified by Eqs.~\eqref{eq: restriction on the phonon Fock space b}.
These two reduced density matrices are connected by the $S$-matrix as follows:
\begin{align}
& \rho^{\text{out}}_{\mu_1\mu_2} = \sum_{\Psi, \Phi}
\Braket{\Psi|\hat{a}^{\text{out}\, \dagger}_{\mu_2}\hat{a}^{\text{out}}_{\mu_1}|\Phi}
\frac{\Braket{\Phi|\hat{\rho}|\Psi}}{\mathrm{Tr}\,[\hat{\rho}]} \nonumber\\
&= \sum_{\Psi, \Phi}\sum_{\nu_1, \nu_2} S^{*}_{\mu_2\nu_2}
\Braket{\Psi|\hat{a}^{\text{in}\, \dagger}_{\nu_2}\hat{a}^{\text{in}}_{\nu_1}|\Phi}S_{\mu_1\nu_1}
\frac{\Braket{\Phi|\hat{\rho}|\Psi}}{\mathrm{Tr}\,[\hat{\rho}]} \nonumber\\
&= \sum_{\nu_1, \nu_2} S_{\mu_1\nu_1} \rho^{\text{in}}_{\nu_1\nu_2} 
\left(S^{\dagger}\right)_{\nu_2\mu_2}.
\label{eq: S mat for rhos}
\end{align}

We now make two assumptions.
First, phonon excitations corresponding to different incident waves are uncorrelated.
Second, those of different scattered waves are also uncorrelated.
Under these assumptions, the reduced density matrices become approximately diagonal~\cite{PeierlsTextbook}:
\begin{equation}
\rho^{\text{in}}_{\nu_1\nu_2}\simeq \delta_{\nu_1\nu_2}f^{\text{in}}_{\nu_1},\qquad
\rho^{\text{out}}_{\mu_1\mu_2}
\simeq \delta_{\mu_1\mu_2}f^{\text{out}}_{\mu_1},
\end{equation}
where $f^{\text{in}}_{\nu}$ and $f^{\text{out}}_{\mu}$ are the phonon distribution functions
for the incident and the scattered waves, respectively.
These functions satisfy the relation
\begin{equation}
 f^{\text{out}}_{\mu}(\omega, \bm{k}_{\parallel})
 = \sum_{\nu} \Abs{S_{\mu\nu} (\omega, \bm{k}_{\parallel})}^2 f^{\text{in}}_{\nu}(\omega, \bm{k}_{\parallel}).
\label{eq: DERIVED b.c. for phonon distribution fnc}
\end{equation}
The squared moduli of the $S$-matrix elements correspond to the power reflectance $\mathcal{R}_{nm}, \mathcal{R}'_{nm}$ 
and transmittance $\mathcal{T}_{nm}, \mathcal{T}'_{nm}$ of classical elastic waves, as shown in 
Eqs.~\eqref{eq: elastic S matrix in circularly pol}--\eqref{eq: R and T definitions d} of Appendix~\ref{sec: elastic plane wave}. 
Therefore, Eq.~\eqref{eq: DERIVED b.c. for phonon distribution fnc} reduces to the boundary condition~\eqref{eq: boundary condition assump} 
introduced earlier~\footnote{
When the off-diagonal elements of the phonon density matrix play a significant role~\cite{Zhong2023}, 
it is more appropriate to use the full boundary condition given by Eq.~\eqref{eq: S mat for rhos}.}.

\subsection{Remarks on the validity}

\subsubsection{Energy conservation law}

The boundary condition~\eqref{eq: DERIVED b.c. for phonon distribution fnc} holds in thermal equilibrium, where  
$f^{\text{out}}_{\mu}(\omega, \bm{k}_{\parallel}) = f^{\text{in}}_{\nu}(\omega, \bm{k}_{\parallel}) 
= \left[e^{\hbar\omega/k_{\text{B}}T} - 1\right]^{-1} \equiv f^{\text{eq}}(\omega, T)$.  
This equality follows from the unitarity of the $S$-matrix~\eqref{eq: R and T sum unity}, as shown in Appendix~\ref{sec: elastic plane wave}.  
We also find that phonon number is conserved even off equilibrium:  
$\sum_{\mu} f^{\text{out}}_{\mu}(\omega, \bm{k}_{\parallel}, z = 0) = \sum_{\nu} f^{\text{in}}_{\nu}(\omega, \bm{k}_{\parallel}, z = 0)$.
This condition can also be rewritten as  
\begin{equation}
 \sum_{s, n} s F_{s,n}(\omega, \bm{k}_{\parallel}, z=-0) 
 = \sum_{s, n} s f_{s,n}(\omega, \bm{k}_{\parallel}, z=+0),\label{eq: ph num conservation}
\end{equation}
where $F$ and $f$ denote the phonon distribution functions in the regions $z<0$ and $z>0$, respectively.
We used the notation $F_{\bm{k}n}(z) \equiv F_{s,n}(\omega, \bm{k}_{\parallel}, z)$ and $f_{\bm{q}n}(z) \equiv f_{s,n}(\omega, \bm{q}_{\parallel}, z)$  
(see Sec.~\ref{subsec: our assump} for details).  

From this relation~\eqref{eq: ph num conservation}, we confirm energy conservation at the interface.
The net energy flux density remains continuous across the boundary: 
\begin{align}
& \int \frac{\rmd^3\bm{k}}{(2\pi)^3}\sum_n
 \hbar\Omega_{\bm{k}n}v^z_{\bm{k}n} F_{\bm{k}n}(z=-0) \nonumber\\
&= \int \rmd\omega \rmd^2\bm{k}_{\parallel} \sum_{s,n} \frac{N_n(\omega, \bm{k}_{\parallel})}{V} \nonumber\\
&\qquad\quad \times \hbar\omega \, s v_n \cos\theta_n F_{s,n}(\omega, \bm{k}_{\parallel}, z=-0) \nonumber\\
&= \int \frac{\rmd\omega \rmd^2\bm{k}_{\parallel}}{(2\pi)^3} \hbar\omega \sum_{s,n} s F_{s,n}(\omega, \bm{k}_{\parallel}, z=-0) \nonumber\\
&= \int \frac{\rmd\omega \rmd^2\bm{k}_{\parallel}}{(2\pi)^3} \hbar\omega \sum_{s,n} s f_{s,n}(\omega, \bm{k}_{\parallel}, z=+0) \nonumber\\
&= \int \frac{\rmd^3\bm{q}}{(2\pi)^3}\sum_n \hbar\omega_{\bm{q}n} c^z_{\bm{q}n} f_{\bm{q}n}(z=+0). \label{eq: net ene flux from coarse b.c.}
\end{align}

\subsubsection{Comparison with previously known results}

The conditions presented in Refs.~\cite{Minnich2011,Hua2016} are similar to Eq.~\eqref{eq: boundary condition assump}
in that both describe linear relations among the phonon distribution functions for incident, reflected, and transmitted waves at each frequency $\omega$.
Meanwhile, Refs.~\cite{Minnich2011,Hua2016} focus on diffuse scattering at the interface. 
In contrast, we consider phonon scattering at a smooth, flat interface in the present study; 
the specular reflection and transmission allow us to formulate the boundary condition incorporating the polarization degree of freedom.

Regarding the specular reflection and transmission, our boundary conditions agree well with
two previously known result for limited cases or targets.
In the case of perfect reflection, Ziman's monograph~\cite{ZimanTextbookElPh} provides a boundary condition for distribution functions
with specularity parameter $p=1$. 
The condition given there coincides with 
the special case (perfect reflection) of our result in Eq.~\eqref{eq: boundary condition assump}. 
Furthermore, when we calculate the interfacial thermal resistance using Eq.~\eqref{eq: boundary condition assump}, 
the result agrees with that obtained from the acoustic mismatch model~\cite{Little1959,Khalatnikov1965} 
(see Supplemental Material~\cite{SupplementalMaterialFullPaper}).
These agreements further support the validity of the present boundary condition.

\subsubsection{Applicability}

The boundary conditions in Eq.~\eqref{eq: boundary condition assump} and the associated distribution functions are independent of the spatial coordinates $x$ and $y$ parallel to the interface. 
In this sense, Eq.~\eqref{eq: boundary condition assump} represents coarse-grained boundary conditions.
This feature follows directly from their derivation:  
we evaluated the energy flux per unit area across the interface and averaged the work done by elastic waves, lattice vibrations, and phonon wave packets over the $xy$ plane.  
Consequently, the spatial variations of the phonon distribution functions  
$F_{\bm{k}n}(x,y,z)$ and $f_{\bm{q}n}(x,y,z)$ along the interface are integrated out, and the boundary conditions apply to the coarse-grained functions  
$F_{\bm{k}n}(z)$ and $f_{\bm{q}n}(z)$.

The coarse-grained boundary conditions~\eqref{eq: boundary condition assump} is a practical basis for studying interfacial transport.  
They apply to a broad class of physical quantities whose definitions do not involve the in-plane coordinates, including the phonon energy and SAM.
In the next section, we formulate the detailed boundary condition that enables us to address physical quantities with explicit dependence on the real coordinates, e.g., the phonon OAM.

\section{Formulation of the detailed boundary conditions}
\label{sec: detailed b.c. formula}

In this section, we introduce the boundary conditions~\eqref{163324_9Jun25} and \eqref{163328_9Jun25}
for the phonon distribution functions that explicitly depend on the lateral coordinates $(x, y)$. 
We justify these conditions by showing that they reproduce the classical transverse shift 
of elastic wave packets at interfacial scattering, 
illustrated as $\Delta r_{\mathrm{IF}}$ in Fig.~\ref{fig: intro schematics}(a).

We consider the case where the wave packet is incident from the region $z < 0$~\footnote{%
The same argument can be straightforwardly extended to cases where wave packets are incident from both sides of the interface.}.
To distinguish the two regions, we label the lateral position of the wave packet center as $\bm{R}_{\parallel}$ for $z < 0$ 
and as $\bm{r}_{\parallel}$ for $z > 0$. 
We impose open boundary conditions, instead of periodic ones, in the $x$ and $y$ directions. 
This choice makes the $x$ and $y$ coordinates well defined. 
{In particular, this is necessary within the present formulation because the extrinsic OAM defined in 
Eqs.~\eqref{eq: OAM definition Lz} and \eqref{eq: OAM definition tildeL z} explicitly 
depends on the wave-packet position coordinates $\bm{R}_{\parallel}$ and $\bm{r}_{\parallel}$. 
Under periodic boundary conditions, the OAM defined in 
Eqs.~\eqref{eq: OAM definition Lz} and \eqref{eq: OAM definition tildeL z} is not well defined within the present formulation. 
We thus focus on open boundary conditions, which are also more directly relevant to experimental situations.}
For clarity, we assume that the junction of the two crystals forms a prism with a uniform cross section in the $xy$ plane and extends along the $z$ axis.
We set the origins of $\bm{R}_{\parallel}$ and $\bm{r}_{\parallel}$ at the center of gravity of this cross section.

\subsection{Classical transverse shifts}

We present expressions for the transverse shifts of a classical elastic beam incident from $z<0$ at an interface.  
For a circularly polarized incident beam, the transverse shifts of the scattered beams are obtained as derived in Appendix~\ref{sec: IF shift of elastic beam}:
\begin{subequations}
 \begin{align}
 &\Delta r^{\text{r-L}}_{\text{IF}}(\overline{\sigma})
 =  \Delta r^{\text{t-L}}_{\text{IF}}(\overline{\sigma}) = - \frac{\overline{\sigma} \cot \theta_{\text{T}}}{k},
\label{170438_23Jan25}\\
 &\Delta r^{\text{r-T}}_{\text{IF}}(\overline{\sigma}) =
 - \frac{\overline{\sigma} \cot\theta_{\text{T}}}{k}\, 
 \frac{2 \mathcal{R}_{\text{RH}, \text{RH}}}{\mathcal{R}_{\text{RH}, \text{RH}} + \mathcal{R}_{\text{LH}, \text{RH}}},\\
 &\Delta r^{\text{t-T}}_{\text{IF}}(\overline{\sigma})\nonumber\\
& =
 - \frac{\overline{\sigma} \cot\theta_{\text{T}}}{k}\, 
 \left(1- 
 \frac{\mathcal{T}_{\text{RH}, \text{RH}} - \mathcal{T}_{\text{LH}, \text{RH}}}{ \mathcal{T}_{\text{RH}, \text{RH}} + \mathcal{T}_{\text{LH}, \text{RH}}}
 \, \frac{\cos\Theta_{\text{T}}}{\cos\theta_{\text{T}}}\right).\label{135819_28Jan25}
 \end{align}
\label{173511_4Nov24}%
\end{subequations}
Here $k = \Abs{\bm{k}}$ is the modulus of central wavevector of the beam. 
The labels $a=\mathrm{r\text{-}L},\mathrm{r\text{-}T},\mathrm{t\text{-}L},\mathrm{t\text{-}T}$ denote 
reflected longitudinal, reflected transverse, transmitted longitudinal, and transmitted transverse waves, respectively.  
For each $a$, $\Delta r^{a}_{\mathrm{IF}}$ is the beam-centroid displacement perpendicular to the incident plane. 
We use $\theta_{\mathrm{T}}$ for the angle of incidence of transverse modes in $z<0$, 
and $\Theta_{\mathrm{T}}$ for the angle of refraction of transmitted transverse modes in $z>0$, as shown in Fig.~\ref{fig: incidentplane}.

The parameter $\overline{\sigma}$ gives the average helicity of the incident beam:
\begin{equation}
 \overline{\sigma}\equiv \frac{\Abs{u_{+, \text{RH}}}^2 - \Abs{u_{+, \text{LH}}}^2}{\Abs{u_{+, \text{RH}}}^2 + \Abs{u_{+, \text{LH}}}^2}, 
\end{equation}
where $u_{+,\mathrm{RH}}$ and $u_{+,\mathrm{LH}}$ are the displacement amplitudes of the right- and left-handed circular components at the beam centroid.  
Equivalently, the helicity can be expressed in terms of the energy fluxes of the RH and LH components.  
For two beams in the RH and LH modes with the same frequency $\omega$ and the same $\bm k_{\parallel}$, the energy-flux ratio equals the squared-amplitude ratio: 
$j_{\mathfrak{u}, +,\text{LH}}/j_{\mathfrak{u}, +,\text{RH}} = \Abs{u_{+, \text{LH}} / u_{+, \text{RH}}}^2$. 
Hence
\begin{equation}
 \overline{\sigma}
 = \frac{j_{\mathfrak{u}, +,\text{RH}} - j_{\mathfrak{u}, +,\text{LH}}}{j_{\mathfrak{u}, +,\text{RH}} + j_{\mathfrak{u}, +,\text{LH}}}.\label{135607_28Jan25}
\end{equation}

\subsection{Prototypical case of the boundary conditions}

Let us consider a specific scattering process of phonon wave packets that incorporates the transverse shift. An incident wave packet, fully polarized in the RH mode ($\overline{\sigma}=+1$), enters from the region $z<0$, and scattered longitudinal wave packets emerge at the interface. 
Since both wave packets are in pure polarization states (as opposed to mixed states), the incident (RH-mode) and scattered (L-mode) wave packets can be directly related to the corresponding phonon distribution functions.
This one-to-one correspondence allows us to formulate the detailed boundary conditions in a straightforward manner.

By naturally introducing $x$, $y$ dependence into the boundary condition~\eqref{eq: boundary condition assump}, we obtain the following expressions:
\begin{subequations}
 \begin{align}
 &F_{-, \text{L}}(\omega, \bm{k}_{\parallel}, \bm{R}_{\parallel, -, \text{L}})\nonumber\\
 &= \mathcal{R}_{\text{L}, \text{RH}}(\omega, \bm{k}_{\parallel})\, 
 F_{+, \text{RH}}(\omega, \bm{k}_{\parallel}, \bm{R}_{\parallel, +, \text{RH}}),\label{175340_23Jan25}\\
 &f_{+, \text{L}}(\omega, \bm{k}_{\parallel}, \bm{r}_{\parallel, +, \text{L}})\nonumber\\
 &= \mathcal{T}_{\text{L}, \text{RH}}(\omega, \bm{k}_{\parallel})\, 
 F_{+, \text{RH}}(\omega, \bm{k}_{\parallel}, \bm{R}_{\parallel, +, \text{RH}}) \label{175346_23Jan25}.
 \end{align}
Here we label the center-of-mass coordinate of the wave packet as $(\bm{R}_{\parallel, s, n}, z)$ in the region $z < 0$ and 
$(\bm{r}_{\parallel, s, n}, z)$ in the region $z > 0$, for mode $(s, n)$. 
The argument $z =\pm 0$ is omitted in the two equations.

The spatial distribution of mechanical work performed at the interface by the incident, reflected, or transmitted wave packet 
should be consistent whether calculated through
classical elasticity or quantum mechanics. 
This consistency indicates that the position at which mechanical work is exerted, as determined by classical elasticity, 
coincides with the position expectation value of the phonon distribution function.
Accordingly, the transverse shift $\Delta r^{\text{r-L}}_{\text{IF}}(\overline{\sigma} = +1)$
in Eq.~\eqref{170438_23Jan25} is equal to 
$\left(\bm{R}_{\parallel, -, \text{L}} - \bm{R}_{\parallel, +, \text{RH}}\right)_{y'}$.
Also, the transverse shift $\Delta r^{\text{t-L}}_{\text{IF}}(\overline{\sigma} = +1)$
in Eq.~\eqref{170438_23Jan25} is equal to 
$\left(\bm{r}_{\parallel, +, \text{L}} - \bm{R}_{\parallel, +, \text{RH}}\right)_{y'}$. 
The subscript $y'$ denotes the component along the unit vector $\hat{\bm y}'=\hat{\bm z}\times\hat{\bm k}_{\parallel}$, which is perpendicular to the incident plane, 
as shown in Fig.~\ref{fig: intro schematics}(c).  
Here $\hat{\bm{k}}_{\parallel} = \bm{k}_{\parallel}/\Abs{\bm{k}_{\parallel}}$. 
Using the result from classical elasticity~\eqref{170438_23Jan25}, we obtain the following relation:
\begin{equation}
 \left(\bm{R}_{\parallel, -, \text{L}}\right)_{y'}
 = \left(\bm{r}_{\parallel, +, \text{L}}\right)_{y'}
 = \left(\bm{R}_{\parallel, +, \text{RH}}\right)_{y'} - \frac{\cot \theta_{\text{T}}}{k}.
 \label{175351_23Jan25}  
\end{equation}
\end{subequations}

Similarly, we consider the case where a fully left-handed circularly polarized (LH) wave with $\overline{\sigma}=-1$ is incident. 
For reflected and transmitted waves of the longitudinal (L) mode, the boundary conditions take the form: 
\begin{subequations}
 \begin{align}
 &F_{-, \text{L}}(\omega, \bm{k}_{\parallel}, \bm{R}_{\parallel, -, \text{L}})\nonumber\\
 &= \mathcal{R}_{\text{L}, \text{LH}}(\omega, \bm{k}_{\parallel})\, 
 F_{+, \text{LH}}(\omega, \bm{k}_{\parallel}, \bm{R}_{\parallel, +, \text{LH}}),\label{175356_23Jan25}\\
 &f_{+, \text{L}}(\omega, \bm{k}_{\parallel}, \bm{r}_{\parallel, +, \text{L}})\nonumber\\
 &= \mathcal{T}_{\text{L}, \text{LH}}(\omega, \bm{k}_{\parallel})\, 
 F_{+, \text{LH}}(\omega, \bm{k}_{\parallel}, \bm{R}_{\parallel, +, \text{LH}})\label{175359_23Jan25}.
 \end{align}
These coordinates satisfy the relation:
 \begin{align}
 \left(\bm{R}_{\parallel, -, \text{L}}\right)_{y'}
 = \left(\bm{r}_{\parallel, +, \text{L}}\right)_{y'}
 = \left(\bm{R}_{\parallel, +, \text{LH}}\right)_{y'} + \frac{\cot \theta_{\text{T}}}{k}.
 \label{175401_23Jan25}
 \end{align}
\end{subequations}

In contrast, when either the incident or scattered wave packets are \emph{mixed states}, the boundary conditions cannot be determined in this way.  
{This situation arises, for example, when the incident wave is 
partially circularly polarized}, i.e., $\overline{\sigma} \neq \pm 1$, or when we examine reflected or transmitted transverse wave packets. 
In such cases, the notion that a single distribution function directly represents a wave packet breaks down. 
We then determine the boundary conditions by imposing AM conservation.

\subsection{General case: TAM conservation}

The interface at $z = 0$ is assumed to be smooth, flat, and invariant under continuous rotational symmetry. 
According to Noether's theorem, this symmetry leads to the conservation of the AM component along the interface normal. 
For the displacement field in an elastic medium, 
the conserved quantity is the sum of the SAM and OAM carried by the phonon wave packet~\cite{Nakane2018,Bliokh2006}.

This conservation of TAM is already embedded in the boundary conditions given in Eqs.~\eqref{175340_23Jan25}--\eqref{175401_23Jan25}. 
For example, by using the relation for transverse wave incidence, $\Abs{\bm{k}_{\parallel}} = k \sin\theta_{\text{T}}$, 
along with the identity $\left(\bm{R}_{\parallel, -, \text{L}}\right)_{y'} = \bm{R}_{\parallel, -, \text{L}} \cdot (\hat{\bm{z}} 
\times \hat{\bm{k}}_{\parallel})$, 
we can rewrite Eq.~\eqref{175351_23Jan25} as
\begin{align}
&\left(\bm{R}_{\parallel, -, \text{L}}\times \hbar\bm{k}_{\parallel}\right)_z
= \left(\bm{r}_{\parallel, +, \text{L}} \times \hbar\bm{k}_{\parallel}\right)_z \nonumber\\
&= \left(\bm{R}_{\parallel, +, \text{RH}} \times \hbar\bm{k}_{\parallel}\right)_z + \hbar \cos \theta_{\text{T}}. \label{180853_23Jan25}
\end{align}
This equation, combined with the absence of SAM in longitudinal waves and the expression for the SAM of the incident RH-mode wave packet, 
$S^z_{+, \text{RH}} = \hbar \cos \theta_{\text{T}}$, leads to the following result. 
Equation~\eqref{180853_23Jan25} shows that the TAM is conserved among the incident RH-mode packet, 
the reflected L-mode packet, and the transmitted L-mode packet. 
Each of these packets carries a $z$ component of AM other than SAM, given respectively by
$\left(\bm{R}_{\parallel, +, \text{RH}} \times \hbar \bm{k}_{\parallel}\right)_z$,
$\left(\bm{R}_{\parallel, -, \text{L}} \times \hbar \bm{k}_{\parallel}\right)_z$,
and
$\left(\bm{r}_{\parallel, +, \text{L}} \times \hbar \bm{k}_{\parallel}\right)_z$. 
As defined in Eqs.~\eqref{eq: OAM definition Lz} and \eqref{eq: OAM definition tildeL z}, these terms correspond to the $z$ components of the OAM 
carried by each wave packet. 
The phonon distribution is thus redistributed among wave packets at the interface in a manner that preserves TAM.  

We generalize this argument and assume that the distribution function is redistributed only among wave packets with the same TAM 
before and after interface scattering. 
We impose the boundary condition~\eqref{163324_9Jun25} for $n = \text{L}, \text{RH}, \text{LH}$ under the requirement 
that the TAM components satisfy the equality~\eqref{163328_9Jun25}. 
These two conditions were summarized in Sec.~\ref{subsec: b.c. summa}. 
With this assumption, TAM conservation during interface scattering follows directly.

\subsection{Consistency with classical transverse shifts}

We now examine the validity of the hypothesis introduced above.  
Our aim is to verify that 
the expectation value of the transverse shift of phonon wave packets, calculated from the boundary conditions in Eqs.~\eqref{163324_9Jun25} and \eqref{163328_9Jun25}, 
coincides with the transverse shift of the beam centroid predicted by elasticity theory in Eqs.~\eqref{170438_23Jan25}--\eqref{135819_28Jan25}.

As in the previous analysis, we consider a transverse wave packet that carries only circular polarization 
and is incident from the region $z<0$. 
For simplicity, we place the center of the incident wave packet at the origin, $\bm{R}_{\parallel}=0$.  

We first note that the RH and LH components of the incident wave packet at the same position $\bm{R}_{\parallel}=0$
carry different TAM; they share the same orbital contribution but differ in spin.  
Hence we impose the boundary conditions~\eqref{163324_9Jun25} and~\eqref{163328_9Jun25} separately for RH and LH incidence: 
\begin{subequations}
\begin{align}
&
\begin{bmatrix}
 F_{-,n}(\omega, \bm{k}_{\parallel}, \Delta\bm{R}^{\text{RH}}_{\parallel, -, n}) \\[6pt]
 f_{+,n}(\omega, \bm{k}_{\parallel}, \Delta\bm{r}^{\text{RH}}_{\parallel, +, n}) 
\end{bmatrix}
\nonumber\\  &\quad = 
\begin{bmatrix}
 \mathcal{R}_{n, \text{RH}}(\omega, \bm{k}_{\parallel}) \\[6pt]
 \mathcal{T}_{n, \text{RH}}(\omega, \bm{k}_{\parallel})
\end{bmatrix}
\,  F_{+, \text{RH}}(\omega, \bm{k}_{\parallel}, \bm{R}_{\parallel} = 0), \label{184635_2Oct25} \\[6pt]
&
\begin{bmatrix}
 F_{-,n}(\omega, \bm{k}_{\parallel}, \Delta\bm{R}^{\text{LH}}_{\parallel, -, n}) \\[6pt]
 f_{+,n}(\omega, \bm{k}_{\parallel}, \Delta\bm{r}^{\text{LH}}_{\parallel, +, n}) 
\end{bmatrix}
\nonumber\\  &\quad = 
\begin{bmatrix}
 \mathcal{R}_{n, \text{LH}}(\omega, \bm{k}_{\parallel}) \\[6pt]
 \mathcal{T}_{n, \text{LH}}(\omega, \bm{k}_{\parallel})
\end{bmatrix}
\,  F_{+, \text{LH}}(\omega, \bm{k}_{\parallel}, \bm{R}_{\parallel} = 0)
\label{161131_6Feb25}
\end{align}
for each $n = \text{L}, \text{RH}, \text{LH}$. 
We express the transverse shifts arising from RH and LH incidence using superscripts RH and LH, respectively,  
as $\Delta\bm{R}^{\text{RH(LH)}}_{\parallel, -, n}$ and $\Delta\bm{r}^{\text{RH(LH)}}_{\parallel, +, n}$.  
These shifts satisfy the conservation of TAM as given by Eq.~\eqref{163328_9Jun25}:
\begin{align}
 &S^z_{+, \text{RH}}(\omega, \bm{k}_{\parallel}) 
 = S^z_{-, n}(\omega, \bm{k}_{\parallel}) + \left(\Delta\bm{R}^{\text{RH}}_{\parallel, -, n}\times \hbar \bm{k}_{\parallel}\right)_z\nonumber\\
& = \widetilde{S}^z_{+, n}(\omega, \bm{k}_{\parallel}) + 
\left(\Delta\bm{r}^{\text{RH}}_{\parallel, +, n}\times \hbar \bm{k}_{\parallel}\right)_z , \label{eq: b.c. IF shift formulation: conservation TAM c}\\
& S^z_{+, \text{LH}}(\omega, \bm{k}_{\parallel}) 
 = S^z_{-, n}(\omega, \bm{k}_{\parallel}) + 
\left(\Delta\bm{R}^{\text{LH}}_{\parallel, -, n}\times \hbar \bm{k}_{\parallel}\right)_z\nonumber\\
& = \widetilde{S}^z_{+, n}(\omega, \bm{k}_{\parallel}) + 
\left(\Delta\bm{r}^{\text{LH}}_{\parallel, +, n}\times \hbar \bm{k}_{\parallel}\right)_z 
\end{align}
\end{subequations}
From this relation, we immediately obtain (for $n = \text{L}, \text{RH}, \text{LH}$ and $m = \text{RH}, \text{LH}$):
\begin{subequations}
 \begin{align}
 \left(\Delta\bm{R}^{m}_{\parallel, -, n}\right)_{y'}
 &= - \frac{S^z_{+, m}(\omega, \bm{k}_{\parallel}) - S^z_{-, n}(\omega, \bm{k}_{\parallel})}{\hbar k\sin\theta_{\text{T}}},\\
 \left(\Delta\bm{r}^{m}_{\parallel, +, n}\right)_{y'}
 &=  - \frac{S^z_{+, m}(\omega, \bm{k}_{\parallel}) - \widetilde{S}^z_{+, n}(\omega, \bm{k}_{\parallel}) }{\hbar k\sin\theta_{\text{T}}},
\label{eq: acoustic IF shift expression}
\end{align}
where we have used the relations 
$\Abs{\bm{k}_{\parallel}} = k \sin\theta_{\text{T}}$, 
$\left(\Delta\bm{R}^m_{\parallel, -, n}\right)_{y'} = -\left(\Delta\bm{R}^m_{\parallel, -, n} \times \hat{\bm{k}}_{\parallel}\right)_z$, and
$\left(\Delta\bm{r}^m_{\parallel, +, n}\right)_{y'} = -\left(\Delta\bm{r}^m_{\parallel, +, n} \times \hat{\bm{k}}_{\parallel}\right)_z$. 
\end{subequations}

We now evaluate the expectation value of the transverse shift of the reflected longitudinal phonons. 
This value is the weighted average of the two shifts $\left(\Delta\bm{R}^{m}_{\parallel, -, \text{L}}\right)_{y'}$
for $m = \text{RH},\,\text{LH}$, 
each corresponding to incidence in the RH or LH mode, with weights determined by the distribution function of the reflected longitudinal phonons: 
\begin{align}
&\Braket{\left(\Delta \bm{R}_{\parallel, -, \text{L}}\right)_{y'}}\nonumber\\
&\equiv \frac{\displaystyle\sum_{m = \text{RH}, \text{LH}}
\left(\Delta \bm{R}^{m}_{\parallel, -, \text{L}}\right)_{y'}
 F_{-, \text{L}}(\omega, \bm{k}_{\parallel}, \Delta\bm{R}^{m}_{\parallel, -, \text{L}})}
{\displaystyle\sum_{m = \text{RH}, \text{LH}}
 F_{-, \text{L}}(\omega, \bm{k}_{\parallel}, \Delta\bm{R}^{m}_{\parallel, -, \text{L}})}.
\end{align}
The denominator on the right-hand side is evaluated using the boundary condition~\eqref{184635_2Oct25}, \eqref{161131_6Feb25}. 
The identity $\mathcal{R}_{\text{L}, \text{RH}} = \mathcal{R}_{\text{L}, \text{LH}}$, shown in Eqs.~\eqref{eq: symmetry of R and T a}--\eqref{eq: symmetry of R and T c}, 
yields
\begin{align}
&\mathcal{R}_{\text{L}, \text{RH}}(\omega, \bm{k}_{\parallel}) 
F_{+, \text{RH}}(\omega, \bm{k}_{\parallel}, \bm{R}_{\parallel} = 0)\nonumber\\
&\qquad  +  \mathcal{R}_{\text{L}, \text{LH}}(\omega, \bm{k}_{\parallel}) 
F_{+, \text{LH}}(\omega, \bm{k}_{\parallel}, \bm{R}_{\parallel} = 0) \nonumber\\
&= \mathcal{R}_{\text{L}, \text{RH}} \left( F_{+, \text{RH}} + F_{+, \text{LH}}\right),  
\end{align}
The numerator is similarly simplified as
\begin{align}
&- \frac{S^z_{+, \text{RH}} - S^z_{-, \text{L}}}{\hbar k\sin\theta_{\text{T}}} 
\mathcal{R}_{\text{L}, \text{RH}}F_{+, \text{RH}} \nonumber\\
&\qquad - \frac{S^z_{+, \text{LH}} - S^z_{-, \text{L}}}{\hbar k\sin\theta_{\text{T}}} 
\mathcal{R}_{\text{L}, \text{LH}}F_{+, \text{LH}}\nonumber\\
& = - \mathcal{R}_{\text{L}, \text{RH}}\frac{\cot\theta_{\text{T}}}{k}
\left(F_{+, \text{RH}}- F_{+, \text{LH}}\right) 
\end{align}
Combining these results, we obtain
\begin{equation}
\Braket{\left(\Delta \bm{R}_{\parallel, -, \text{L}}\right)_{y'}}
= - \frac{\cot\theta_{\text{T}}}{k}\,
\frac{F_{+, \text{RH}}- F_{+, \text{LH}}}{F_{+, \text{RH}}+ F_{+, \text{LH}}}.
\label{132836_6Nov24}
\end{equation}
The transverse shift of the transmitted L-mode is obtained in the same manner:
\begin{equation}
\Braket{\left(\Delta \bm{r}_{\parallel, +, \text{L}}\right)_{y'}}
 = - \frac{\cot\theta_{\text{T}}}{k}\,
\frac{F_{+, \text{RH}}- F_{+, \text{LH}}}{F_{+, \text{RH}}+ F_{+, \text{LH}}}.
\label{132839_6Nov24}
\end{equation}

We now evaluate the expectation value of the transverse shift for the reflected transverse wave. 
This value is given by the weighted average of the four shifts $\left(\Delta\bm{R}^{m}_{\parallel, -, n}\right)_{y'}$ 
for $n, m = \text{RH}, \text{LH}$, where the weights are determined by the distribution functions of the corresponding reflected transverse phonons. 
Explicitly, we have
\begin{align}
&\Braket{\left(\Delta \bm{R}_{\parallel, -, \text{T}}\right)_{y'}}\nonumber\\
&\equiv \frac{\displaystyle \sum_{n, m = \text{RH}, \text{LH}}
\left(\Delta\bm{R}^{m}_{\parallel, -, n}\right)_{y'}
F_{-, n}(\omega, \bm{k}_{\parallel}, \Delta\bm{R}^{m}_{\parallel, -, n})
}{\displaystyle \sum_{n, m = \text{RH}, \text{LH}}
F_{-, n}(\omega, \bm{k}_{\parallel}, \Delta\bm{R}^{m}_{\parallel, -, n})}.
\label{135259_28Jan25} 
\end{align}
Using the boundary condition~\eqref{161131_6Feb25} and the symmetry relations 
$\mathcal{R}_{\text{RH}, \text{RH}} = \mathcal{R}_{\text{LH}, \text{LH}}$ and 
$\mathcal{R}_{\text{RH}, \text{LH}} = \mathcal{R}_{\text{LH}, \text{RH}}$, 
we find that the denominator of Eq.~\eqref{135259_28Jan25} becomes
\begin{equation}
\left(\mathcal{R}_{\text{RH}, \text{RH}} + \mathcal{R}_{\text{LH}, \text{RH}}\right)\, 
\left(F_{+, \text{RH}} + F_{+, \text{LH}}\right),
\end{equation}
and the numerator simplifies to
\begin{align}
&\sum_{n, m = \text{RH}, \text{LH}}
\left(
- \frac{S^z_{+, m} - S^z_{-, n}}{\hbar k\sin\theta_{\text{T}}} \mathcal{R}_{nm} F_{+, m}
\right)\nonumber\\
&= - \mathcal{R}_{\text{RH}, \text{RH}} \frac{2 \cot\theta_{\text{T}}}{k}
\left(F_{+, \text{RH}} - F_{+, \text{LH}}\right). 
\end{align}
Combining these results, we obtain the expectation value of the transverse shift for the reflected transverse wave:
\begin{align}
&\Braket{\left(\Delta \bm{R}_{\parallel, -, \text{T}}\right)_{y'}}\nonumber\\
&= - \frac{2 \mathcal{R}_{\text{RH}, \text{RH}} }{\mathcal{R}_{\text{RH}, \text{RH}} + \mathcal{R}_{\text{LH}, \text{RH}} } 
\, \frac{\cot\theta_{\text{T}}}{k}\,
\frac{F_{+, \text{RH}} - F_{+, \text{LH}}}{F_{+, \text{RH}} + F_{+, \text{LH}}}.
\label{132846_6Nov24} 
\end{align}
The corresponding result for the transmitted transverse wave is obtained similarly:
\begin{align}
& \Braket{\left(\Delta \bm{r}_{\parallel, +, \text{T}}\right)_{y'}}\nonumber\\
& = -\frac{\cot\theta_{\text{T}}}{k}\, \frac{F_{+, \text{RH}} - F_{+, \text{LH}}}{F_{+, \text{RH}} + F_{+, \text{LH}}}\nonumber\\
&\times \left(
1- \frac{\cos\Theta_{\text{T}}}{\cos\theta_{\text{T}}}\, 
\frac{\mathcal{T}_{\text{RH},\text{RH}} - \mathcal{T}_{\text{LH},\text{RH}}}{
\mathcal{T}_{\text{RH},\text{RH}} + \mathcal{T}_{\text{LH},\text{RH}}}\right).
\label{132851_6Nov24} 
\end{align}

Next, we consider the energy flux density carried by two incident transverse wave packets in the RH and LH modes.  
These wave packets share the same central frequency and in-plane wavevector $(\omega, \bm{k}_{\parallel})$, 
and have a spread in wave-vector space of $\Delta\omega \Delta^2\bm{k}_{\parallel}$.  
Following the calculation in Eq.~\eqref{eq: net ene flux from coarse b.c.}, their energy fluxes are given by
\begin{subequations}
 \begin{align}
 j_{\mathfrak{u}, +, \text{RH}}
 &= \hbar\omega v^z_{\bm{k}, \text{RH}} F_{\bm{k}, \text{RH}} \frac{\Delta^3 \bm{k}}{(2\pi)^3}\nonumber\\
 &= \hbar\omega F_{+, \text{RH}}(\omega, \bm{k}_{\parallel}) \frac{\Delta \omega \Delta^2\bm{k}_{\parallel}}{(2\pi)^3},  \\
 j_{\mathfrak{u}, +, \text{LH}}
 &= \hbar\omega F_{+, \text{LH}}(\omega, \bm{k}_{\parallel}) \frac{\Delta \omega \Delta^2\bm{k}_{\parallel}}{(2\pi)^3}. 
\end{align}
\end{subequations}
Thus, the ratio
\begin{equation}
 \frac{F_{+, \text{RH}} - F_{+, \text{LH}}}{F_{+, \text{RH}} + F_{+, \text{LH}}}
= \frac{j_{\mathfrak{u}, +, \text{RH}} - j_{\mathfrak{u}, +, \text{LH}}}
{j_{\mathfrak{u}, +, \text{RH}} + j_{\mathfrak{u}, +, \text{LH}}} = \overline{\sigma}
\end{equation}
reduces to the average helicity in classical elasticity, as shown in Eq.~\eqref{135607_28Jan25}.
Therefore, the expressions for the transverse shifts of the reflected and transmitted wave packets obtained from the phonon framework, 
Eqs.~\eqref{132836_6Nov24}, \eqref{132839_6Nov24}, \eqref{132846_6Nov24}, and \eqref{132851_6Nov24}, 
exactly match the corresponding expressions derived from classical elasticity theory, Eqs.~\eqref{170438_23Jan25}--\eqref{135819_28Jan25}.
This agreement supports the validity of the detailed boundary conditions proposed in Eqs.~\eqref{163324_9Jun25} and~\eqref{163328_9Jun25}.

\section{Generation of OAM at the interface}
\label{sec: Generation of orbital angular momentum at the interface}

Using the detailed boundary conditions formulated in the previous section, 
we show the spatial distributions of spin, orbital, and total AM near the interface.  
As in our companion {Letter}~\cite{SuzukiSumitaKato2024a}, we study a junction of a chiral crystal under a temperature gradient and an achiral crystal.  
We introduce the OAM density function and adopt a junction model in which phonon wave packets are weakly bound to the system.  
Within the Boltzmann theory, we demonstrate that interfacial reflection and transmission modify the phonon ensemble and generate net OAM density and flux.  

{As in Sec.~\ref{sec: detailed b.c. formula}, we impose open boundary conditions in the $x$ and $y$ directions to ensure that the extrinsic OAM associated with the wave-packet positions is well defined within the present formulation.
Although we use a finite cross section to formulate the OAM density function, the interfacial OAM generation discussed below is not a finite-size edge effect. 
It originates from the discontinuity of the SAM flux at the interface, and the resulting cross-section-averaged OAM densities and fluxes are independent 
of the cross-sectional area and the specific form of the OAM density function, as shown below. 
The effects of the detailed edge structure and confinement are discussed separately in Secs.~\ref{140701_6Oct26} and \ref{140706_6Oct26}.} 
Throughout this section, we use lowercase ``$j$'' for fluxes, and uppercase ``$J$'' or roman ``$\mathrm{J}$'' for TAM.

\subsection{OAM density function}
\label{subsec: OAM dens fnc}

We introduce a function that characterizes the distribution of OAM carried by phonon wave packets in the system, which we refer to as the OAM density function.  

\begin{figure}[tbp]
\centering
\includegraphics[pagebox=artbox,width=0.99\columnwidth]{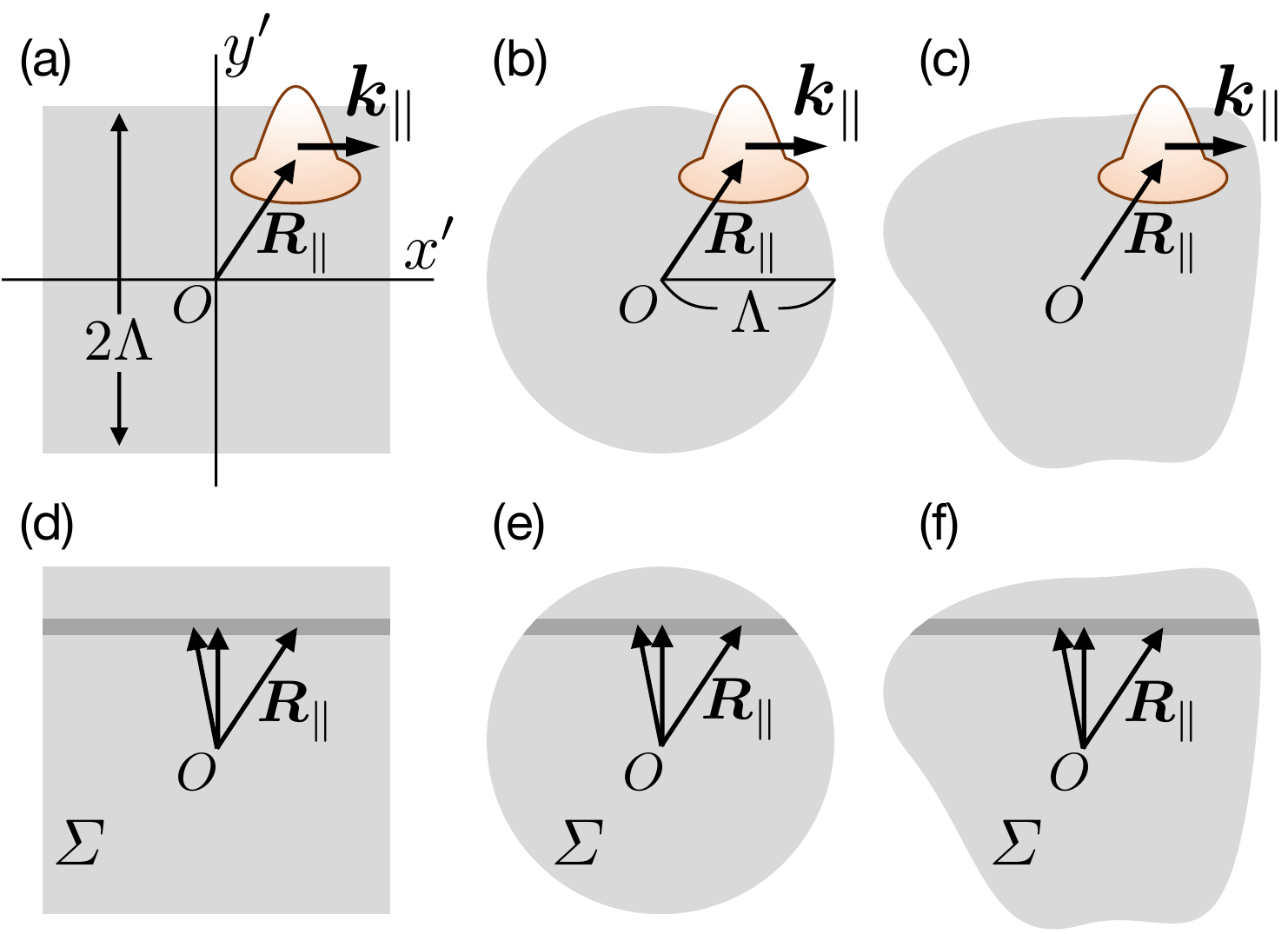}
\caption{
Schematic illustration of the OAM distribution carried by phonon wave packets (orange bell shapes) with central wavevector $\bm{k}_{\parallel}$ for cross sections with square [(a),(d)], circular [(b),(e)], and arbitrary [(c),(f)] shapes.
The origin $O$ is located at the centroid of each cross section.
The OAM of a wave packet,
$L^{z}=(\bm{R}_{\parallel}\times\hbar\bm{k}_{\parallel})_{z}$,
varies with its center-of-mass coordinate $\bm{R}_{\parallel}$.
Within the narrow strip parallel to $\bm{k}_{\parallel}$, shown as the dark gray regions in (d)–(f), $L^{z}$ remains constant.
The ratio of the strip area to the total cross-sectional area $\varSigma$ defines the OAM density function $W_{\bm{k}_{\parallel}}(L^{z})$.}
\label{fig:crosssection}
\end{figure}

As an example, we consider a junction with a square cross section parallel to the $xy$ plane and side length $2\Lambda$, illustrated in Figs.~\ref{fig:crosssection}(a) and \ref{fig:crosssection}(d).
We place the origin of the in-plane coordinate $\bm{R}_{\parallel}$ at the center of the square.  
When the in-plane wavevector $\bm{k}_{\parallel}$ is aligned with one side of the square, the $z$ component of the OAM,
$L^z = (\bm{R}_{\parallel}\times \hbar \bm{k}_{\parallel})_z = - \hbar \Abs{\bm{k}_{\parallel}} \left(\bm{R}_{\parallel}\right)_{y'}$,
is uniformly distributed over the interval 
$[- \hbar \Abs{\bm{k}_{\parallel}} \Lambda,\ + \hbar \Abs{\bm{k}_{\parallel}} \Lambda]$.  
Therefore, for the wave packets in the state labeled by the 4-tuple $(\omega, \bm{k}_{\parallel}, s, n)$, 
the fraction of wave packets with OAM within an infinitesimal interval $[L^z_0,~L^z_0+\Delta L^z]$ is given by
\begin{equation}
W_{\bm{k}_{\parallel}}(L^z_0)\, \Delta L^z = 
\Delta L^z \, 
\begin{cases}
\displaystyle
\frac{1}{2 \hbar \Abs{\bm{k}_{\parallel}} \Lambda} & \frac{\Abs{L^z_0}}{\hbar \Abs{\bm{k}_{\parallel}} \Lambda} < 1 \\
 0 & \text{otherwise}
\end{cases}.\label{125112_25Jan25}
\end{equation}

As a second example, we consider a circular cross section of radius $\Lambda$, 
shown in Figs.~\ref{fig:crosssection}(b) and \ref{fig:crosssection}(e).
In this case, the fraction of phonon wave packets in the state $(\omega, \bm{k}_{\parallel}, s, n)$  with OAM in the interval $[L^z_0,~L^z_0+\Delta L^z ]$ is given by
\begin{align}
&W_{\bm{k}_{\parallel}}(L^z_0)\, \Delta L^z \nonumber\\
&= \Delta L^z\, 
\begin{cases}
\displaystyle \frac{1}{\hbar \Abs{\bm{k}_{\parallel}} \Lambda}\, 
\frac{2}{\pi}\sqrt{1- \left(\frac{L^z_0}{\hbar \Abs{\bm{k}_{\parallel}} \Lambda}\right)^2}
& \frac{\Abs{L^z_0}}{\hbar \Abs{\bm{k}_{\parallel}} \Lambda} < 1 \\
 0 & \text{otherwise}
\end{cases}.\label{125118_25Jan25} 
\end{align}

To extend the notion of the OAM density beyond the previous examples, 
we consider phonon wave packets at a fixed state $(\omega, \bm{k}_{\parallel}, s, n)$ and 
vary only their centroid positions $\bm{R}_{\parallel}$ within
an arbitrary cross section 
[Figs.~\ref{fig:crosssection}(c) and \ref{fig:crosssection}(f)].
We then define $W_{\bm{k}_{\parallel}}(L^z_0)\, \Delta L^z \geq 0$ as the fraction of $\bm{R}_{\parallel}$ 
for which the $z$ component of the OAM carried by the wave packets lies within the interval $[L^z_0,~L^z_0 + \Delta L^z]$.
That is,
\begin{equation}
 W_{\bm{k}_{\parallel}}(L^z)
\equiv \int \frac{\rmd^2 \bm{R}_{\parallel}}{\varSigma}~\delta \bm{(} L^z(\bm{k}_{\parallel}, \bm{R}_{\parallel}) - L^z \bm{)}, 
\end{equation}
where 
$L^z(\bm{k}_{\parallel}, \bm{R}_{\parallel}) = (\bm{R}_{\parallel} \times \hbar \bm{k}_{\parallel})_z$
is given in Eq.~\eqref{eq: OAM definition Lz}, 
and the integration extends over the entire cross section. The cross-sectional area is denoted by $\varSigma \equiv \int \rmd^2 \bm{R}_{\parallel}$.
Since the shape of the cross section is independent of $z$, the function $W_{\bm{k}_{\parallel}}(L^z)$ also does not depend on $z$.

By construction, the function $W$ satisfies
\begin{equation}
 \int^{\infty}_{-\infty} \rmd L^z~W_{\bm{k}_{\parallel}}(L^z) = 1.\label{182450_28Jan25}
\end{equation}
We also place the origin of $\bm{R}_{\parallel}$ at the centroid of the cross section, so that $\int \rmd^2 \bm{R}_{\parallel}~\bm{R}_{\parallel} = 0$, 
which leads to
\begin{align}
&\int^{\infty}_{-\infty} \rmd L^z~L^z\,  W_{\bm{k}_{\parallel}}(L^z) \nonumber\\
&= \int^{\infty}_{-\infty} \rmd L^z~L^z 
\int \frac{\rmd^2 \bm{R}_{\parallel}}{\varSigma}~\delta \bm{(} \left(\bm{R}_{\parallel}\times \hbar \bm{k}_{\parallel}\right)_z - L^z \bm{)}\nonumber\\
& = \int \frac{\rmd^2 \bm{R}_{\parallel}}{\varSigma}~\left(\bm{R}_{\parallel}\times \hbar \bm{k}_{\parallel}\right)_z
= 0. \label{195758_28Jan25}
 \end{align}
Combining these properties, the following relation also holds for any $L'$:
\begin{align}
& \int^{\infty}_{-\infty} \rmd L^z~L^z \, W_{\bm{q}_{\parallel}} (L^z - L')\nonumber\\
&= \int^{\infty}_{-\infty} \rmd L^z~(L^z + L') \, W_{\bm{q}_{\parallel}} (L^z)
= L'. \label{eq: infty shift}
\end{align}

\subsection{Reformulation of the boundary conditions}

We introduce averaged distribution functions that explicitly depend on the OAM:
\begin{subequations}
 \begin{align}
 &\breve{F}_{s, n}(\omega, \bm{k}_{\parallel}, L^z, z)\nonumber\\
 &\equiv \frac{\int \rmd^2 \bm{R}_{\parallel}~{F}_{s, n}(\omega, \bm{k}_{\parallel}, \bm{R}_{\parallel}, z)\, 
 \delta \bm{(} \left(\bm{R}_{\parallel}\times \hbar \bm{k}_{\parallel}\right)_z - L^z \bm{)}}
 {\varSigma\, W_{\bm{k}_{\parallel}}(L^z)}, \label{181855_25Jan25}\\ 
 &\breve{f}_{s, n}(\omega, \bm{q}_{\parallel}, \widetilde{L}^z, z)\nonumber\\
 &\equiv \frac{\int \rmd^2 \bm{r}_{\parallel}~{f}_{s, n}(\omega, \bm{q}_{\parallel}, \bm{r}_{\parallel}, z)\, 
 \delta \bm{(} \left(\bm{r}_{\parallel}\times \hbar \bm{q}_{\parallel}\right)_z - \widetilde{L}^z \bm{)}}
 {\varSigma\, W_{\bm{q}_{\parallel}}(\widetilde{L}^z)}\label{181858_25Jan25}.
\end{align}
\end{subequations}
The functions $\breve{F}_{s, n}$ and $\breve{f}_{s, n}$ denote spatial averages of the detailed distribution functions ${F}_{s, n}$ and ${f}_{s, n}$, respectively. 
The averages are taken over the coordinate $x'$, which is aligned with either $\bm{k}_{\parallel}$ or $\bm{q}_{\parallel}$
(as indicated by the dark gray regions in Figs.~\ref{fig:crosssection}(d)--\ref{fig:crosssection}(f)).
In particular, the cross-sectional average of the distribution function is given by
\begin{align}
& \int \frac{\rmd^2 \bm{R}_{\parallel}}{\varSigma}~F_{s, n}(\omega, \bm{k}_{\parallel}, \bm{R}_{\parallel}, z)\nonumber\\
&= \int \rmd L^z~W_{\bm{k}_{\parallel}}(L^z) \breve{F}_{s, n}(\omega, \bm{k}_{\parallel}, L^z, z).\label{182513_28Jan25} 
\end{align}
An analogous relation holds between $f_{s, n}$ and $\breve{f}_{s, n}$.

We now derive the boundary conditions satisfied by the averaged distribution functions. 
We multiply the detailed boundary conditions~\eqref{163324_9Jun25} by the $\delta$ function
\begin{equation}
\delta \bm{(} \left(\bm{R}_{\parallel, -, n}\times \hbar \bm{k}_{\parallel}\right)_z + S^z_{-,n}(\omega, \bm{k}_{\parallel}) - J^z \bm{)}, 
\end{equation}
and then integrate over the in-plane coordinate $\bm{R}_{\parallel, -, n}$ across the cross section. The integration yields
\begin{widetext}
 \begin{subequations}
 \begin{align}
& \begin{bmatrix}
W_{\bm{k}_{\parallel}}( L^z_{-, n})\, \breve{F}_{-, n}(\omega, \bm{k}_{\parallel}, L^z_{-, n}, z = 0)\\[8pt]  
W_{\bm{k}_{\parallel}}( \widetilde{L}^z_{+, n})\, \breve{f}_{+, n}(\omega, \bm{k}_{\parallel}, \widetilde{L}^z_{+, n}, z = 0)
  \end{bmatrix}
\nonumber\\
&= \sum_{m = \text{L}, \text{RH}, \text{LH}}
\begin{bmatrix}
 \mathcal{R}_{nm}(\omega, \bm{k}_{\parallel}) & \mathcal{T}'_{nm}(\omega, \bm{k}_{\parallel}) \\[8pt]
 \mathcal{T}_{nm}(\omega, \bm{k}_{\parallel}) & \mathcal{R}'_{nm}(\omega, \bm{k}_{\parallel})
\end{bmatrix}
\begin{bmatrix}
W_{\bm{k}_{\parallel}}( L^z_{+, m})\, \breve{F}_{+, m}(\omega, \bm{k}_{\parallel}, L^z_{+, m}, z= 0)\\[8pt]
W_{\bm{k}_{\parallel}}( \widetilde{L}^z_{-, m})\, \breve{f}_{-, m} (\omega, \bm{k}_{\parallel}, \widetilde{L}^z_{-, m}, z = 0)
\end{bmatrix}. 
\label{130705_25Jan25}
\end{align}
According to Eq.~\eqref{163328_9Jun25}, the above result holds only when the TAM is conserved:
\begin{equation}
S^z_{-,n} (\omega, \bm{k}_{\parallel}) + L^z_{-,n}
= \widetilde{S}^z_{+, n} (\omega, \bm{k}_{\parallel}) + \widetilde{L}^z_{+,n}
= S^z_{+,m} (\omega, \bm{k}_{\parallel}) + L^z_{+, m}
= \widetilde{S}^z_{-, m} (\omega, \bm{k}_{\parallel}) + \widetilde{L}^z_{-, m}
\label{182214_28Jan25} 
\end{equation}
This condition applies to $n, m = \mathrm{L}, \mathrm{RH}, \mathrm{LH}$.
\end{subequations}
\end{widetext}

\subsection{Wave packets under weak cross-sectional confinement}

The center positions $\bm{R}_{\parallel}$ and $\bm{r}_{\parallel}$ of phonon wave packets are confined to a finite cross-sectional area in the $xy$ plane.  
Meanwhile, tight confinement of the wave packets complicates the TAM conservation at the interface; 
reflection or transmission induces transverse shifts of their centers, which can cause part of the wave packet to extend beyond the defined cross section.  

We address the difficulty by 
adopting a weak-confinement approximation, in which the center positions of incident and reflected/transmitted wave packets can lie beyond the nominal cross section.  
Physically, this relaxed confinement corresponds to a gradual decay of the mass density and elastic constants outside the cross-sectional region. 
This approximation allows the OAM density function $W_{\bm{k}_{\parallel}} (L^z)$ to remain positive even for large $L^z$.  

Even under this weak-confinement approximation, most wave packets remain well localized within the nominal cross section. 
In other words, we assume that $W_{\bm{k}_{\parallel}}(L^z)$ decays rapidly for $\Abs{L^z}$ beyond a certain threshold, 
reflecting the finite size of the cross section.

For example, we reconsider the case where the nominal cross section is a square of side length $2\Lambda$, which we addressed in Sec.~\ref{subsec: OAM dens fnc}.
We modify the OAM density function~\eqref{125112_25Jan25} by introducing a dimensionless parameter $b>0$:
\begin{equation}
W_{\bm{k}_{\parallel}}(L^z, b)
= \frac{1}{2\hbar \Abs{\bm{k}_{\parallel}} \Lambda}\,
\frac{b/\ln (1 + e^b)}{\displaystyle\exp\left[b\left(\frac{\Abs{L^z}}{\hbar \Abs{\bm{k}_{\parallel}} \Lambda} -1\right)\right] + 1}.
\label{131000_25Jan25}
\end{equation}
This expression resembles the Fermi distribution at finite temperature. 
In the limit $b \to \infty$, it reduces to the step-like behavior of the original form~\eqref{125112_25Jan25} with respect to $\Abs{L^z}$. 
The modified OAM density in Eq.~\eqref{131000_25Jan25} is strictly positive for all $-\infty < L^z < \infty$, 
satisfies the conditions in Eqs.~\eqref{182450_28Jan25} and \eqref{195758_28Jan25}, and decays rapidly near $\Abs{L^z} = \hbar \Abs{\bm{k}_{\parallel}} \Lambda$. 
The characteristic width of the smooth decay is about $\hbar \Abs{\bm{k}_{\parallel}} \Lambda / b$. 
We introduce the corresponding length scale
\begin{equation}
\xi_{\text{con}} \equiv \frac{\Lambda}{b},
\label{174048_6Feb25}
\end{equation}
which sets the spatial extent over which weakly confined phonon wave packets spread beyond the nominal boundary.

The choice of $W_{\bm{k}_{\parallel}}$ in Eq.~\eqref{131000_25Jan25} serves only as an example. 
In the subsequent analysis, we permit arbitrary forms of the OAM density function. 
As shown later, statistical averages of physical observables do not depend on the explicit form of $W_{\bm{k}_{\parallel}}$.

\subsection{Boltzmann theory}
\label{subsec: Boltzmann theory}

We now focus on a situation in which phonon OAM can emerge. 
Consider a junction composed of a chiral crystal under a temperature gradient ${\partial T}/{\partial z}$
in the region $z < 0$ and an achiral crystal in the region $z > 0$, as discussed in our companion {Letter}~\cite{SuzukiSumitaKato2024a}.  
In the chiral region ($z < 0$), the temperature gradient induces SAM of phonons polarized along the $z$ axis~\cite{Hamada2018,Zhang2025}.  
The achiral region ($z > 0$), in contrast, is free from external fields and absorbs the SAM diffusing across the interface. 
Experimental realizations of such chiral/achiral junctions have recently been reported~\cite{Kim2023,Ohe2024}.
Reference~\cite{SuzukiSumitaKato2024a} shows that the SAM flux exhibits a discontinuity at the interface. 
Since a smooth interface cannot be a source or sink of TAM, 
the generation of OAM at the interface, which compensates for the discontinuity of the SAM flux, is necessary. 

Our remaining task is to determine the magnitude and spatial distribution of the OAM density and flux along the $z$ direction. 
We formulate the phonon Boltzmann equation for the junction system, where $n = \text{L}, \text{RH}, \text{LH}$:  
\begin{subequations}  
\begin{equation}  
 v^z_{\bm{k}n}\frac{\partial F^{(1)}_{\bm{k}n} (\bm{R}_{\parallel}, z)}{\partial z}  
 + v^z_{\bm{k}n}\frac{\partial T}{\partial z} \frac{\partial F^{(0)}_{\bm{k}n}}{\partial T}  
 = - \frac{F^{(1)}_{\bm{k}n}(\bm{R}_{\parallel}, z)}{\tau} \label{122352_26Jan25}  
\end{equation}  
for the chiral crystal occupying the region $z<0$, and  
\begin{equation}  
 c^z_{\bm{q}n}\frac{\partial f^{(1)}_{\bm{q}n}(\bm{r}_{\parallel}, z)}{\partial z}  
 = - \frac{f^{(1)}_{\bm{q}n}(\bm{r}_{\parallel}, z)}{\widetilde{\tau}} \label{122357_26Jan25}  
\end{equation}  
for the achiral crystal in the region $z>0$.  
\end{subequations}  
The sound velocities in the regions $z<0$ and $z>0$ are denoted as $\bm{v}_{\bm{k}n}$ and $\bm{c}_{\bm{q}n}$, respectively.
The equilibrium distribution function is given by
\begin{equation}
F^{(0)}_{\bm{k}n} = \frac{1}{\exp\left(\hbar\Omega_{\bm{k}n}/k_{\text{B}}T\right) - 1}. 
\end{equation}
The functions $F^{(1)}$ and $f^{(1)}$ describe deviations from equilibrium that are linear in the temperature gradient $\partial T/\partial z$
in $z< 0$.

{In Eqs.~\eqref{122352_26Jan25} and \eqref{122357_26Jan25}, 
the relaxation times $\tau$ and $\widetilde{\tau}$ are introduced phenomenologically
for the chiral and achiral crystals, respectively; different values are allowed because
the two crystals generally have different scattering environments.
The present theory assumes low temperatures where long-wavelength acoustic phonons dominate;
impurity and boundary scattering are therefore expected to provide the primary relaxation mechanisms.
The corresponding relaxation times are typically of order $\si{ns}$,
which corresponds to mean free path of order $\si{\micro m}$
for typical sound velocities, as estimated from Table~8.3 and Eq.~(8.6.9) of Ref.~\cite{ZimanTextbookElPh}.
The relaxation times affect the quantitative values of the SAM and OAM distributions.
However, the interfacial generation of OAM itself follows directly from TAM conservation;
therefore, the qualitative conclusions regarding the generation and decay of OAM are not sensitive
to the precise values of $\tau$ and $\widetilde{\tau}$.}

In the Boltzmann equations~\eqref{122352_26Jan25} and~\eqref{122357_26Jan25}, we neglect four drift terms involving $x$ and $y$ derivatives:  
\begin{equation}
v^x_{\bm{k}n} \frac{\partial F^{(1)}_{\bm{k}n}}{\partial x}, \quad
v^y_{\bm{k}n} \frac{\partial F^{(1)}_{\bm{k}n}}{\partial y}, \quad
c^x_{\bm{q}n} \frac{\partial f^{(1)}_{\bm{q}n}}{\partial x}, \quad
c^y_{\bm{q}n} \frac{\partial f^{(1)}_{\bm{q}n}}{\partial y}. \label{eq: 4 drift terms}
\end{equation}
We assume these contributions are sufficiently small, 
while we retain the dependence of $F^{(1)} = F^{(1)}_{\bm{k}n}(\bm{R}_{\parallel}, z)$ on the in-plane coordinates $\bm{R}_{\parallel}$.  
In Appendix~\ref{sec: appendix validity of xy derivatives in BTE}, we show that this approximation holds when the confinement length $\xi_{\text{con}}$ of the phonon wave packet 
is much larger than the scale $\sqrt{\lambda l}$, where $\lambda = 2\pi/k$ is the phonon wavelength and $l$ is the mean free path.

From the Boltzmann equations~\eqref{122352_26Jan25} and~\eqref{122357_26Jan25}, $F^{(1)}$ and $f^{(1)}$ take the form~\cite{ZimanTextbookElPh}
\begin{subequations}
 \begin{align}
 &F^{(1)}_{\bm{k}n}(\bm{R}_{\parallel}, z) \nonumber\\
&= B_{\bm{k}n} 
 + \Theta_{\text{H}}(-v^z_{\bm{k}n}) \, C_{\bm{k}n}(\bm{R}_{\parallel})\exp\left[z\Big/({-\tau v^z_{\bm{k}n}})\right], \label{100545_30Sep25}\\ 
 &f^{(1)}_{\bm{q}n}(\bm{r}_{\parallel}, z) = \Theta_{\text{H}}(c^z_{\bm{q}n}) \, D_{\bm{q}n}(\bm{r}_{\parallel})
 \exp\left[-z\Big/{\widetilde{\tau} c^z_{\bm{q}n}}\right].\label{100548_30Sep25}
 \end{align}
\end{subequations}
Here we introduced the deviation of the distribution in the bulk of the chiral crystal as
\begin{equation}
 B_{\bm{k}n}= - \tau  {v}^z_{\bm{k}n} \frac{\partial T}{\partial z}\,\frac{\partial F^{(0)}_{\bm{k}n}}{\partial T}, 
\end{equation}
which is spatially uniform. 
The Heaviside step function $\Theta_{\text{H}}$ is defined by 
\begin{equation}
 \Theta_{\text{H}}(w) = 
\begin{cases}
1 & w >0\\ 0 & w <0
\end{cases}.
\label{eq: Heaviside step fnc}
\end{equation}
The amplitudes $C_{\bm{k} n}(\bm{R}_\parallel)$ 
and $D_{\bm{q} n}(\bm{r}_\parallel)$ near the interface are determined by the boundary conditions, as shown below.

We apply the averaging procedures in Eqs.~\eqref{181855_25Jan25} and \eqref{181858_25Jan25} to the solutions in Eqs.~\eqref{100545_30Sep25} and \eqref{100548_30Sep25} 
and obtain
\begin{subequations}
 \begin{align}
 &\breve{F}^{(1)}_{s, n}(\omega, \bm{k}_{\parallel}, L^z_{s, n}, z) \nonumber\\
&= B_{s, n}(\omega, \bm{k}_{\parallel})
 + \delta_{s, -} \, \breve{C}_{-, n}(\omega, \bm{k}_{\parallel}, L^z_{-, n}) \exp\left(\frac{z/\tau v_n}{\cos\theta_n}\right),
\label{182532_28Jan25}\\ 
 &\breve{f}^{(1)}_{s, n}(\omega, \bm{q}_{\parallel}, \widetilde{L}^z_{s, n}, z)\nonumber\\
 &= \delta_{s, +} \, \breve{D}_{+, n}(\omega, \bm{q}_{\parallel}, \widetilde{L}^z_{s, n})
 \exp\left(-\frac{z/\widetilde{\tau} c_n}{\cos\Theta_n}\right).\label{172201_29Jan25}
 \end{align}
\end{subequations}
We substitute these solutions into the boundary condition in Eq.~\eqref{130705_25Jan25} and obtain  
\begin{subequations}
 \begin{align}
 &W_{\bm{k}_{\parallel}}( L^z_{-, n}) 
 \left[ B_{-,n}(\omega, \bm{k}_{\parallel})
 + \breve{C}_{-,n}(\omega, \bm{k}_{\parallel}, L^z_{-, n}) \right]\nonumber\\
& = \sum_m \mathcal{R}_{nm}(\omega, \bm{k}_{\parallel})\,  W_{\bm{k}_{\parallel}}( L^z_{+, m})
 \, B_{+, m}(\omega, \bm{k}_{\parallel}), \label{101057_30Sep25}\\
 &W_{\bm{k}_{\parallel}}(\widetilde{L}^z_{+, n}) \breve{D}_{+,n}(\omega, \bm{k}_{\parallel}, \widetilde{L}^z_{+, n})\nonumber\\
 &= \sum_m \mathcal{T}_{nm}(\omega, \bm{k}_{\parallel})\, 
 W_{\bm{k}_{\parallel}}({L}^z_{+,m})\, B_{+, m}(\omega, \bm{k}_{\parallel}).\label{101102_30Sep25}
 \end{align}
According to Eq.~\eqref{182214_28Jan25}, TAM must be conserved:
\begin{align}
& S^z_{+,m}(\omega, \bm{k}_{\parallel}) + L^z_{+, m}
 = S^z_{-,n}(\omega, \bm{k}_{\parallel}) + L^z_{-, n} \nonumber\\
& = \widetilde{S}^z_{+,n}(\omega, \bm{k}_{\parallel}) + \widetilde{L}^z_{+, n},  
\end{align}
\end{subequations}
for $n, m = \text{L},\, \text{RH},\, \text{LH}$. 
Consequently, the amplitudes $\breve{C}_{-, n}$ and $\breve{D}_{+, n}$ 
in Eqs.~\eqref{101057_30Sep25} and \eqref{101102_30Sep25} depend on OAM:
\begin{widetext}
 \begin{subequations}
 \begin{align}
\breve{C}_{-,n}(\omega, \bm{k}_{\parallel}, L^z)
 &= - B_{-,n}(\omega, \bm{k}_{\parallel}) + \sum_m \mathcal{R}_{nm}(\omega, \bm{k}_{\parallel}) B_{+, m}(\omega, \bm{k}_{\parallel})\, 
 \frac{W_{\bm{k}_{\parallel}}\bm{(}L^z + S^z_{-,n}(\omega, \bm{k}_{\parallel}) - S^z_{+,m}(\omega, \bm{k}_{\parallel})\bm{)}
 }{W_{\bm{k}_{\parallel}}(L^z) },\label{143309_6Nov24}\\
  \breve{D}_{+,n}(\omega, \bm{k}_{\parallel}, L^z)
&= \sum_m \mathcal{T}_{nm}(\omega, \bm{k}_{\parallel}) B_{+, m}(\omega, \bm{k}_{\parallel})\, 
 \frac{
 W_{\bm{k}_{\parallel}}\bm{(}L^z + \widetilde{S}^z_{+,n}(\omega, \bm{k}_{\parallel}) - S^z_{+,m}(\omega, \bm{k}_{\parallel})\bm{)}
 }{W_{\bm{k}_{\parallel}}(L^z)}.\label{143312_6Nov24}
 \end{align}
 \end{subequations}
\end{widetext}
Within our approximation, the denominator $W_{\bm{k}_{\parallel}}(L^z)$ is positive for all values of $L^z$.

\subsection{SAM near the interface}

Before discussing OAM generation, we evaluate the SAM density $S_z$ polarized along the $z$ direction and the SAM flux $j^{\text{S}}_z$ flowing in the $z$ direction. 
Our objective is to verify that 
the SAM derived from the distribution function above matches the SAM obtained through the coarse-grained boundary condition; 
the latter SAM has been shown in our companion {Letter}~\cite{SuzukiSumitaKato2024a}.

For $z < 0$, using Eqs.~\eqref{182450_28Jan25}, \eqref{182513_28Jan25}, \eqref{182532_28Jan25}, and \eqref{143309_6Nov24} we obtain
\begin{widetext}
 \begin{align}
 & 
 \left\{
 \begin{aligned}
 & S_z (z < 0) \\
 & j^{\text{S}}_z (z < 0)
 \end{aligned}
 \right\}
 = \sum_n \int \frac{\rmd^3 \bm{k}}{(2\pi)^3}\int\frac{\rmd^2 \bm{R}_{\parallel}}{\varSigma}  
 \left\{
 \begin{aligned}
 & S^z_{\bm{k}n} \\
 & S^z_{\bm{k}n} v^z_{\bm{k}n}
 \end{aligned}
 \right\} F^{(1)}_{\bm{k}n}(\bm{R}_{\parallel}, z)\nonumber\\
 & = \sum_{s, n} \int 
 \frac{\rmd \omega \rmd^2\bm{k}_{\parallel} }{(2\pi)^3 v_n \cos\theta_n(\omega, \bm{k}_{\parallel})}
 \int \rmd L^z~W_{\bm{k}_{\parallel}}(L^z)
 S^z_{s, n} (\omega, \bm{k}_{\parallel})\, \left\{
 \begin{aligned}
 & 1 \\ & s\, v_n\cos\theta_n(\omega, \bm{k}_{\parallel}) 
 \end{aligned}
 \right\}\nonumber\\
 &\qquad \qquad \times\left\{
 B_{s,n}(\omega, \bm{k}_{\parallel})
 + \delta_{s, -} \breve{C}_{-,n}(\omega, \bm{k}_{\parallel}, L^z) 
 \exp\left[\frac{z/\tau v_n}{\cos\theta_n (\omega, \bm{k}_{\parallel})}\right] \right\}\nonumber\\
 & = \sum_{s,n}\int \frac{\rmd \omega \rmd^2\bm{k}_{\parallel} }{(2\pi)^3}
 \left\{
 \begin{aligned}
 & (v_n \cos\theta_n)^{-1} \\ & s
 \end{aligned}
 \right\}
 S^z_{s, n}
 B_{s,n} - \sum_n \int \frac{\rmd \omega \rmd^2\bm{k}_{\parallel} }{(2\pi)^3}
 \left\{
 \begin{aligned}
 & (v_n \cos\theta_n)^{-1} \\ & -1
 \end{aligned}
 \right\}
 S^z_{-, n}
 B_{-,n} \exp\left(\frac{z/\tau v_n}{\cos\theta_n}\right) \nonumber\\
 & + \sum_n \int \frac{\rmd \omega \rmd^2\bm{k}_{\parallel} }{(2\pi)^3}
 \left\{
 \begin{aligned}
 & (v_n \cos\theta_n)^{-1} \\ & -1
 \end{aligned}
 \right\}
 S^z_{-, n}
 \exp\left(\frac{z/\tau v_n}{\cos\theta_n}\right)
 \sum_m \mathcal{R}_{nm} B_{+,m}
 \underbrace{\int^{\infty}_{-\infty} \rmd L^z~W_{\bm{k}_{\parallel}}(L^z + S^z_{-, n} -S^z_{+,m})}_{\displaystyle =1 }. 
 \label{185434_5Nov24}
 \end{align}
\end{widetext}
Here we list the density and flux vertically in braces for clarity. 
For simplicity, we omit the explicit dependence on $(\omega, \bm{k}_{\parallel})$ in the final expression. 
The resulting expressions for the SAM density and flux, averaged over the $xy$ plane, are independent of both the cross-sectional area $\varSigma$ 
and the OAM density function $W$.

To confirm the consistency between Eq.~\eqref{185434_5Nov24} and the analysis in our companion {Letter}~\cite{SuzukiSumitaKato2024a}, we substitute $S^z_{s,n}$ 
from Eqs.~\eqref{eq: SAM for T mode def z<0} and \eqref{eq: SAM for L mode def z<0}, together with 
the distribution deviation in the RH and LH modes in the bulk region ($z < 0$) under a temperature gradient~\cite{SuzukiSumitaKato2024a},
 \begin{equation}
 B_{\bm{k}, \text{RH}} - B_{\bm{k}, \text{LH}}
\simeq \frac{\tau\chi k_z}{T}\frac{\partial T}{\partial z}
\left.\frac{\coth w - \frac{3}{2w}}{\sinh^2 w/w^2}\right|_{w=\frac{\hbar v_{\text{T}} k}{2k_{\text{B}}T}},
\label{205628_28Jan25}
\end{equation}
into Eq.~\eqref{185434_5Nov24}. 
Here $\chi$ is a constant that characterizes the $k^2$ splitting of the transverse-mode dispersion (see Refs.~\cite{Tsunetsugu2022,Tsunetsugu2026,SuzukiSumitaKato2024a}). 
We define the SAM $S_0$ induced by the temperature gradient in the chiral bulk ($z<0$) as $S_0 \propto \partial T/\partial z$~\footnote{
Since the region $z < 0$ consists of a chiral crystal, SAM is induced parallel to the temperature gradient~\cite{Hamada2018,Oiwa2022}.
According to Eq.~(9) of our companion {Letter}~\cite{SuzukiSumitaKato2024a}, the induced SAM density is 
$S_0 = (4\pi^2/45)\,\hbar \tau\chi \left(k_{\text{B}}/\hbar v_{\text{T}}\right)^4 T^3 \,\partial T/\partial z$.}.
The SAM flux then becomes
\begin{subequations}
 \begin{equation}
 \frac{j^{\text{S}}_z (z < 0)}{(3/2)S_0 v_{\text{T}}} = I_{3,1}\left[1 + \Delta \mathcal{R}(\theta); 1, \frac{z}{\tau v_{\text{T}}}\right]. \label{201937_28Jan25}
 \end{equation}
 Here we introduced $\Delta\mathcal{R}(\theta)
 \equiv \mathcal{R}_{\text{RH},\text{RH}}(\omega, \bm{k}_{\parallel}) - \mathcal{R}_{\text{LH},\text{RH}}(\omega, \bm{k}_{\parallel})$
 with $\theta$ the angle of incidence. The integral
 \begin{align}
 &I_{3,1}[g(\theta); \gamma, w]
 \equiv \int^{\pi/2}_0 \rmd\theta~\sin\theta\cos^2\theta \nonumber\\
 &\times \sqrt{1- \gamma^2\sin^2\theta} \, g(\theta)
 \exp\left[- \frac{\abs{w}}{\sqrt{1- \gamma^2\sin^2\theta}}\right],\label{eq: integral func 1.} %
 \end{align}
 includes exponential factors that describe the spatial decay of the SAM flux away from the interface. 
Equation~\eqref{201937_28Jan25} shows that the decay length is about the mean free path $\tau v_{\text{T}}$. 
 We obtain the SAM density ${S}_z(z<0)$ from the constitutive relation derived from Eq.~\eqref{185434_5Nov24}:
 \begin{equation}
 \frac{\partial {j}_z^{\text{S}} (z < 0)}{\partial z} = - \frac{{S}_z(z<0) - {S}_{0}}{\tau}.\label{110209_30Sep25} %
 \end{equation}
\end{subequations}

In the same manner, we determine the SAM flux and density in the region $z>0$:
\begin{subequations}
 \begin{align}
\frac{j^{\text{S}}_z (z > 0)}{(3/2)S_0 v_{\text{T}}} &= 
I_{3,1}\left[\Delta \mathcal{T}(\theta); \frac{c_{\text{T}}}{v_{\text{T}}}, \frac{z}{\widetilde{\tau} c_{\text{T}}}\right],  \label{110134_30Sep25}
\\[4pt]
 \frac{\partial {j}_z^{\text{S}} (z > 0)}{\partial z} &= - \frac{{S}_z(z > 0)}{\widetilde{\tau}}, \label{110137_30Sep25}
 \end{align}
\end{subequations}
where $\Delta\mathcal{T}(\theta)
\equiv \mathcal{T}_{\text{RH},\text{RH}}(\omega, \bm{k}_{\parallel}) - \mathcal{T}_{\text{LH},\text{RH}}(\omega, \bm{k}_{\parallel})$.
Equations~\eqref{201937_28Jan25}, \eqref{110209_30Sep25}, \eqref{110134_30Sep25} and \eqref{110137_30Sep25}
coincide with Eqs.~(5a), (5b), and {(10a)} in 
our companion {Letter}~\cite{SuzukiSumitaKato2024a}.

\subsection{OAM near the interface}
\label{153615_6Oct26}

We now evaluate the OAM density $L_z(z)$ and its flux $j^{\mathrm{L}}_z(z)$ with both propagation and polarization along the $z$ axis. 
Using Eqs.~\eqref{195758_28Jan25}, \eqref{eq: infty shift}, \eqref{182513_28Jan25}, \eqref{182532_28Jan25}, \eqref{172201_29Jan25}, \eqref{143309_6Nov24}, 
and \eqref{143312_6Nov24}, we obtain 
\begin{widetext}
\begin{subequations}
  \begin{align}
 & j^{\text{L}}_z(z < 0)
 = \sum_n \int\frac{\rmd^3 \bm{k}}{(2\pi)^3}
 \int \frac{\rmd^2 \bm{R}_{\parallel}}{\varSigma}
 \left(\bm{R}_{\parallel} \times \hbar \bm{k}_{\parallel}\right)_z v^z_{\bm{k}n} F^{(1)}_{\bm{k}n}(\bm{R}_{\parallel}, z)\nonumber\\
 & = \sum_{s, n} \int 
 \frac{\rmd \omega \rmd^2\bm{k}_{\parallel} }{(2\pi)^3 v_n \cos\theta_n(\omega, \bm{k}_{\parallel})}
 \int \rmd L^z~W_{\bm{k}_{\parallel}}(L^z) L^z\, s\, v_n\cos\theta_n(\omega, \bm{k}_{\parallel}) \nonumber\\
 &\qquad \qquad \times\left[
 B_{s,n}(\omega, \bm{k}_{\parallel})
 + \delta_{s, -} \breve{C}_{-,n}(\omega, \bm{k}_{\parallel}, L^z) 
 \exp\left(\frac{z/\tau v_n}{\cos\theta_n}\right) \right]\nonumber\\
 & = -\sum_n \int \frac{\rmd \omega \rmd^2\bm{k}_{\parallel} }{(2\pi)^3}
 \exp\left(\frac{z/\tau v_n}{\cos\theta_n}\right)
 \sum_m \mathcal{R}_{nm} B_{+,m}
 \underbrace{\int^{\infty}_{-\infty} \rmd L^z~L^z\, 
 W_{\bm{k}_{\parallel}}(L^z + S^z_{-, n} -S^z_{+,m})}_{\displaystyle = S^z_{+,m} - S^z_{-, n}},\label{200521_28Jan25}
 \end{align}
 \begin{align}
 & j^{\text{L}}_z(z> 0)
 =\sum_n \int \frac{\rmd^3 \bm{q}}{(2\pi)^3}\int\frac{\rmd^2 \bm{r}_{\parallel}}{\varSigma}
 \left(\bm{r}_{\parallel} \times \hbar\bm{q}_{\parallel}\right)_z c^z_{\bm{q}n}
 f^{(1)}_{\bm{q}n}(\bm{r}_{\parallel}, z)\nonumber\\
 & = \sum_{n} \int \frac{\rmd \omega \rmd^2\bm{q}_{\parallel} }{(2\pi)^3}
 \exp\left(-\frac{z/\widetilde{\tau} c_n}{\cos\Theta_n}\right)
 \int^{\infty}_{-\infty} \rmd L^z~L^z\, W_{\bm{q}_{\parallel}}(L^z)\breve{D}_{+,n}(\omega, \bm{q}_{\parallel}, L^z)\nonumber\\
 &= \sum_{n} \int \frac{\rmd \omega \rmd^2\bm{q}_{\parallel} }{(2\pi)^3}
 \exp\left(-\frac{z/\widetilde{\tau} c_n}{\cos\Theta_n}\right)\sum_m \mathcal{T}_{nm} B_{+, m} \underbrace{\int^{\infty}_{-\infty} \rmd L^z~L^z \, W_{\bm{q}_{\parallel}}
 (L^z + \widetilde{S}^z_{+, n} - S^z_{+,  m})}_{\displaystyle = 
 S^z_{+,  m} - \widetilde{S}^z_{+, n}}.\label{200528_28Jan25}
 \end{align}
\end{subequations}
\end{widetext}
For both $z < 0$ and $z > 0$, the final expressions are independent of the cross-sectional area $\varSigma$ 
and the OAM density function $W$.
In particular, using Eq.~\eqref{eq: b.c. IF shift formulation: conservation TAM c}--\eqref{eq: acoustic IF shift expression}, the OAM flux at the interface 
is rewritten as
\clearpage
\begin{subequations}
 \begin{align}
 & j^{\text{L}}_z(z= -0)\nonumber\\
 &= - \sum_{n,m}\int \frac{\rmd \omega \rmd^2\bm{k}_{\parallel} }{(2\pi)^3}
 \left(\Delta\bm{R}^m_{\parallel, -, n}\times \hbar \bm{k}_{\parallel}\right)_z \mathcal{R}_{nm} B_{+,m}, \label{eq: jLz -0 }\\
 & j^{\text{L}}_z(z= +0)\nonumber\\
 & =  \sum_{n,m}\int \frac{\rmd \omega \rmd^2\bm{q}_{\parallel} }{(2\pi)^3}
 \left(\Delta\bm{r}^m_{\parallel, +, n}\times \hbar \bm{q}_{\parallel}\right)_z \mathcal{T}_{nm} B_{+,m}. \label{eq: jLz +0 }
 \end{align}
\end{subequations}
Here $\Delta\bm{R}^m_{\parallel, -, n}$ and $\Delta\bm{r}^m_{\parallel, +, n} $ are transverse shifts of the reflected and transmitted phonons, respectively.
The functions $\mathcal{R}_{nm} B_{+,m}$ and $\mathcal{T}_{nm} B_{+,m}$ 
give the nonequilibrium distribution of the reflected and transmitted phonons, respectively, while $B_{+,m}$ specifies the nonequilibrium distribution of the incident phonons.

\begin{figure}[tbp]
\centering
\includegraphics[pagebox=artbox,width=0.99\columnwidth]{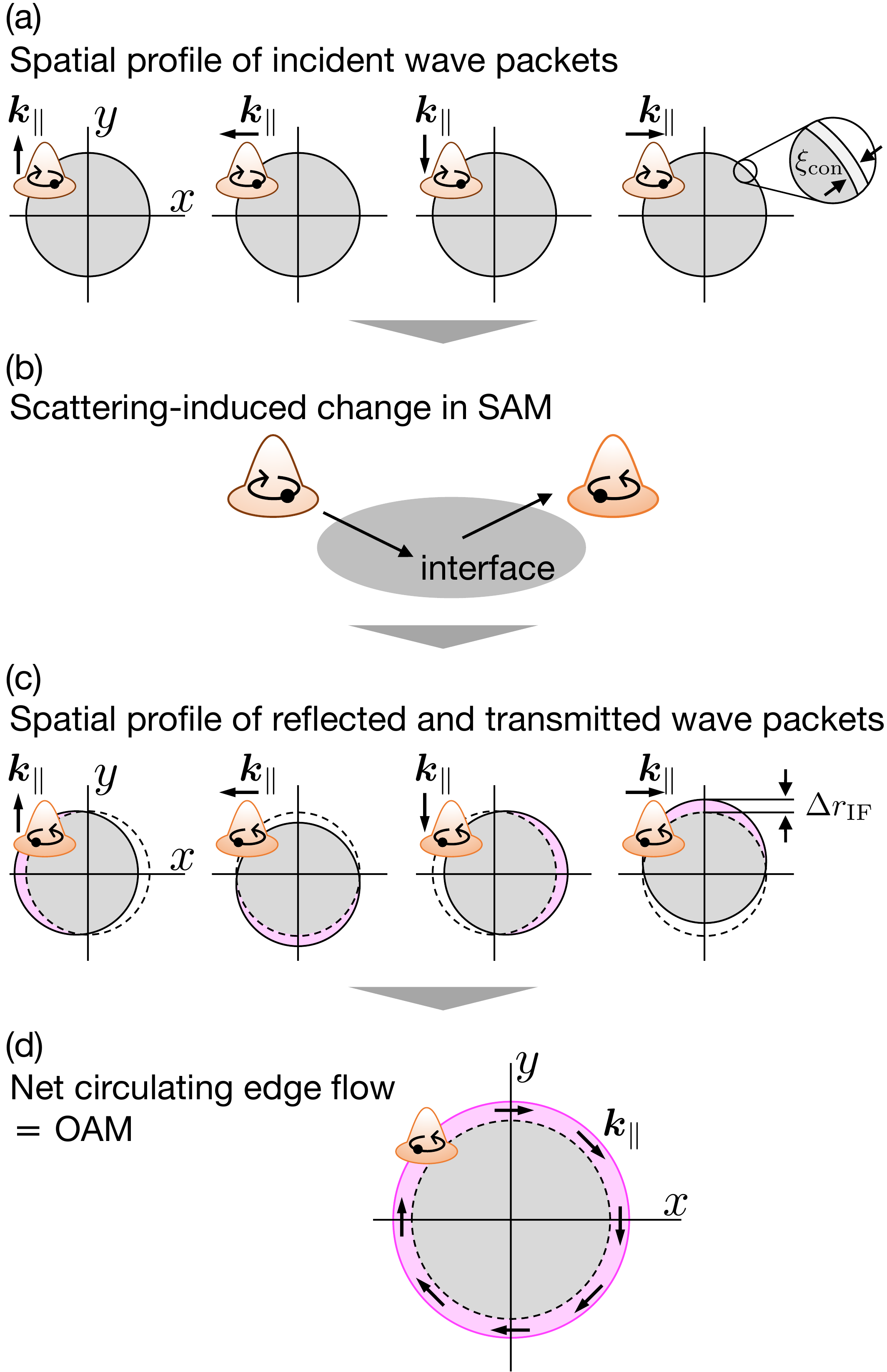}
\caption{
Mechanism of OAM generation at a smooth interface, shown in four stages.  
(a)~A set of phonon ensembles incident from the bulk, representing deviations from equilibrium, spreads uniformly across the cross section (gray circles) 
with no dependence on $\bm{k}_{\parallel}$; 
these ensembles therefore carry no OAM. The confinement length at the cross-sectional edge, $\xi_{\mathrm{con}}$, is indicated in the inset.  
(b)~Phonon wave packets in these ensembles (orange bell shapes) are scattered at the interface, and part of each ensemble undergoes 
circular polarization reversal, changing the SAM.  
(c)~To compensate for this SAM change, the scattered wave packets are displaced by $\Delta r_{\mathrm{IF}}$, 
shifting the ensembles in the direction perpendicular to $\bm{k}_{\parallel}$.  
(d)~The remaining OAM appears as a mechanical AM: 
an imbalance of $\hbar \bm{k}_{\parallel}$ at the edges, interpreted as a circulating phonon flow along the boundary (pink shaded region).
Note that the width of the region is exaggerated for clarity; 
under the weak-confinement condition the correct relation is $\Delta r_{\mathrm{IF}} \ll \xi_{\mathrm{con}}$~(see also 
Appendix~\ref{sec: appendix validity of xy derivatives in BTE}). 
}
\label{165750_6Nov24}
\end{figure}

We interpret the OAM generation described by Eqs.~\eqref{eq: jLz -0 } and \eqref{eq: jLz +0 } with the aid of Figs.~\ref{165750_6Nov24}(a)--\ref{165750_6Nov24}(d).  
Phonon wave packets in the ensemble $B_{+, m}(\omega, \bm{k}_{\parallel})$ are scattered at the interface; 
part of each ensemble reverses its circular polarization and changes the SAM from $S^z_{+, m}$ to $S^z_{-, n}$ or $\widetilde{S}^z_{+, n}$.  
To compensate for this SAM change, the scattered wave packets shift by $\Delta r_{\mathrm{IF}} = \Delta\bm{R}^m_{\parallel, -, n}, \,\Delta\bm{r}^m_{\parallel, +, n}$. 
This displacement moves the ensembles perpendicular to $\bm{k}_{\parallel}$ or $\bm{q}_{\parallel}$.  
As a result, the scattered ensembles acquire net OAM, 
as a statistical average of $\left(\Delta\bm{R}^m_{\parallel, -, n}\times \hbar \bm{k}_{\parallel}\right)_z$
or $\left(\Delta\bm{r}^m_{\parallel, +, n}\times \hbar \bm{q}_{\parallel}\right)_z$. 
This net OAM gives rise to a circulating phonon flow along the edge, 
as illustrated in Figs.~\ref{fig: intro schematics}(d) and \ref{165750_6Nov24}(d). 
The circulating edge flow arises from bulk modes; it differs qualitatively from the energy transport associated with surface modes, such as whispering gallery modes.

We return to Eqs.~\eqref{200521_28Jan25} and \eqref{200528_28Jan25}. 
We simplify these equations by substituting the expressions for $S^z_{s,n}$, shown in Eqs.~\eqref{eq: SAM for T mode def z<0} and \eqref{eq: SAM for L mode def z<0}, 
and the circularly polarized phonon population~\eqref{205628_28Jan25} under the temperature gradient $\partial T/\partial z$ in the region $z<0$. 
After algebraic simplification, we derive an analytical integral expression that explicitly involves the reflectance 
$\Delta \mathcal{R}(\theta)$ and transmittance $\Delta \mathcal{T}(\theta)$: 
\begin{subequations}
\begin{align}
&\frac{j^{\mathrm{L}}_z(z<0)}{(3/2) S_0 v_{\mathrm{T}}} = \widetilde{I}\left[- \mathcal{R}_{\mathrm{L},\mathrm{RH}}(\theta); 
\frac{v_{\mathrm{L}}}{v_{\mathrm{T}}}, \frac{z}{\tau v_{\mathrm{L}}}\right]\nonumber\\
&\quad  + \widetilde{I}\left[1 - \mathcal{R}_{\mathrm{RH},\mathrm{RH}}(\theta)
- \mathcal{R}_{\mathrm{LH},\mathrm{RH}}(\theta); 1, \frac{z}{\tau v_{\mathrm{T}}}\right]\nonumber\\
&\quad - I_{3,1}\left[1 + \Delta \mathcal{R}(\theta); 1, \frac{z}{\tau v_{\text{T}}}\right],\label{201219_7Oct25}\\
&\frac{j^{\mathrm{L}}_z(z>0)}{(3/2) S_0 v_{\mathrm{T}}}
= \widetilde{I}\left[\mathcal{T}_{\mathrm{L},\mathrm{RH}}(\theta); 
\frac{c_{\mathrm{L}}}{v_{\mathrm{T}}}, \frac{z}{\widetilde{\tau} c_{\mathrm{L}}}\right]\nonumber\\
&\quad + \widetilde{I}\left[\mathcal{T}_{\mathrm{RH},\mathrm{RH}}(\theta)
+ \mathcal{T}_{\mathrm{LH},\mathrm{RH}}(\theta); \frac{c_{\mathrm{T}}}{v_{\mathrm{T}}}, 
\frac{z}{\widetilde{\tau} c_{\mathrm{T}}}\right]\nonumber\\
& - I_{3,1}\left[\Delta \mathcal{T}(\theta); \frac{c_{\text{T}}}{v_{\text{T}}}, \frac{z}{\widetilde{\tau} c_{\text{T}}}\right].\label{201224_7Oct25}
\end{align} 
\end{subequations}
Here we introduce the following integral in addition to Eq.~\eqref{eq: integral func 1.}:
\begin{align}
\widetilde{I}[g(\theta); \gamma, w] &\equiv
\int^{\pi/2}_0\rmd\theta~\sin\theta\cos^3\theta\nonumber\\
&\times g(\theta) \exp\left[-\frac{\Abs{w}}{\sqrt{1- \gamma^2\sin^2\theta}}\right]. 
\end{align}
The OAM density polarized along the $z$ direction, $L_z(z)$, is obtained through the constitutive relation:
\begin{subequations}
 \begin{align}
 \frac{\partial j^{\text{L}}_z}{\partial z} &= - \frac{L_z}{\tau}, &z&< 0,\label{201136_7Oct25}\\
 \frac{\partial j^{\text{L}}_z}{\partial z}& = - \frac{L_z}{\widetilde{\tau}}, &z&>0.\label{201139_7Oct25}
 \end{align}
\end{subequations}

\subsection{TAM near the interface}

The TAM flux along the $z$ axis, $j^{\text{J}}_z$, with both propagation and polarization aligned with $z$, 
is given as the sum of the spin and orbital components:
\begin{equation}
j^{\text{J}}_z(z ) = j^{\text{S}}_z(z) + j^{\text{L}}_z(z). 
\end{equation}
The explicit expression for $j^{\text{J}}_z$ is summarized as:
\begin{subequations}
 \begin{align}
 \frac{j^{\mathrm{J}}_z(z<0)}{(3/2) S_0 v_{\mathrm{T}}}
 &= \widetilde{I}\left[- \mathcal{R}_{\mathrm{L},\mathrm{RH}}(\theta); 
 \frac{v_{\mathrm{L}}}{v_{\mathrm{T}}}, \frac{z}{\tau v_{\mathrm{L}}}\right]\nonumber\\
 &\!\!\!\!\!\!\!\!\!\!\!\!\!\!\!\!\!\!\!\! + \widetilde{I}\left[1 - \mathcal{R}_{\mathrm{RH},\mathrm{RH}}(\theta)
 - \mathcal{R}_{\mathrm{LH},\mathrm{RH}}(\theta); 1, \frac{z}{\tau v_{\mathrm{T}}}\right],\label{185446_6Feb25}\\
 \frac{j^{\mathrm{J}}_z(z>0)}{(3/2) S_0 v_{\mathrm{T}}}
 &= \widetilde{I}\left[\mathcal{T}_{\mathrm{L},\mathrm{RH}}(\theta); 
 \frac{c_{\mathrm{L}}}{v_{\mathrm{T}}}, \frac{z}{\widetilde{\tau} c_{\mathrm{L}}}\right]\nonumber\\
 &\!\!\!\!\!\!\!\!\!\!\!\! + \widetilde{I}\left[\mathcal{T}_{\mathrm{RH},\mathrm{RH}}(\theta)
 + \mathcal{T}_{\mathrm{LH},\mathrm{RH}}(\theta); \frac{c_{\mathrm{T}}}{v_{\mathrm{T}}}, 
 \frac{z}{\widetilde{\tau} c_{\mathrm{T}}}\right].\label{185449_6Feb25}
 \end{align}
\end{subequations}
This result satisfies the conservation of TAM at the interface and ensures the continuity condition:
\begin{equation}
j^{\mathrm{J}}_z(z = -0) = j^{\mathrm{J}}_z(z = +0). 
\end{equation}
This equality follows directly from applying the energy conservation condition~\eqref{eq: R and T sum unity 1} 
to Eqs.~\eqref{185446_6Feb25} and \eqref{185449_6Feb25}.

The TAM density polarized along the $z$ direction, $J_z$, is also obtained from the constitutive relation:
\begin{subequations}
 \begin{align}
 \frac{\partial {j}_z^{\text{J}} (z)}{\partial z} &= - \frac{{J}_z(z) - {S}_{0}}{\tau},  &z& < 0, \label{110849_7Feb25}\\
 \frac{\partial {j}_z^{\text{J}} (z)}{\partial z} &= - \frac{{J}_z(z)}{\widetilde{\tau}}, &z& > 0. \label{110851_7Feb25}
 \end{align}
\end{subequations}

\subsection{Generation of OAM in selected junction systems}

We demonstrate OAM generation at crystal interfaces induced by a local temperature gradient, using representative junctions.
Quartz is chosen as the chiral crystal, following the setup of the interfacial AM transport experiment in Ref.~\cite{Ohe2024}.  
The calculated densities and fluxes of SAM, OAM, and TAM, obtained from  
Eqs.~\eqref{201937_28Jan25}, \eqref{110209_30Sep25}, \eqref{110134_30Sep25}, \eqref{110137_30Sep25}, 
\eqref{201219_7Oct25}, \eqref{201224_7Oct25}, \eqref{201136_7Oct25}, \eqref{201139_7Oct25}, 
\eqref{185446_6Feb25}, \eqref{185449_6Feb25}, \eqref{110849_7Feb25} and \eqref{110851_7Feb25},  
are presented in Figs.~\ref{fig: AM spatial distribution}(a) and \ref{fig: AM spatial distribution}(b).
In particular, Fig.~\ref{fig: AM spatial distribution}(b) demonstrates the continuity of the TAM flux across the interface.

{We highlight three key features of the results shown in Figs.~\ref{fig: AM spatial distribution}(a) and \ref{fig: AM spatial distribution}(b).}
First, both the OAM density and its flux reach their maximum near the interface.  
This behavior arises because the amplitudes of the reflected and transmitted phonon distribution functions, $C_{\bm{k}n}$ and $D_{\bm{q}n}$, which govern OAM transport, 
are localized near the interface, as indicated by Eqs.~\eqref{100545_30Sep25} and \eqref{100548_30Sep25}.

Second, in all junction systems studied, the OAM density tends to be positive, while its flux is negative in the region $z < 0$. 
This behavior implies that reflected wave packets acquire positive OAM at the interface, but their flux becomes negative 
due to their negative group velocity.

Third, the OAM can reach magnitudes comparable to those of the SAM. 
In the quartz/vacuum and quartz/platinum junctions, the OAM density and flux on the $z < 0$ side near the interface 
become comparable to the spin components. 
This result highlights that, in studying phonon AM transport across an interface, 
one must account not only for circular polarization but also for the contribution from OAM.

\section{Discussion}
\label{sec: Discussion}

\subsection{Strict confinement of wave packets}
\label{140701_6Oct26}

In the previous section, we assumed that the cross-sectional boundary in the $xy$ plane is smooth and not sharply defined. 
{In particular, we neglected phonon reflection at the edge. 
A more rigorous treatment includes this reflection; it also restores the drift terms in Eq.~\eqref{eq: 4 drift terms}. 
This treatment requires boundary conditions based on the acoustic power reflectance along the edge. 
This regime corresponds to the strict-confinement limit, $\xi_{\mathrm{con}}\to0$.}

{We can clarify the range of applicability of the weak-confinement approximation.
Let $\lambda$, $l$, and $\xi_{\mathrm{con}}$ denote the phonon wavelength, mean free path, and confinement length near the cross-sectional edge, respectively.
We can neglect edge reflection when $\xi_{\mathrm{con}}\gg\lambda$.
Moreover, we can neglect phonon accumulation near the edge when $\xi_{\mathrm{con}}\gg\sqrt{\lambda l}$; 
this is precisely the condition under which the in-plane drift terms in Eq.~\eqref{eq: 4 drift terms} can be neglected, 
as derived in Appendix~\ref{sec: appendix validity of xy derivatives in BTE}.
Since we assume the regime $l\gg\lambda$, the condition $\xi_{\mathrm{con}}\gg\sqrt{\lambda l}$ also guarantees $\xi_{\mathrm{con}}\gg\lambda$.
Therefore, the weak-confinement approximation is applicable when
\begin{equation}
  \xi_{\mathrm{con}} \gg \sqrt{\lambda l}.
\end{equation}
Otherwise, we must consider a formulation for strictly confined wave packets.}

Meanwhile, we encounter an apparent inconsistency near the edge in the strict-confinement regime.  
The transverse shift of phonon wave packets is strongly suppressed, 
yet the conservation of TAM and phonon energy, which would otherwise escape, must still be maintained. 
Thus, the phonon distribution function near the boundary must undergo a redistribution that the detailed boundary conditions 
in Eqs.~\eqref{163324_9Jun25} and \eqref{163328_9Jun25} fail to capture.  
How this redistribution occurs under strict confinement and affects OAM generation remains an open issue.

\subsection{Dependence of OAM on edge and confinement}
\label{140706_6Oct26}

In Sec.~\ref{sec: Generation of orbital angular momentum at the interface}, we demonstrated the generation of OAM under weak confinement of phonon wave packets.  
{The interfacial generation mechanism itself does not rely on the detailed structure of the cross-sectional edge. 
Nevertheless, the edge and confinement can give additional contributions to the OAM through their modification of the phonon wave-packet motion.}

{The motion of phonon wave packets along the boundary is sensitive to the edge structure and to the nature of the confinement. }
A variation in this motion produces a momentum shift $\hbar \Delta \bm{k}_{\parallel}$. 
This shift contributes an OAM of order $\hbar |\Delta \bm{k}_{\parallel}| \Lambda$, where $\Lambda$ denotes the typical length of the cross section.  
Integration over the perimeter, of order $\Lambda$, gives a macroscopic OAM variation proportional to the cross-sectional area $\varSigma \sim \Lambda^2$.
{This scaling refers to these additional edge- and confinement-related contributions and should not be confused with the interfacial OAM generation discussed 
in Sec.~\ref{sec: Generation of orbital angular momentum at the interface}. 
Within the regime of validity of our weak-confinement treatment, the cross-sectional average of the interfacial OAM is independent of 
the cross-sectional area, as shown in Sec.~\ref{153615_6Oct26}. 
Thus, reducing the cross-sectional area reduces the OAM proportionally to the area, 
but does not by itself selectively suppress the interfacial OAM contribution relative to the SAM contribution.}

The variation in the motion of phonon wave packets can arise from elastic spin-orbit coupling near the edge~\cite{Bliokh2006}.  
When the ratio $\lambda / \xi_{\text{con}}$, with $\lambda$ the wavelength and $\xi_{\text{con}}$ the characteristic length of spatial variation 
in mass density or elastic constants, becomes of order unity, the anomalous velocity becomes comparable to the group velocity.  
The phonon trajectory then bends along the cross-sectional edge.
{Such effects correspond to sharp spatial variations near the boundary and lie beyond the weak-confinement regime considered 
in Sec.~\ref{sec: Generation of orbital angular momentum at the interface}.
A related dependence of OAM on confinement has also been discussed for chiral superfluids, where} 
the OAM varies with the relation between the coherence length 
and the length associated with the curvature of the confining potential near the boundary~\cite{Tada2018}.

{\subsection{Experimental signatures of phonon SAM and OAM transfer}}

{Experimental detection of phonon SAM transfer has already been reported using the inverse spin Hall effect~\cite{Ohe2024}. 
Phonon SAM transmitted across the interface is expected to transfer AM to electron spins in an adjacent heavy metal, generating an inverse spin Hall voltage.
By contrast, experimental observation of phonon OAM transfer has not yet been established. 
One possible experimental signature is a circulating heat current near the edge of the interface between two dissimilar crystals. 
When phonon SAM accumulates on one side of the interface, for example because only one crystal is subject to a thermal gradient, 
the leakage of phonon SAM generates phonon OAM near the interface. 
As depicted in Fig.~\ref{165750_6Nov24}, the generated phonon OAM may manifest itself as a circulating heat current.}

{We finally note a limitation of the present theoretical framework concerning macroscopic rigid-body rotation. 
The TAM considered in this study is defined for the phonon displacement field with the equilibrium lattice positions held fixed, 
as discussed by Nakane and Kohno~\cite{Nakane2018}.
The dynamical degree of freedom associated with rigid-body rotation of the entire crystal is not included in our formulation.
Accordingly, we do not identify the phonon TAM considered here with the AM associated with rigid-body rotation.}

{In a finite sample, whether the relaxation of phonon TAM is accompanied by a transfer of AM to macroscopic mechanical rotation remains an open question. 
The present framework does not allow us to address this issue, because it does not include the rigid-body rotational degree of freedom or its coupling 
to the phonon degrees of freedom. 
A treatment of this process would require an extension of the theory to include the mechanical dynamics of the sample and 
the corresponding AM balance. We leave this issue for future work.}

{\subsection{Effect of discrete rotational symmetry on the boundary conditions}}

{In addition to the SAM and OAM discussed in the present study, 
phonons in crystals with discrete rotational symmetry can carry pseudo spin and pseudo 
orbital AM~\cite{Zhang2015,AKato2023,Ishito2023a,Ishito2023b,Oishi2024,Tateishi2025,Ishizuka2025}. 
These quantities correspond to quantum numbers and phase factors associated with discrete rotations. 
They are distinct from the SAM originating from circular polarization and the OAM associated with the real-space motion of phonon wave packets.}

{Pseudo AM can constrain interface scattering through symmetry-based selection rules 
when both the incident and scattered phonon states lie on high-symmetry points or lines shared by the two crystals. 
For generic interface scattering processes with general incident angles, however, these selection rules do not apply. 
Therefore, the general formulation developed in the present study remains applicable.}

\subsection{Mechanisms of TAM relaxation}
\label{111306_7Feb25}

We have incorporated finite lifetimes $\tau$ and $\widetilde{\tau}$ in the constitutive relations~\eqref{110849_7Feb25} and \eqref{110851_7Feb25} of the TAM.
These time constants arise from 
{the relaxation-time approximation adopted in the Boltzmann equations~\eqref{122352_26Jan25} and \eqref{122357_26Jan25}.
Within the present constant-relaxation-time treatment, the relaxation of the TAM is therefore governed by the same phenomenological relaxation times 
as the phonon distribution functions. 
In particular, the characteristic spatial scales are set by the corresponding phonon mean free paths, 
such as $\tau v_{\mathrm{T}}$ and $\widetilde{\tau}c_{\mathrm{T}}$, as also evident from Eqs.~\eqref{185446_6Feb25} and \eqref{185449_6Feb25}.}

{The relaxation times $\tau$ and $\widetilde{\tau}$ should not, however, be regarded as microscopic TAM lifetimes obtained from a specific scattering mechanism. 
Rather, they phenomenologically characterize the net relaxation of the nonequilibrium phonon distributions on the two sides of the interface. 
A microscopic treatment based on a more general collision integral of the Boltzmann equation, 
together with its eigenmode analysis~\cite{Hardy1970,Cepellotti2016}, could in principle yield a distinct effective relaxation time for TAM.
Such a distinction is not resolved within the present constant-relaxation-time approximation, and we do not introduce an additional independent TAM relaxation time.}

{Understanding the microscopic mechanisms contributing to the relaxation encoded in $\tau$ and $\widetilde{\tau}$ is nevertheless important 
for a microscopic description of TAM transport.} Several dissipation mechanisms may contribute, and we highlight three examples.

The first mechanism involves the reversal of phonon TAM through phonon--phonon scattering. 
This process arises from lattice anharmonicity and reflects the discrete rotational symmetry of the crystal. 
As a result, the scattering itself reverses the TAM.

The second mechanism is dephasing driven by random forces that violate TAM conservation. 
In covalent crystals, non-central interatomic forces, such as bond-bending interactions~\cite{Keating1966a,Martin1970,Stillinger1985,PhillipsTextbook}, 
induce incoherent precession of phonon TAM, leading to a gradual loss of the net TAM. 
A similar mechanism for phonon SAM relaxation has been proposed in Ref.~\cite{Suzuki2025}.

A third mechanism involves the transfer of TAM to other elementary excitations. 
Previous studies~\cite{Korenev2016,Nova2017,Holanda2018,Sasaki2021,Jeong2022,Kim2023,Ueda2023,Ohe2024,Choi2024, 
Juraschek2019,Hamada2020,Juraschek2022,Tauchert2022,Fransson2023,Chaudhary2024,Funato2024,Yao2024,Li2024,Yokoyama2024,Nishimura2025}
suggest that phonon AM can couple to other degrees of freedom, 
such as electron spins in metals or magnons in magnetic materials.

{Quantitatively determining the relaxation times from these microscopic mechanisms remains an open problem.}
{Related relaxation processes have nevertheless started to be explored experimentally.
In particular, rotational Umklapp processes caused by phonon--phonon scattering have recently been observed 
as a relaxation mechanism of pseudo AM associated with discrete rotational symmetry~\cite{Minakova2026}. }

\subsection{Application of boundary conditions to phonon and boson transport}

The boundary conditions in Eq.~\eqref{eq: boundary condition assump} and Eqs.~\eqref{163324_9Jun25}--\eqref{163328_9Jun25} 
are expected to be relevant not only for phonon AM transport but also for phonon-drag effects observed at interfaces of stacked thin films~\cite{Wang2013,Kimura2021}.  
The transmission of phonon AM across domain boundaries between right- and left-handed chiral crystals~\cite{Matsubara2025}, 
also provides a promising ground for applying the present framework.  
Furthermore, these boundary conditions can be extended to other long-wavelength, low-energy bosons with linear dispersion---such as 
photons or magnons in antiferromagnets---and they thereby establish a unified theoretical framework for polarization-dependent transport at boundaries.

\section{Conclusions}
\label{sec: summary}

We have established boundary conditions for phonon wave packets at a smooth interface between two crystals, 
which consistently incorporate both polarization and center-of-mass degrees of freedom. 
Based on these conditions, we predict the generation of phonon OAM, 
namely a circulating edge flow, induced by a local thermal gradient.  

We have identified two distinct levels of boundary conditions for the phonon distribution function. 
The first one is a coarse-grained condition, derived analytically from elastic-wave scattering and energy conservation without \emph{ad hoc} assumptions. 
The second one is a detailed condition that explicitly retains the center-of-mass information of each phonon wave packet.  
The latter naturally reproduces the classical transverse shift of elastic beams, consistent with continuum elasticity.

Using these boundary conditions, we have clarified the spatial distribution and amplitude of interfacial phonon OAM 
in a junction between a crystal under a temperature gradient and another kept without external driving.
The predicted OAM can reach a magnitude comparable to the SAM associated with circularly polarized phonons, 
highlighting the substantial role of interfacial OAM in phonon transport.

\begin{acknowledgments}
{We wish to thank M.~Kato, H.~Matsuura, S.~Murakami, E.~Saitoh, H.~Shishido, J.~Kishine, Y.~Togawa,
and H.~Kusunose for helpful discussions.
In particular, we thank J.~Kishine for information on earlier phonon angular momentum studies, 
H.~Matsuura for communication on boundary condition and interfacial phonon drag~\cite{Wang2013,Kimura2021},
and S.~Murakami for his comments on total angular momentum conservation at the interface.
This work was supported by JSPS KAKENHI Grants No.~JP20K03855, No.~JP21H01032, No.~JP22KJ0856, No.~JP23K03333, No.~24KJ1036, {and No.~JP25H02113}. 
This research was also supported by Joint Research by the Institute for Molecular Science (IMS program No.~23IMS1101). 
This research was also supported by the grant of OML Project by the National Institutes of Natural Sciences (NINS program No.~OML012301).}
\end{acknowledgments}

\appendix

\section{Review: Reflection and refraction of elastic waves}
\label{sec: elastic plane wave}

We give an overview of the boundary scattering of elastic plane waves at an interface plane between
two elastic bodies in contact.
The interface is schematically shown in Fig.~\ref{fig: incidentplane}.
We assume that the whole medium vibrates with a single frequency $\omega$
and a wavevector component $\bm{k}_{\parallel}$ parallel to the interface $z =0$.

\subsection{Linear polarizations}
\label{subsec: elastic plane wave setup}

In the main part of the paper, we adopt circularly polarized modes as the degenerate transverse modes;  
however, we use linearly polarized modes in this section because the linear basis simplifies the description of the boundary conditions.  
The linear polarization comprises two modes: one polarized normal to the incident plane and one polarized within the incident plane.  
We call the former as the SH mode and the latter the SV mode~\footnote{
The SH/SV modes refer to horizontally/vertically polarized shear waves, respectively, following geophysical conventions.}.

Let us focus on the region $z <0$.
The polarization vectors of the linear basis 
\begin{equation}
\bm{e}_{\bm{k}, n} = \bm{e}_{s, n} =
\begin{cases}
\hat{\bm{x}'}\sin\theta_{\text{L}} + s\hat{\bm{z}}\cos\theta_{\text{L}} & n=\text{L}\\[6pt]
\hat{\bm{y}'} & n=\text{SH}\\[6pt]
-\hat{\bm{z}}\sin\theta_{\text{T}} + s\hat{\bm{x}'}\cos\theta_{\text{T}} & n=\text{SV}
\end{cases}
\label{eq: L SH SV polarization vec} 
\end{equation}
are related with those of the circular one, given in Eq.~\eqref{eq: basis L RH LH}, as follows:
$\bm{e}_{s,\text{RH}}\equiv (\bm{e}_{s,\text{SV}} + \iu\,\bm{e}_{s,\text{SH}})/\sqrt{2}$ and $\bm{e}_{s,\text{LH}}\equiv \bm{e}^*_{s,\text{RH}}$.
The amplitudes of the linear polarization and those of the circular one are also related by
\begin{equation}
\left[
\renewcommand\arraystretch{1.3}
\begin{array}{c}
u_{s,\text{L}}\\
u_{s,\text{RH}}\\
u_{s,\text{LH}}
\end{array} 
\right]
= \begin{bmatrix}
 1 &  &  \\[4pt]
 & -\iu/\sqrt{2} & 1/\sqrt{2}\\[4pt]
 & +\iu/\sqrt{2} & 1/\sqrt{2}
\end{bmatrix}
\left[
\renewcommand\arraystretch{1.3}
\begin{array}{c}
u_{s,\text{L}}\\
u_{s,\text{SH}}\\
u_{s,\text{SV}}
\end{array} 
\right].
\label{eq: lin circ conversion}
\end{equation}

The displacement field for the region $z <0$,  
$\bm{u}^{\text{ela}} (\bm{r}, t)$, is a sum of the plane wave components $(s, n)$ with amplitudes $u_{s, n}$, expressed as
\begin{align}
& \bm{u}^{\text{ela}}(\omega, \bm{k}_{\parallel}, \bm{r},t)\nonumber \\
& = \sum_{s=\pm}\sum_{n=\text{L}, \text{SH}, \text{SV}}
 u_{s, n} \bm{e}_{s, n} e^{ \iu\left(\bm{k}_{\parallel}\cdot \bm{R}_{\parallel} + k^z_{s, n} z  -\omega t \right) } 
+ \text{c.c.}
 \label{eq: deformation amp of elastic body}      
\end{align}
The stress tensor takes the form
$\sigma_{ij} = \sum_{k, l} a_{ijkl} \partial u^{\text{ela}}_{k}/\partial r_l$
with $a_{ijkl}= \rho v^2_{\text{T}}(\delta_{ik}\delta_{jl} + \delta_{il}\delta_{jk})
+\rho (v^2_{\text{L}}- 2 v^2_{\text{T}})\delta_{ij}\delta_{kl}$
for the isotropic elastic body in $z<0$~\cite{LandauLifshitzTextbookVol7,Bliokh2006}.

In the same way, elastic waves in the region $z>0$
are characterized by the sign of propagating direction $s = \pm$ and the mode $n=\text{L},~\text{SH},~\text{SV}$.
We define the polarization vector $\widetilde{\bm{e}}_{s, n}$, amplitude of each plane wave component $\widetilde{u}_{s, n}$,
and stress tensor $\widetilde{\sigma}_{ij}$ in this region.

\subsection{Boundary conditions}

In the case of perfect contact, the displacements and the stresses are continuous at $z=0$ 
for arbitrary time~\cite{LandauLifshitzTextbookVol7,SommerfeldTextbookVol2}.
The equalities
$\sum_{s,n} u_{s,n} \bm{e}_{s,n} = \sum_{s,n} \widetilde{u}_{s,n} \widetilde{\bm{e}}_{s,n}$,
$\sigma_{zz} = \widetilde{\sigma}_{zz}$,
$\sigma_{zx'} = \widetilde{\sigma}_{zx'}$, and $\sigma_{zy'} = \widetilde{\sigma}_{zy'}$
yield the boundary conditions:
\begin{widetext}
\begin{subequations}
  \begin{align}
 &
 \begin{bmatrix}
 \cos\theta_{\text{L}} & -\sin\theta_{\text{T}} & -\cos\theta_{\text{L}} & -\sin\theta_{\text{T}} \\[6pt]
 \sin\theta_{\text{L}} & \cos\theta_{\text{T}} & \sin\theta_{\text{L}} & -\cos\theta_{\text{T}}\\[6pt]
 Z_{\text{L}}\cos 2\theta_{\text{T}} & -Z_{\text{T}}\sin 2\theta_{\text{T}} &
 Z_{\text{L}}\cos 2\theta_{\text{T}} & Z_{\text{T}}\sin 2\theta_{\text{T}} \\[6pt]
 Z_{\text{L}}
 \left(\dfrac{\sin\theta_{\text{T}}}{\sin\theta_{\text{L}}}\right)^2
 \sin 2\theta_{\text{L}} & Z_{\text{T}}\cos 2\theta_{\text{T}} &
 -Z_{\text{L}}\left(\dfrac{\sin\theta_{\text{T}}}{\sin\theta_{\text{L}}}\right)^2
 \sin 2\theta_{\text{L}} & Z_{\text{T}}\cos 2\theta_{\text{T}}
 \end{bmatrix}
 \begin{bmatrix}
 u_{+,\text{L}} \\[12pt]
 u_{+,\text{SV}} \\[12pt]
 u_{-,\text{L}} \\[12pt]
 u_{-,\text{SV}}
 \end{bmatrix}
 \nonumber \\ &\quad = 
 \begin{bmatrix}
 \cos\Theta_{\text{L}} & -\sin\Theta_{\text{T}} & -\cos\Theta_{\text{L}} & -\sin\Theta_{\text{T}} \\[6pt]
 \sin\Theta_{\text{L}} & \cos\Theta_{\text{T}} & \sin\Theta_{\text{L}} & -\cos\Theta_{\text{T}}\\[6pt]
 \zeta_{\text{L}}\cos 2\Theta_{\text{T}} & -\zeta_{\text{T}}\sin 2\Theta_{\text{T}} &
 \zeta_{\text{L}}\cos 2\Theta_{\text{T}} & \zeta_{\text{T}}\sin 2\Theta_{\text{T}} \\[6pt]
 \zeta_{\text{L}}
 \left(\dfrac{\sin\Theta_{\text{T}}}{\sin\Theta_{\text{L}}}\right)^2
 \sin 2\Theta_{\text{L}} & \zeta_{\text{T}}\cos 2\Theta_{\text{T}} &
 -\zeta_{\text{L}}\left(\dfrac{\sin\Theta_{\text{T}}}{\sin\Theta_{\text{L}}}\right)^2
 \sin 2\Theta_{\text{L}} & \zeta_{\text{T}}\cos 2\Theta_{\text{T}}
 \end{bmatrix}
 \begin{bmatrix}
 \widetilde{u}_{+,\text{L}} \\[12pt]
 \widetilde{u}_{+,\text{SV}} \\[12pt]
 \widetilde{u}_{-,\text{L}} \\[12pt]
 \widetilde{u}_{-,\text{SV}}
 \end{bmatrix}, 
 \label{eq: b.c. parallel}
 \end{align}
 \begin{equation}
 \begin{bmatrix}
 1 & 1\\ Z_{\text{T}}\cos\theta_{\text{T}} & -Z_{\text{T}}\cos\theta_{\text{T}}
 \end{bmatrix}
 \begin{bmatrix}
  {u}_{+,\text{SH}} \\
 {u}_{-,\text{SH}}
 \end{bmatrix}
 = \begin{bmatrix}
 1 & 1\\ \zeta_{\text{T}}\cos\Theta_{\text{T}} & -\zeta_{\text{T}}\cos\Theta_{\text{T}}
 \end{bmatrix}
 \begin{bmatrix}
  \widetilde{u}_{+,\text{SH}} \\
 \widetilde{u}_{-,\text{SH}}
 \end{bmatrix} 
 \label{eq: b.c. normal}
 \end{equation}
\label{eq: b.c. total equations}%
\end{subequations}
\end{widetext}
with acoustic impedances
\begin{equation}
 Z_{\text{L}}= \rho v_{\text{L}}, \quad
Z_{\text{T}}= \rho v_{\text{T}}, \quad
\zeta_{\text{L}}= \widetilde{\rho} c_{\text{L}}, \quad
\zeta_{\text{T}}= \widetilde{\rho} c_{\text{T}}.
\label{eq: acoustic imp}
\end{equation}
The first equation~\eqref{eq: b.c. parallel} is known as the Zoeppritz equations~\cite{Knott1899,Zoeppritz1919}.
Analytical solution of it is available in geophysics textbooks~\cite{Ewing1957}.

The boundary condition~\eqref{eq: b.c. parallel} and \eqref{eq: b.c. normal} and Eq.~\eqref{eq: lin circ conversion}
yield the $S$-matrix relation~\eqref{eq: S matrix in circular pol 1} in the circular basis, with a straightforward calculation.
When there is no incident wave from $z > 0$, i.e., $\widetilde{u}_{-, n} =0$,
we can transform Eqs.~\eqref{eq: b.c. parallel} and \eqref{eq: b.c. normal} as another matrix
\begin{equation}
 \left[
\renewcommand\arraystretch{1.3}
\begin{array}{c}
u_{-,\text{L}}\\
u_{-,\text{SH}}\\
u_{-,\text{SV}}\\ \hline
\widetilde{u}_{+,\text{L}}\\
\widetilde{u}_{+,\text{SH}}\\
\widetilde{u}_{+,\text{SV}}
\end{array} 
\right]
=  \left[
 \renewcommand\arraystretch{1.3}
 \begin{array}{ccc}
 \rho_{\text{L},\text{L}} & 0 & \rho_{\text{L},\text{SV}} \\
 0 & \rho_{\text{SH},\text{SH}} & 0 \\
 \rho_{\text{SV},\text{L}} & 0 & \rho_{\text{SV},\text{SV}} \\ \hline
 \tau_{\text{L},\text{L}} & 0 & \tau_{\text{L},\text{SV}}\\
 0 & \tau_{\text{SH},\text{SH}} & 0 \\
 \tau_{\text{SV},\text{L}} & 0 & \tau_{\text{SV},\text{SV}} \\
 \end{array}
 \right] \left[
\renewcommand\arraystretch{1.3}
\begin{array}{c}
u_{+,\text{L}}\\
u_{+,\text{SH}}\\
u_{+,\text{SV}}
\end{array} 
\right],
\label{eq: rho tau coeff definition}
\end{equation}
elements of which we will use in Appendix~\ref{sec: IF shift of elastic beam}.

\subsection{Power reflectance and transmittance}

We interpret the elements of the $S$-matrix in Eq.~\eqref{eq: S matrix in circular pol 1}.
As shown in Eq.~\eqref{eq: ene fl}, 
the energy flux per unit area for each mode $(s, n)$ with $n=\text{L}, \text{RH}, \text{LH}$ is given as 
$2s \omega^2 Z_n\cos\theta_n \Abs{u_{s, n}}^2$ in $z<0$ and 
$2s \omega^2 \zeta_n\cos\Theta_n \Abs{\widetilde{u}_{s, n}}^2$ in $z>0$~\cite{Synge1956}.
Thus, the conservation of energy at the interface reduces to an equality
\begin{equation}
 \sum_{s, n} 2s\omega^2 Z_n\cos\theta_n \Abs{u_{s, n}}^2 
= \sum_{s, n} 2s\omega^2 \zeta_n\cos\Theta_n \Abs{\widetilde{u}_{s, n}}^2. 
\label{eq: ene conservation law}
\end{equation}
This equality requires the $S$-matrix to be unitary: $S^{\dagger}S =1$.
In particular, once we write down its elements as 
\begin{align}
& S(\omega, \bm{k}_{\parallel}) \nonumber\\
& =
 \left[
 \renewcommand\arraystretch{1.3}
 \begin{array}{lll|lll}
 r_{\text{L},\text{L}} & r_{\text{L},\text{RH}} & r_{\text{L},\text{LH}} 
 & t'_{\text{L},\text{L}} & t'_{\text{L},\text{RH}} & t'_{\text{L},\text{LH}} \\
 r_{\text{RH},\text{L}} & r_{\text{RH},\text{RH}} & r_{\text{RH},\text{LH}} 
 & t'_{\text{RH},\text{L}} & t'_{\text{RH},\text{RH}} & t'_{\text{RH},\text{LH}} \\
 r_{\text{LH},\text{L}} & r_{\text{LH},\text{RH}} & r_{\text{LH},\text{LH}} 
 & t'_{\text{LH},\text{L}} & t'_{\text{LH},\text{RH}} & t'_{\text{LH},\text{LH}} \\ \hline
 t_{\text{L},\text{L}} & t_{\text{L},\text{RH}} & t_{\text{L},\text{LH}} 
 & r'_{\text{L},\text{L}} & r'_{\text{L},\text{RH}} & r'_{\text{L},\text{LH}} \\
 t_{\text{RH},\text{L}} & t_{\text{RH},\text{RH}} & t_{\text{RH},\text{LH}} 
 & r'_{\text{RH},\text{L}} & r'_{\text{RH},\text{RH}} & r'_{\text{RH},\text{LH}} \\
 t_{\text{LH},\text{L}} & t_{\text{LH},\text{RH}} & t_{\text{LH},\text{LH}} 
 & r'_{\text{LH},\text{L}} & r'_{\text{LH},\text{RH}} & r'_{\text{LH},\text{LH}}
 \end{array}
 \right],
 \label{eq: elastic S matrix in circularly pol}
\end{align}
and we obtain
\begin{subequations}
 \begin{align}
 \sum_{n = \text{L}, \text{RH}, \text{LH}}\left(\Abs{r_{nm}}^2 + \Abs{t_{nm}}^2\right)
 &= 1 \label{eq: pow re re sum-1}\\
 \sum_{n = \text{L}, \text{RH}, \text{LH}}\left(\Abs{t'_{nm}}^2 + \Abs{r'_{nm}}^2\right)
 &= 1 \label{eq: pow re re sum-2}
 \end{align}
\end{subequations}
for any mode $m = \text{L}, \text{RH}, \text{LH}$.
Based on this relation, square of the elements of the $S$-matrix correspond to the fraction of power from the incident wave
to the reflected and transmitted waves.

Strictly speaking, we must account for grazing incidence, where evanescent waves can appear.  
We define the power reflectances $\mathcal{R}_{nm}$ and $\mathcal{R}'_{nm}$ and the transmittances $\mathcal{T}_{nm}$ and $\mathcal{T}'_{nm}$
as follows: 
\begin{subequations}
 \begin{align}
 \mathcal{R}_{nm} &= \Abs{r_{nm}}^2\, \Theta_{\text{H}}(1-\sin\theta_{n})\Theta_{\text{H}}(1-\sin\theta_{m}), \\
 \mathcal{R}'_{nm} &= \Abs{r'_{nm}}^2\, \Theta_{\text{H}}(1-\sin\Theta_{n})\Theta_{\text{H}}(1-\sin\Theta_{m}), \\
 \mathcal{T}_{nm} &= \Abs{t_{nm}}^2\, \Theta_{\text{H}}(1-\sin\Theta_{n})\Theta_{\text{H}}(1-\sin\theta_{m}), \\
 \mathcal{T}'_{nm} &= \Abs{t'_{nm}}^2\, \Theta_{\text{H}}(1-\sin\theta_{n})\Theta_{\text{H}}(1-\sin\Theta_{m}),\label{eq: R and T definitions d}
\end{align}
\label{eq: R and T definitions}%
\end{subequations}
where $\Theta_{\text{H}}(w)$ is the Heaviside step function defined in Eq.~\eqref{eq: Heaviside step fnc}.
These coefficients are the ratios of the energy flux of an incident wave in mode $m$ to the energy flux of reflected and transmitted waves in mode $n$. 
Their angular dependence for several parameter sets is presented in Supplemental Material~\cite{SupplementalMaterialFullPaper}.

Similarly to Eqs.~\eqref{eq: pow re re sum-1} and \eqref{eq: pow re re sum-2}, the coefficients satisfy the following constraints for each mode $m$:
\begin{subequations}
 \begin{align}
 \sum_n (\mathcal{R}_{nm} + \mathcal{T}_{nm}) &= 1, \label{eq: R and T sum unity 1}\\
 \sum_n (\mathcal{T}'_{nm} + \mathcal{R}'_{nm}) &= 1, 
\end{align}
\label{eq: R and T sum unity}%
\end{subequations}
which holds for arbitrary angle of incidence.
We also notice that the coefficients satisfy two symmetries:
one is the Helmholtz reciprocity $\mathcal{R}_{nm} =\mathcal{R}_{mn},~\mathcal{R}'_{nm} = \mathcal{R}'_{mn}$ 
and $\mathcal{T}_{nm}=\mathcal{T}'_{mn}$; the other is an invariance 
under a replacement of subscripts $(n, m)$ from (L, RH, LH) into (L, LH, RH):
\begin{subequations}
 \begin{align}
 \mathcal{R}_{\text{RH}, \text{RH}}&= \mathcal{R}_{\text{LH}, \text{LH}},
 &\mathcal{T}_{\text{RH}, \text{RH}}&= \mathcal{T}_{\text{LH}, \text{LH}}, \label{eq: symmetry of R and T a}\\
 \mathcal{R}_{\text{RH}, \text{LH}}&= \mathcal{R}_{\text{LH}, \text{RH}},
 &\mathcal{T}_{\text{RH}, \text{LH}}&= \mathcal{T}_{\text{LH}, \text{RH}},\\
 \mathcal{R}_{\text{RH}, \text{L}}&= \mathcal{R}_{\text{LH}, \text{L}},
 &\mathcal{T}_{\text{RH}, \text{L}}&= \mathcal{T}_{\text{LH}, \text{L}}. \label{eq: symmetry of R and T c}
 \end{align}
\label{eq: symmetry of R and T}%
\end{subequations}
Note that these symmetries hold only for the reflectance and transmittance with the same $\omega$ and $\bm{k}_{\parallel}$.

\section{Acoustic analog of the Imbert--Fedorov shift}
\label{sec: IF shift of elastic beam}

In this Appendix, we analyze interfacial scattering of successive elastic wave packets, i.e., acoustic beams~\footnote{
We assume successive scattering of wave packets and equate the long-term average of their transverse shift with that of the corresponding beam. 
For a general discussion of time delays (Wigner time delay) associated with wave packet scattering, see Ref.~\cite{Bliokh2025}.
}. We derive the transverse shift of the reflected and transmitted beams, given in Eqs.~\eqref{170438_23Jan25}--\eqref{135819_28Jan25},  
for circularly polarized incidence; the shift $\Delta r_{\text{IF}}$ is illustrated in Fig.~\ref{fig: intro schematics}(a).  
Our formulation follows the optical review by Ref.~\cite{Bliokh2013} and extends it to acoustics by including both longitudinal and transverse modes.

\subsection{Plane wave components in the incident beam}
\label{subsec: Plane wave components in the incident beam}
\begin{figure}[tbp]
 \centering
\includegraphics[pagebox=artbox,width=0.8\columnwidth]{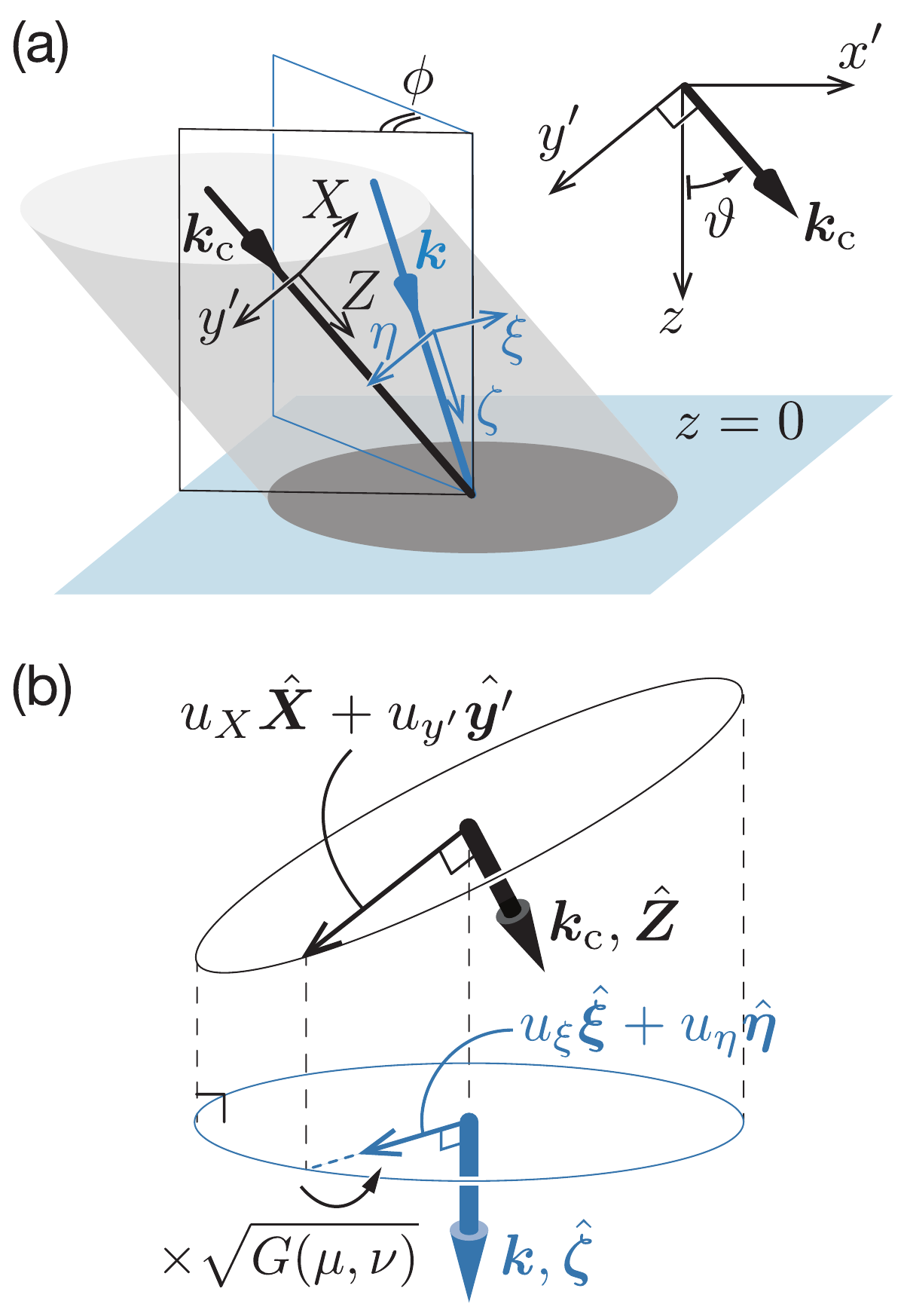}
\caption{Schematic view of (a)~three coordinate frames 
$(x', y', z)$, $(X, y', Z)$, and $(\xi, \eta, \zeta)$, and (b)~displacements 
$u_X \hat{\bm{X}} + u_{y'}\hat{\bm{y'}}$ and $u_{\xi} \hat{\bm{\xi}} + u_{\eta}\hat{\bm{\eta}}$
for the incident beam.
The central wavevector $\bm{k}_{\text{c}}$ is represented by the black bold arrow, 
while the wavevector $\bm{k}$ of an off-center plane wave component inside the beam 
is represented by the blue bold arrow.
}
\label{fig: acoustic IF shift inci}
\end{figure}

Let us consider a paraxial acoustic beam of transverse mode incident upon the interface plane $z=0$ from the region $z<0$, 
as illustrated in Fig.~\ref{fig: acoustic IF shift inci}(a).
We first specify the central wavevector of the incident beam as
$\bm{k}_{\text{c}} = \bm{k}_{\text{c},\parallel} + k_{\text{c},\perp} \hat{\bm{z}}$.
Here $\bm{k}_{\text{c},\parallel}$ is parallel to the interface.
We take a $(x', y', z)$ coordinate with $x'$ axis in the direction of $\bm{k}_{\text{c},\parallel}$ 
and $y'$ axis in the direction of $\hat{\bm{z}}\times \bm{k}_{\text{c},\parallel}$,
in the same way as Fig.~\ref{fig: incidentplane}. 
We also define the center coordinate of the incident beam $(X, y', Z)$, 
by rotating the original frame $(x', y', z)$ 
around the $y'$ axis by the angle of incidence $\vartheta \equiv \theta_{\text{T}}$.
The construction is illustrated in Fig.~\ref{fig: acoustic IF shift inci}(a).
We write unit vectors 
spanning the $(X, y', Z)$ coordinate system as $(\hat{\bm{X}}, \hat{\bm{y'}}, \hat{\bm{Z}})$.
The polarization vector of the central plane wave lies in the $Xy'$ plane, since transverse wave is incident.

We then describe an off-center plane wave component within the incident beam.
The component is characterized by a wavevector with small orthogonal deflection from the central wavevector
\begin{equation}
 \bm{k} \simeq \bm{k}_{\text{c}} + k_{\text{c}} (\mu \hat{\bm{X}} + \nu \hat{\bm{y'}}).
\label{eq: off center k vector }
\end{equation}
Here we put $k_{\text{c}} \equiv \Abs{\bm{k}_{\text{c}}}$.
Let us take a coordinate $(\xi, \eta, \zeta)$ 
for the off-center plane wave, as shown in Fig.~\ref{fig: acoustic IF shift inci}(a),
with $\zeta$  and $\eta$ axes in the direction of $\bm{k}$ and $\hat{\bm{z}}\times \bm{k}$, respectively.
We take unit vectors $(\hat{\bm{\xi}}, \hat{\bm{\eta}}, \hat{\bm{\zeta}})$ that span the coordinate system.
Since the off-center plane wave is also a transverse wave,
its polarization lies in the $\xi\eta$ plane.

We now write down the displacement field of the incident beam as a superposition of all the plane wave components:
\begin{align}
 \bm{u} (\bm{r},t) &= 2\mathrm{Re}\int^{\infty}_{-\infty} \frac{k_{\text{c}}\rmd \mu}{2\pi}
\int^{\infty}_{-\infty} \frac{k_{\text{c}}\rmd \nu}{2\pi}\cos\vartheta\, 
e^{\iu (\bm{k}\cdot \bm{r} -\omega t)} \nonumber\\
&\quad \times \left(u_{\xi} \hat{\bm{\xi}} + u_{\eta} \hat{\bm{\eta}}\right).
\label{eq: incident beam displacement field} 
\end{align}
Here $u_{\xi} \hat{\bm{\xi}} + u_{\eta} \hat{\bm{\eta}}$ denotes 
the displacement of the off-center plane wave component with deflection $(\mu, \nu)$.
The wavevector $\bm{k}$ is given in Eq.~\eqref{eq: off center k vector }.
We inserted a cosine factor into the integrand~\eqref{eq: incident beam displacement field},
since the area element $k^2_{\text{c}} \cos\vartheta \rmd \mu \rmd \nu$
is invariant during the reflection and transmission at the interface:
not only the wavevector component parallel to the interface 
$\bm{k}_{\parallel} = k_{ x'}\hat{\bm{x'}} + k_{ y'}\hat{\bm{y'}}$ but also its deviation 
from $\bm{k}_{\text{c},\parallel}$, given by
$\rmd k_{x'} = k_{\text{c}} \rmd \mu \cos\vartheta$ and $\rmd k_{y'} = k_{\text{c}} \rmd \nu$,
are conserved during the reflection or transmission.

We assume that the incident beam is a simple
paraxial beam characterized by a definite polarization, as illustrated in Fig.~\ref{fig: acoustic IF shift inci}(b):
the displacement of the off-center plane wave $u_{\xi} \hat{\bm{\xi}} + u_{\eta} \hat{\bm{\eta}}$
points along the projection of the central plane wave displacement, 
denoted as $u_{X} \hat{\bm{X}} + u_{y'} \hat{\bm{y'}}$, onto the $\xi\eta$ plane.
The amplitudes $u_{\xi}$ and $u_{\eta}$, at the same time, attenuate as the degree of deflection $\mu^2 + \nu^2$ increases.
The displacement of the off-center plane wave component is thus expressed as
\begin{align}
&
\begin{bmatrix}
 u_{\xi} \\ u_{\eta} \\ 0
\end{bmatrix} 
= \sqrt{G(\mu, \nu)}\nonumber\\
&\quad  \times \underbrace{
\begin{bmatrix}
 1 & & \\ & 1 & \\ & & 0
\end{bmatrix}
}_{\equiv P}
\underbrace{
\begin{bmatrix}
 \hat{\bm{\xi}} & \hat{\bm{\eta}} & \hat{\bm{\zeta}}
\end{bmatrix}^{-1}
\begin{bmatrix}
 \hat{\bm{X}} & \hat{\bm{y'}} & \hat{\bm{Z}}
\end{bmatrix}
}_{\equiv U(\mu,\nu,\vartheta)}
\begin{bmatrix}
 u_X \\ u_{y'} \\ 0
\end{bmatrix}. 
\label{eq: incident xi eta X y' displacement transfer}
\end{align}
Here we introduced an arbitrary envelope function $G(\mu,\nu) = G(\Abs{\mu}, \Abs{\nu})$
that represents the attenuation in the radial direction,
a projection matrix $P$ onto the transverse displacement,
and a rotation matrix $U(\mu,\nu,\vartheta)$ to transform the coordinates between $(X, y',Z)$ and $(\xi, \eta, \zeta)$.

We could consider an asymmetric beam with $G(\mu,\nu) \neq G(\Abs{\mu}, \Abs{\nu})$
or include a $\bm{k}$-dependent phase factor 
in Eq.~\eqref{eq: incident xi eta X y' displacement transfer}, 
as in the case of the acoustic vortex beam~\cite{Hefner1999,Thomas2003,Ayub2011,Wang2021}, 
but we omit these contributions in the present paper. In particular, we assume 
\begin{equation}
 G(\mu, \nu) \propto \delta (\mu)  
\label{eq: G assump}
\end{equation}
and neglect the deviation $\mu \neq 0$, since we focus only on the transverse shift of beams
and $\mu$ does not contribute to the lowest order of the transverse shift~\cite{Bliokh2013}. 

We then calculate $U(\nu,\vartheta)\equiv U(\mu = 0,\nu,\vartheta)$ as a product of rotation matrices:
\begin{align}
U(\nu,\vartheta) &= 
\begin{bmatrix}
 \cos\vartheta & 0 & -\sin\vartheta \\
0 & 1 & 0 \\
\sin \vartheta & 0 & \cos\vartheta
\end{bmatrix}
\begin{bmatrix}
\cos \phi & \sin\phi & 0 \\
-\sin \phi & \cos\phi & 0 \\
0 & 0 & 1
\end{bmatrix}
\nonumber \\ & \qquad \times
\begin{bmatrix}
 \cos\vartheta & 0 & \sin\vartheta \\
0 & 1 & 0 \\
-\sin \vartheta & 0 & \cos\vartheta
\end{bmatrix}
\nonumber \\ &\simeq
\begin{bmatrix}
 1 & \nu\cot\vartheta & 0 \\
- \nu\cot\vartheta & 1 & -\nu \\
0 & \nu & 1
\end{bmatrix}.
\label{eq: within incident wave coordinate transf}
\end{align}
Here $\vartheta = \theta_{\text{T}}$ represents 
the polar angle of both $\bm{k}_{\text{c}}$ and $\bm{k}$ up to the first order in $\nu$,
while $\phi$ denotes the azimuthal angle of $\bm{k}$, as shown in Fig.~\ref{fig: acoustic IF shift inci}(a).
We take terms up to the first order in $\nu$
and approximate $\phi$ by $\phi \simeq \tan\phi = \nu /\sin \vartheta$.

Note that the coordinates $u_{\xi}$, $u_{\eta}$ and bases $\hat{\bm{\xi}}$, $\hat{\bm{\eta}}$
depend on the off-center deflection $\nu$, i.e.,
the direction of $\bm{k}$, while $u_{X}, u_{y'}, \hat{\bm{X}}$, and $\hat{\bm{y'}}$ do not.

\subsection{Plane wave components in the scattered beam}
\label{subsec: Plane wave components in the reflected/transmitted beam}
\begin{figure}[tbp]
 \centering
\includegraphics[pagebox=artbox,width=0.8\columnwidth]{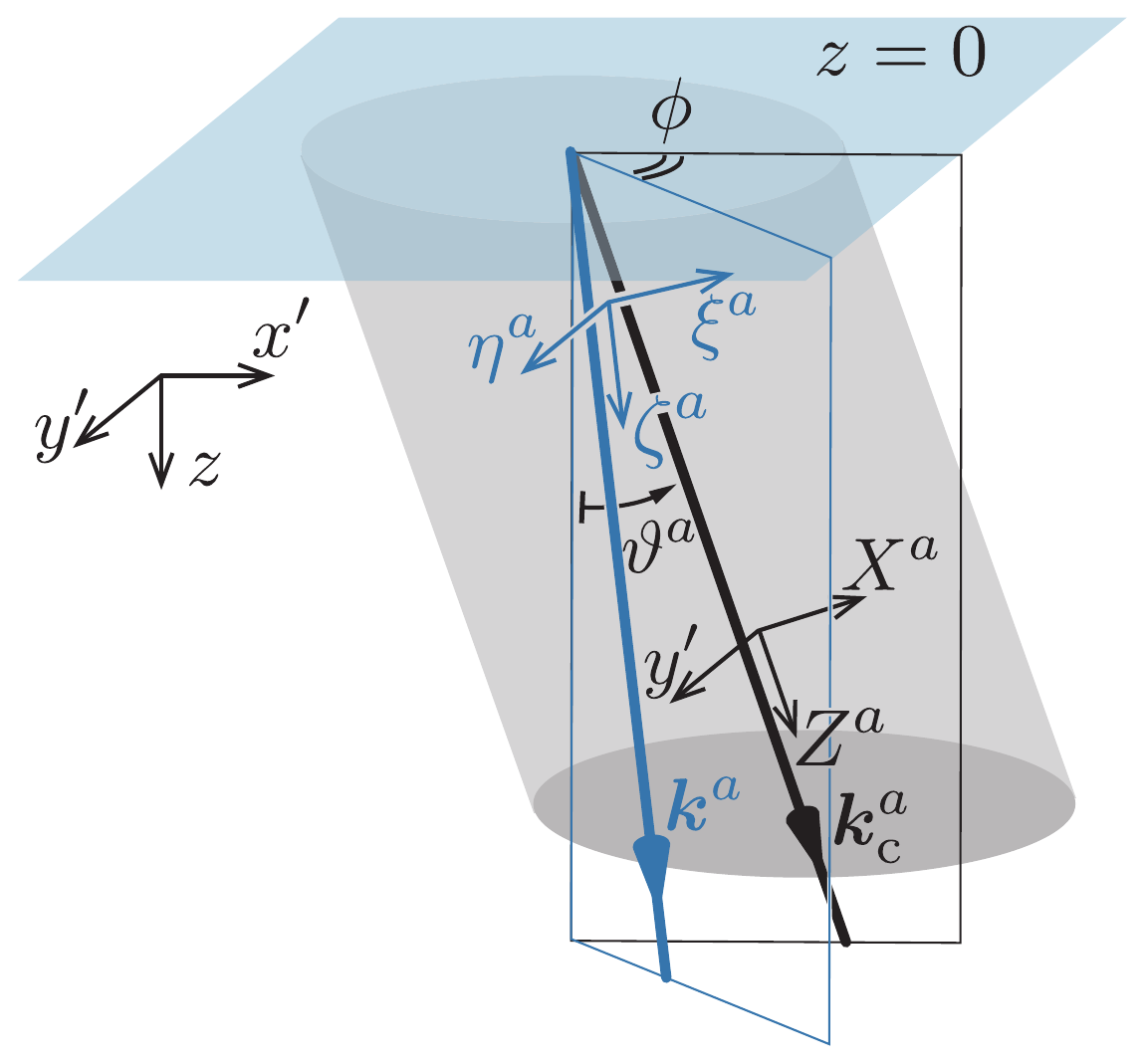}
\caption{Schematic view of coordinate frames for the transmitted beam.
The coordinate frames for the reflected beam is also introduced in the same way;
the only difference is that the $z$ components of the wavevectors are negative with obtuse angle $\vartheta^a$.
The central wavevector is represented by the black bold arrow, 
while the wavevector of an off-center plane wave component is represented by the blue bold arrow.
The reflected and transmitted waves are labeled by $a$ for each mode 
(see the beginning of Appendix~\ref{subsec: Plane wave components in the reflected/transmitted beam}).
}
\label{fig: acoustic IF shift scat}
\end{figure}

We then address reflected and transmitted beams
when the beam prepared in Appendix~\ref{subsec: Plane wave components in the incident beam} is incident.
We first consider the incidence of the central plane wave with the wavevector $\bm{k}_{\text{c}}$.
We refer to the reflected and transmitted waves of it
as the central plane waves of the reflected and transmitted beams.
We write their wavevectors as
$\bm{k}^{\text{r-L}}_{\text{c}}$, $\bm{k}^{\text{r-T}}_{\text{c}}$,
$\bm{k}^{\text{t-L}}_{\text{c}}$, and $\bm{k}^{\text{t-T}}_{\text{c}}$, or simply $\bm{k}^{a}_{\text{c}}$.
Here $a = \text{r-L}, \text{r-T}, \text{t-L}, \text{t-T}$
stand for reflected longitudinal, reflected transverse, transmitted 
longitudinal, and transmitted transverse waves, respectively. 
For convenience, we also write the angles that these wavevectors form with the $z$ axis as
\begin{subequations}
\begin{align}
 \vartheta^{\text{r-L}} &\equiv \pi - \theta_{\text{L}},
 &\vartheta^{\text{r-T}}& \equiv \pi - \theta_{\text{T}},\\
 \vartheta^{\text{t-L}} &\equiv \Theta_{\text{L}},
 &\vartheta^{\text{t-T}} &\equiv \Theta_{\text{T}} 
\end{align}
\label{eq: angles that vec k forms with the z axis}%
\end{subequations}
with $\theta_{\text{L}}$, $\theta_{\text{T}}$, $\Theta_{\text{L}}$, and $\Theta_{\text{T}}$
given in Fig.~\ref{fig: incidentplane}.
The amplitudes and angles of the wavevectors follow Snell's law
$k^a_{\text{c}}\sin \vartheta^{a} = k_{\text{c}}\sin \vartheta$
with $k^a_{\text{c}} = \Abs{\bm{k}^{a}_{\text{c}}}$.
We further define a center coordinate of the reflected and transmitted beam
$(X^a, y', Z^a)$ as shown in Fig.~\ref{fig: acoustic IF shift scat}.
We take unit vectors $(\hat{\bm{X}}^a, \hat{\bm{y'}}, \hat{\bm{Z}}^a)$ that span this coordinate system.

We then consider the incidence of the off-center plane wave component
with the wavevector $\bm{k}$ or the deflection $(\mu, \nu)$ given in Eq.~\eqref{eq: off center k vector }.
The reflected and transmitted plane waves are also characterized by 
the modes $a = \text{r-L}, \text{r-T}, \text{t-L}, \text{t-T}$ and wavevectors
\begin{subequations}
\begin{equation}
 \bm{k}^{a} \simeq \bm{k}^{a}_{\text{c}} + {k}^{a}_{\text{c}} (\mu^a  \hat{\bm{X^a}} + \nu^a  \hat{\bm{y'}}),\label{eq: deflection relation(a)}
\end{equation}
where the small deflection is given by Snell's law as
\begin{equation}
k^a_{\text{c}}\mu^a \Abs{\cos\vartheta^a}
= k_{\text{c}}\mu\cos\vartheta =0,\qquad 
k^{a}_{\text{c}} \nu^a = k_{\text{c}} \nu.
\label{eq: deflection relation(b)}
\end{equation}
\label{eq: deflection relation}%
\end{subequations}
We omitted $\mu$-dependence of $\bm{k}^{a}$, since we have assumed $G(\mu,\nu)\propto \delta (\mu)$ in Eq.~\eqref{eq: G assump}.
We, again, define a coordinate $(\xi^a, \eta^a, \zeta^a)$ with $\zeta^a$ along the wavevector $\bm{k}^{a}$
as shown in Fig.~\ref{fig: acoustic IF shift scat}.
We take unit vectors
$(\hat{\bm{\xi}}^a, \hat{\bm{\eta}}^a, \hat{\bm{\zeta}}^a)$ that span this coordinate system.
The polarization of the transverse (longitudinal) waves lies in the $\xi^a\eta^a$ plane ($\zeta^a$ direction).

In the same way as Eq.~\eqref{eq: within incident wave coordinate transf},
we introduce a rotation matrix $U^a$
that transforms the coordinates between $(X^a, y', Z^a)$ and $(\xi^a, \eta^a, \zeta^a)$:
\begin{align}
 U^a (\nu^a, \vartheta^a)& \equiv
\begin{bmatrix}
 \hat{\bm{\xi}}^a & \hat{\bm{\eta}}^a & \hat{\bm{\zeta}}^a
\end{bmatrix}^{-1}
\begin{bmatrix}
 \hat{\bm{X}}^a & \hat{\bm{y'}} & \hat{\bm{Z}}^a
\end{bmatrix} \nonumber \\
& \simeq
\begin{bmatrix}
 1 & \nu^a\cot\vartheta^a & 0 \\
- \nu^a\cot\vartheta^a & 1 & -\nu^a \\
0 & \nu^a & 1
\end{bmatrix}.
\label{eq: coordinate transf within the reflected transmitted beam}
\end{align}

We now write down the reflected and transmitted beam as superposition of plane wave components.
In the same way as Eq.~\eqref{eq: incident beam displacement field}, this is expressed as
\begin{align}
 \bm{u}^a (\bm{r},t) &= 2\mathrm{Re}\int \frac{{k^a_{\text{c}}}^2\Abs{\cos\vartheta^a}
\rmd \mu^a \rmd \nu^a}{(2\pi)^2}
e^{\iu (\bm{k}^a\cdot \bm{r} -\omega t)} \nonumber\\
&\quad \times \left(u^a_{\xi}\hat{\bm{\xi}^a} + u^a_{\eta}\hat{\bm{\eta}}^a + u^a_{\zeta}\hat{\bm{\zeta}^a}\right).
\label{eq: refl transmi superposition amp}
\end{align}
Here $u^a_{\xi} = u^a_{\eta} = 0$ for longitudinal waves $a = \text{r-L}, \text{t-L}$
and $u^a_{\zeta} = 0$ for transverse waves $a = \text{r-T}, \text{t-T}$.
In the integral of Eq.~\eqref{eq: refl transmi superposition amp},
we have introduced an area element in common with the incident and scattered beams
\begin{align}
&{k^a_{\text{c}}}^2\Abs{\cos\vartheta^a} \rmd \mu^a \rmd \nu^a =
\rmd^2 \bm{k}^a_{\parallel} = \rmd k^a_{x'} \rmd k^a_{y'} \nonumber \\
&= \rmd k_{x'} \rmd k_{y'} = \rmd^2 \bm{k}_{\parallel}
= {k_{\text{c}}}^2 \cos\vartheta \rmd \mu \rmd \nu
\label{eq: invariant area element}
\end{align}
with $\bm{k}_{\parallel}$ and $\bm{k}^a_{\parallel}$ being
components of $\bm{k}$ and $\bm{k}^a$ parallel to the interface,
respectively.

Finally, we use reflection and transmission coefficients in terms of amplitude
$\rho_{nm}$ and $\tau_{nm}$, shown in Eq.~\eqref{eq: rho tau coeff definition},
in order to relate the displacement of incident off-center plane waves $(u_{\xi}, u_{\eta}, u_{\zeta})$
with that of reflected and transmitted waves $(u^a_{\xi}, u^a_{\eta}, u^a_{\zeta})$:
\begin{subequations}
 \begin{align}
 \begin{bmatrix}
 u^{\text{r-L}}_{\xi}\\ u^{\text{r-L}}_{\eta}\\ u^{\text{r-L}}_{\zeta} 
 \end{bmatrix}
 &= 
 \begin{bmatrix}
 0 & 0 & 0 \\  0 & 0 & 0 \\
 \rho_{\text{L}, \text{SV}} & 0 & 0
 \end{bmatrix}
 \begin{bmatrix}
 u_{\xi}\\ u_{\eta}\\ 0 
 \end{bmatrix}
 \equiv  {F}^{\text{r-L}}
 \begin{bmatrix}
 u_{\xi}\\ u_{\eta}\\ 0 
 \end{bmatrix}, \label{eq: elastic Fresnel coeff matrices ini}
 \\
 \begin{bmatrix}
 u^{\text{r-T}}_{\xi}\\ u^{\text{r-T}}_{\eta}\\ u^{\text{r-T}}_{\zeta} 
 \end{bmatrix}
 &= 
 \begin{bmatrix}
 \rho_{\text{SV}, \text{SV}} & 0 & 0 \\  0 & \rho_{\text{SH}, \text{SH}} & 0 \\ 0 & 0 & 0
 \end{bmatrix}
 \begin{bmatrix}
 u_{\xi}\\ u_{\eta}\\ 0 
 \end{bmatrix}
 \equiv  {F}^{\text{r-T}}
 \begin{bmatrix}
 u_{\xi}\\ u_{\eta}\\ 0 
 \end{bmatrix},
 \\
 \begin{bmatrix}
 u^{\text{t-L}}_{\xi}\\ u^{\text{t-L}}_{\eta}\\ u^{\text{t-L}}_{\zeta} 
 \end{bmatrix}
 &= 
 \begin{bmatrix}
 0 & 0 & 0 \\  0 & 0 & 0 \\
 \tau_{\text{L}, \text{SV}} & 0 & 0
 \end{bmatrix}
 \begin{bmatrix}
 u_{\xi}\\ u_{\eta}\\ 0 
 \end{bmatrix}
 \equiv  {F}^{\text{t-L}}
 \begin{bmatrix}
 u_{\xi}\\ u_{\eta}\\ 0 
 \end{bmatrix},
 \\
 \begin{bmatrix}
 u^{\text{t-T}}_{\xi}\\ u^{\text{t-T}}_{\eta}\\ u^{\text{t-T}}_{\zeta} 
 \end{bmatrix}
 &= 
 \begin{bmatrix}
 \tau_{\text{SV}, \text{SV}} & 0 & 0 \\  0 & \tau_{\text{SH}, \text{SH}} & 0 \\ 0 & 0 & 0
 \end{bmatrix}
 \begin{bmatrix}
 u_{\xi}\\ u_{\eta}\\ 0 
 \end{bmatrix}
 \equiv  {F}^{\text{t-T}}
 \begin{bmatrix}
 u_{\xi}\\ u_{\eta}\\ 0 
 \end{bmatrix}. \label{eq: elastic Fresnel coeff matrices fin}
\end{align}
\label{eq: elastic Fresnel coeff matrices}%
\end{subequations}
Here we have defined each $3\times 3$ matrix in Eqs.~\eqref{eq: elastic Fresnel coeff matrices ini}--\eqref{eq: elastic Fresnel coeff matrices fin}
by $F^a(\vartheta)$.
We neglect $\nu$ dependence of the coefficients $F^a(\vartheta)$,
since the angle of incidence of the off-center plane waves
deviates from that of the central plane wave $\vartheta$ only of the second order of $\nu$.

Note that the coordinates $(u^a_{\xi}, u^a_{\eta}, u^a_{\zeta})$ and bases $(\hat{\bm{\xi}}^a, \hat{\bm{\eta}}^a, \hat{\bm{\zeta}}^a)$
depend on the off-center deflection $\nu$, i.e., the direction of $\bm{k}$, 
while $(\hat{\bm{X}}^a, \hat{\bm{y'}}, \hat{\bm{Z}}^a)$ do not.

\subsection{Transverse shift of beams}
\label{subsec: transverse shift}

We derive the shift of the reflected or transmitted beam in the $y'$ direction.  
Recall that the transverse shift is defined with weighting by the energy flux across the interface.  
Using the expression for the energy flux, the shift is given by
\begin{equation}
  \Delta r^a_{\text{IF}} =\left.
\frac{\displaystyle
\int \rmd x'\rmd y' \omega^2 Z^a \cos\vartheta^a \overline{\Abs{\bm{u}^a(\bm{r},t)}^2}\, y'}
{\displaystyle
\int \rmd x'\rmd y' \omega^2 Z^a \cos\vartheta^a \overline{\Abs{\bm{u}^a(\bm{r},t)}^2}}
\right|_{z=0}.
\label{eq: IF shift weighted by energy flux}
\end{equation}
Here acoustic impedance $Z^a$ is given as $Z^{\text{r-L}} = Z_{\text{L}}$, 
$Z^{\text{r-T}} = Z_{\text{T}}$, $Z^{\text{t-L}} = \zeta_{\text{L}}$, and $Z^{\text{t-T}} = \zeta_{\text{T}}$.
The long-time average is denoted by a bar over a symbol, as defined in Eq.~\eqref{eq: l time average}.

We first consider the denominator of Eq.~\eqref{eq: IF shift weighted by energy flux},
which provides total energy flux of a beam labeled $a$.
According to Eq.~\eqref{eq: refl transmi superposition amp} we expand it as
\begin{widetext}
 \begin{align}
 & \omega^2 Z^a \cos\vartheta^a\int \rmd x'\rmd y' 
 \overline{
 \Abs{
 \int \frac{\rmd^2 \bm{k}^a_{\parallel}}{(2\pi)^2}
 e^{\iu (\bm{k}^a_{\parallel}\cdot \bm{r} - \omega t)} 
 \left(u^a_{\xi}\hat{\bm{\xi}^a} + u^a_{\eta}\hat{\bm{\eta}}^a + u^a_{\zeta}\hat{\bm{\zeta}^a}\right)
 + \text{c.c.}
 }^2} \nonumber\\
 &= \omega^2 Z^a \cos\vartheta^a\int \rmd x'\rmd y' 
 \int \frac{\rmd^2 \bm{k}^a_{1, \parallel}}{(2\pi)^2}
 \int \frac{\rmd^2 \bm{k}^a_{2, \parallel}}{(2\pi)^2}
 e^{\iu (-\bm{k}^a_{1, \parallel} + \bm{k}^a_{2, \parallel})\cdot \bm{r}} \nonumber\\ &\qquad \times
 \left[\left.
 \left(u^a_{\xi}\hat{\bm{\xi}^a} + u^a_{\eta}\hat{\bm{\eta}}^a + u^a_{\zeta}\hat{\bm{\zeta}^a}\right)
 \right|_{\bm{k}^a_{\parallel} = \bm{k}^a_{1, \parallel}}\right]^{\dagger}\cdot \left[\left.
 \left(u^a_{\xi}\hat{\bm{\xi}^a} + u^a_{\eta}\hat{\bm{\eta}}^a + u^a_{\zeta}\hat{\bm{\zeta}^a}\right)
 \right|_{\bm{k}^a_{\parallel} = \bm{k}^a_{2, \parallel}}\right]
 + \text{c.c.} \nonumber \\
 &= 2 \omega^2 Z^a \cos\vartheta^a
 \int \frac{\rmd^2 \bm{k}^a_{1, \parallel}}{(2\pi)^2}\int \frac{\rmd^2 \bm{k}^a_{2, \parallel}}{(2\pi)^2}
 (2\pi)^2 \delta (\bm{k}^a_{1, \parallel} - \bm{k}^a_{2, \parallel})
 \left.\left(
 \Abs{u^a_{\xi}}^2 + \Abs{u^a_{\eta}}^2 + \Abs{u^a_{\zeta}}^2
 \right)\right|_{\bm{k}^a_{\parallel} = \bm{k}^a_{1, \parallel} = \bm{k}^a_{2, \parallel}},
 \end{align}
\end{widetext}
where $\text{c.c.}$ denotes the complex conjugate of the previous terms.

With the help of Eqs.~\eqref{eq: incident xi eta X y' displacement transfer} and \eqref{eq: elastic Fresnel coeff matrices ini}--\eqref{eq: elastic Fresnel coeff matrices fin}, 
we have
\begin{align}
& 2 \omega^2 Z^a \cos\vartheta^a
\int \frac{\rmd^2 \bm{k}^a_{\parallel}}{(2\pi)^2} G(\mu, \nu)\nonumber\\
& \times
\begin{bmatrix}
 u_{X} \\ u_{y'} \\ 0 
\end{bmatrix}^{\dagger}
U^{\dagger} (\nu, \vartheta) P^{\dagger}
{F^a}^{\dagger}(\vartheta) F^a(\vartheta)
P U(\nu, \vartheta)
\begin{bmatrix}
 u_{X} \\ u_{y'} \\ 0 
\end{bmatrix} \nonumber \\
& \simeq 
2\omega^2 Z^a \cos\vartheta^a
\begin{bmatrix}
 u_{X} \\ u_{y'} \\ 0 
\end{bmatrix}^{\dagger}
{F^a}^{\dagger}(\vartheta) F^a(\vartheta)
\begin{bmatrix}
 u_{X} \\ u_{y'} \\ 0 
\end{bmatrix}
N^{\text{beam}}.
\end{align}
In the above integral, the variables $\bm{k}^a_{\parallel}$, $\mu$, and $\nu$ follow 
relations~\eqref{eq: deflection relation(a)}, \eqref{eq: deflection relation(b)} and \eqref{eq: invariant area element}.
In other words, the deflections of $\bm{k}$ in the reflected and transmitted beams are related with
those in the incident beam $\mu$ and $\nu$.
In the last line we introduce a constant of the incident beam
\begin{equation}
N^{\text{beam}} \equiv \int \frac{k^2_{\text{c}} \cos\vartheta \rmd \mu\rmd \nu}{(2\pi)^2} G(\mu, \nu),
\end{equation}
and neglect second or higher order terms in $\nu$, stemming from $F^a(\vartheta)$ and $U(\nu, \vartheta)$.
The first order terms in $\nu$ also vanish in the integral over the symmetric interval.

We then simplify the numerator of Eq.~\eqref{eq: IF shift weighted by energy flux}
using integration by parts as follows:
\begin{widetext}
 \begin{align}
 &\omega^2 Z^a \cos\vartheta^a\int \rmd x'\rmd y' 
 \int \frac{\rmd^2 \bm{k}^a_{1, \parallel}}{(2\pi)^2}
 \int \frac{\rmd^2 \bm{k}^a_{2, \parallel}}{(2\pi)^2}
 \left[-\iu \frac{\partial}{\partial k^a_{2, y'}}
 e^{\iu (-\bm{k}^a_{1, \parallel} + \bm{k}^a_{2, \parallel})\cdot \bm{r}}
 \right]\nonumber\\ & \qquad \times 
 \left[\left.
 \left(u^a_{\xi}\hat{\bm{\xi}^a} + u^a_{\eta}\hat{\bm{\eta}}^a + u^a_{\zeta}\hat{\bm{\zeta}^a}\right)
 \right|_{\bm{k}^a_{\parallel} = \bm{k}^a_{1, \parallel}}\right]^{\dagger} \cdot \left[\left.
 \left(u^a_{\xi}\hat{\bm{\xi}^a} + u^a_{\eta}\hat{\bm{\eta}}^a + u^a_{\zeta}\hat{\bm{\zeta}^a}\right)
 \right|_{\bm{k}^a_{\parallel} = \bm{k}^a_{2, \parallel}}\right]
 + \text{c.c.} \nonumber \\ 
 &= 
 2\omega^2 Z^a \cos\vartheta^a \int \frac{\rmd^2 \bm{k}^a_{\parallel}}{(2\pi)^2}
 \mathrm{Re}\left[
 \left(u^a_{\xi}\hat{\bm{\xi}^a} + u^a_{\eta}\hat{\bm{\eta}}^a + u^a_{\zeta}\hat{\bm{\zeta}^a}\right)^{\dagger}
 \iu\frac{\partial}{\partial k^a_{y'}}
 \left(u^a_{\xi}\hat{\bm{\xi}^a} + u^a_{\eta}\hat{\bm{\eta}}^a + u^a_{\zeta}\hat{\bm{\zeta}^a}\right)
 \right].
\label{eq: numerator on the way}
 \end{align}
\end{widetext}
Here we discarded the surface term resulting from the integration by parts, since the beam has a finite spatial extent.

According to the property~\eqref{eq: deflection relation(b)} we will rewrite the $k$ derivative as
\begin{equation}
\frac{\partial }{\partial k^a_{y'}}
= \frac{1}{k^a_{\text{c}}}
\frac{\partial }{\partial \nu^a}
= \frac{1}{k_{\text{c}}}
\frac{\partial }{\partial \nu}.
\label{eq: ky deriv transf}
\end{equation}
We also note that the basis $(\hat{\bm{\xi}}^a, \hat{\bm{\eta}}^a, \hat{\bm{\zeta}}^a)$
depends on the deflection of the beam $\nu$.
We thus transform the displacement in the $(\xi^a, \eta^a, \zeta^a)$ coordinate
into that in the $(X^a, y', Z^a)$ coordinate by using $U^a$.
From Eqs.~\eqref{eq: incident xi eta X y' displacement transfer}, \eqref{eq: coordinate transf within the reflected transmitted beam}
and \eqref{eq: elastic Fresnel coeff matrices ini}--\eqref{eq: elastic Fresnel coeff matrices fin}, we obtain
\begin{align}
& u^a_{\xi}\hat{\bm{\xi}^a} + u^a_{\eta}\hat{\bm{\eta}}^a + u^a_{\zeta}\hat{\bm{\zeta}^a} 
= 
\begin{bmatrix}
 \hat{\bm{\xi}}^a & \hat{\bm{\eta}}^a & \hat{\bm{\zeta}}^a
\end{bmatrix}
\begin{bmatrix}
 u^a_{\xi} \\ u^a_{\eta} \\ u^a_{\zeta}
\end{bmatrix} \nonumber\\
&=
\begin{bmatrix}
 \hat{\bm{X}}^a & \hat{\bm{y'}} & \hat{\bm{Z}}^a
\end{bmatrix}
{U^a}^{\dagger}(\nu^a, \vartheta^a)\nonumber\\
& \quad \times \sqrt{G(\mu, \nu)} F^a(\vartheta) P U(\nu, \vartheta)
\begin{bmatrix}
 u_{X} \\ u_{y'} \\ 0
\end{bmatrix},
\label{eq: coord transf used at the end of the day}
\end{align}
and introduce a transfer matrix 
\begin{equation}
 T^a(\nu)\equiv 
{U^a}^{\dagger}(\nu^a, \vartheta^a) F^a(\vartheta) U(\nu, \vartheta).
\label{eq: T^a mat def}
\end{equation}
Since $F^a P = F^a$, we omit the projection matrix $P$.

With the help of Eqs.~\eqref{eq: ky deriv transf}, \eqref{eq: coord transf used at the end of the day},
\eqref{eq: T^a mat def},
and the property $G(\mu,\nu) = G(\Abs{\mu}, \Abs{\nu})$
we transform Eq.~\eqref{eq: numerator on the way}, i.e.,
the numerator of Eq.~\eqref{eq: IF shift weighted by energy flux}, as
\begin{widetext}
\begin{align}
& 2\omega^2 Z^a\cos\vartheta^a \int \frac{\rmd^2 \bm{k}_{\parallel}}{(2\pi)^2}
 \mathrm{Re}\left\{
\begin{bmatrix}
 u_{X} \\ u_{y'} \\ 0
\end{bmatrix}^{\dagger}
{T^a}^{\dagger}(\nu)\sqrt{G(\mu, \nu)}
\frac{\iu}{k_{\text{c}}}\frac{\partial}{\partial \nu}
\left( \sqrt{G(\mu, \nu)}
{T^a}(\nu)
\begin{bmatrix}
 u_{X} \\ u_{y'} \\ 0
\end{bmatrix}
\right)\right\}\nonumber\\
&= 2\omega^2 Z^a\cos\vartheta^a \int \frac{\rmd^2 \bm{k}_{\parallel}}{(2\pi)^2}
G(\mu, \nu)
\mathrm{Re}\left\{
\begin{bmatrix}
 u_{X} \\ u_{y'} \\ 0
\end{bmatrix}^{\dagger}
{T^a}^{\dagger}(\nu)
\frac{\iu}{k_{\text{c}}}\frac{\partial {T^a}(\nu)}{\partial \nu}
\begin{bmatrix}
 u_{X} \\ u_{y'} \\ 0
\end{bmatrix}
\right\} \nonumber\\
&\simeq 2\omega^2 Z^a\cos\vartheta^a \mathrm{Re}\left\{
\begin{bmatrix}
 u_{X} \\ u_{y'} \\ 0
\end{bmatrix}^{\dagger}
{T^a}^{\dagger}(0)
\frac{\iu}{k_{\text{c}}}
\left.\frac{\partial {T^a}(\nu)}{\partial \nu}\right|_{\nu = 0}
\begin{bmatrix}
 u_{X} \\ u_{y'} \\ 0
\end{bmatrix}
\right\} N^{\text{beam}}.
\label{eq: y shift weighted by energy flux}
\end{align} 
\end{widetext}
Here we omit the term including $\partial \sqrt{G(\mu, \nu)}/\partial \nu$,
since it vanishes in the integral over the symmetric interval.
In the last line we also neglect second or higher order terms in $\nu$ of $T^a(\nu)$.
The first order terms in $\nu$ also vanish in the integral over the symmetric interval.

The factors $N^{\text{beam}}$ in both numerator and denominator
of Eq.~\eqref{eq: IF shift weighted by energy flux} cancel each other out.
We can also replace the matrix $F^a(\vartheta)$ in the denominator 
with $T^a(0)$.
We thus obtain the transverse shift independent of the detail of the envelope function of the beam:
\begin{equation}
 \Delta r^a_{\text{IF}} 
= \frac{\displaystyle
\mathrm{Re}~\left\{
\begin{bmatrix}
 u_X \\ u_{y'} \\ 0
\end{bmatrix}^{\dagger}
\left.
{{T}^{a}}^{\dagger}(0) \iu \frac{\partial {T}^{a}(\nu)}{\partial \nu} 
\right|_{\nu = 0}
\begin{bmatrix}
 u_X \\ u_{y'} \\ 0
\end{bmatrix}
\right\}
}{\displaystyle k_{\text{c}}\, 
\begin{bmatrix}
 u_X \\ u_{y'} \\ 0
\end{bmatrix}^{\dagger}
{{T}^{a}}^{\dagger}(0){T}^{a}(0)
\begin{bmatrix}
 u_X \\ u_{y'} \\ 0
\end{bmatrix}
}.\label{eq: delta rIF }
\end{equation}

In contrast to the transverse-wave incidence,  
longitudinal-wave incidence produces no transverse shift in either the reflected or transmitted waves.  
One can show this absence of a transverse shift in the same manner as above.

\subsection{Case of purely circular polarization}
\label{151957_6Feb25}

To describe transport phenomena using the phonon distribution function, 
we assume that the incident wave is purely circularly polarized~\footnote{
The restriction to pure circular polarization is unnecessary when we employ the $3 \times 3$ phonon density matrix 
instead of the distribution function for the three modes $n = \text{L},\, \text{RH},\, \text{LH}$.}.  
Under this assumption, only two of the four Stokes parameters, those corresponding to intensity and circular polarization, remain finite: 
\begin{subequations}
\begin{align}
 &\Abs{u_{X}}^2 + \Abs{u_{y'}}^2, 
&\overline{\sigma} = \frac{2\mathrm{Im}\left[u^*_X u_{y'} \right]}{\Abs{u_X}^2 + \Abs{u_{y'}}^2}&
\end{align}
The two associated with linear polarization vanish: 
\begin{equation}
 \frac{2\mathrm{Re}\left[u^*_X u_{y'} \right]}{\Abs{u_X}^2 + \Abs{u_{y'}}^2} = \frac{\Abs{u_X}^2 - \Abs{u_{y'}}^2}{\Abs{u_X}^2 + \Abs{u_{y'}}^2} = 0. 
\label{120707_6Nov24}
\end{equation}
\end{subequations}
The condition in Eq.~\eqref{120707_6Nov24} allows us to simplify the expression for the transverse shift:
\begin{subequations}
 \begin{align}
 \Delta r^{\text{r-L}}_{\text{IF}} &=  \Delta r^{\text{t-L}}_{\text{IF}} = - \frac{\overline{\sigma} \cot \theta_{\text{T}}}{k_{\text{c}}},\label{eq: trans shift rho and tau a}\\
 \Delta r^{\text{r-T}}_{\text{IF}} &=
 - \frac{\overline{\sigma} \cot\theta_{\text{T}}}{k_{\text{c}}}\, 
 \frac{\Abs{\rho_{\text{SV}, \text{SV}} + \rho_{\text{SH}, \text{SH}}}^2}{\Abs{\rho_{\text{SV}, \text{SV}}}^2 + \Abs{\rho_{\text{SH}, \text{SH}}}^2},\\
 \Delta r^{\text{t-T}}_{\text{IF}} &=
 - \frac{\overline{\sigma} \cot\theta_{\text{T}}}{k_{\text{c}}}\, 
 \left(1- \frac{2 \mathrm{Re}\left[\tau^*_{\text{SV},\text{SV}} \tau_{\text{SH}, \text{SH}}\right]}{
 \Abs{\tau_{\text{SV}, \text{SV}}}^2 + \Abs{\tau_{\text{SH}, \text{SH}}}^2}
 \, \frac{\cos\Theta_{\text{T}}}{\cos\theta_{\text{T}}}\right).\label{eq: trans shift rho and tau c}
 \end{align}
\end{subequations}

We rearrange the transverse shift using the $S$-matrix elements in the circular basis, shown in Eq.~\eqref{eq: elastic S matrix in circularly pol}.  
We convert the amplitudes from the linear to the circular basis using Eq.~\eqref{eq: lin circ conversion}:
\begin{equation}
\left[
\renewcommand\arraystretch{1.3}
\begin{array}{c}
u_{\text{RH}}\\ u_{\text{LH}}
\end{array} 
\right] = \frac{1}{\sqrt{2}}
\begin{bmatrix}
 1 & -\iu\\[4pt]
 1 & +\iu
\end{bmatrix}
\left[
\renewcommand\arraystretch{1.3}
\begin{array}{c}
u_{X}\\ u_{y'}
\end{array} 
\right]. 
\end{equation}
With straightforward calculations, we obtain
\begin{subequations}
\begin{align}
& \frac{\Abs{\rho_{\text{SV}, \text{SV}} + \rho_{\text{SH}, \text{SH}}}^2}{\Abs{\rho_{\text{SV}, \text{SV}}}^2 + \Abs{\rho_{\text{SH}, \text{SH}}}^2}
 = \frac{2 \Abs{r_{\text{RH}, \text{RH}}}^2}{\Abs{r_{\text{RH}, \text{RH}}}^2 + \Abs{r_{\text{LH}, \text{RH}}}^2}\nonumber \\ &
 = \frac{2 \mathcal{R}_{\text{RH}, \text{RH}}}{\mathcal{R}_{\text{RH}, \text{RH}} + \mathcal{R}_{\text{LH}, \text{RH}}},\\
& \frac{2 \mathrm{Re}\left[\tau^*_{\text{SV},\text{SV}} \tau_{\text{SH}, \text{SH}}\right]}{
 \Abs{\tau_{\text{SV}, \text{SV}}}^2 + \Abs{\tau_{\text{SH}, \text{SH}}}^2}
=  \frac{\Abs{t_{\text{RH}, \text{RH}}}^2 - \Abs{t_{\text{LH}, \text{RH}}}^2}{\Abs{t_{\text{RH}, \text{RH}}}^2 + \Abs{t_{\text{LH}, \text{RH}}}^2}
\nonumber\\ & 
= \frac{\mathcal{T}_{\text{RH}, \text{RH}} - \mathcal{T}_{\text{LH}, \text{RH}}}{ \mathcal{T}_{\text{RH}, \text{RH}} + \mathcal{T}_{\text{LH}, \text{RH}}}.
 \end{align}
\end{subequations}
Substituting these results 
into Eqs.~\eqref{eq: trans shift rho and tau a}--\eqref{eq: trans shift rho and tau c}, we finally arrive at Eqs.~\eqref{170438_23Jan25}--\eqref{135819_28Jan25}.

\section{Omission of in-plane drift terms under weak confinement}
\label{sec: appendix validity of xy derivatives in BTE}

In Sec.~\ref{subsec: Boltzmann theory}, we analyzed the spatial distribution of OAM by neglecting the $x$- and $y$-derivative terms 
in the Boltzmann equations~\eqref{122352_26Jan25} and \eqref{122357_26Jan25}.  
This approximation is valid when the boundary confinement varies smoothly compared with characteristic length scales 
such as the phonon mean free path or the transverse shift.  
In this Appendix, we examine the validity of this approximation by estimating a lower bound on the confinement width $\xi_{\mathrm{con}}$ near the boundary.

For the following analysis, we examine the $y'$ dependence of the phonon distribution function.  
The weak-confinement condition requires that
\begin{equation}
 \Abs{v^{y'}_{\bm{k}n}\frac{\partial F^{(1)}_{\bm{k}n}}{\partial 
\left(\bm{R}_{\parallel}\right)_{y'}}} \ll \Abs{\frac{F^{(1)}_{\bm{k}n}}{\tau}},\qquad 
\Abs{c^{y'}_{\bm{q}n}\frac{\partial f^{(1)}_{\bm{q}n}}{\partial 
\left(\bm{r}_{\parallel}\right)_{y'}}} \ll \Abs{\frac{f^{(1)}_{\bm{q}n}}{\widetilde{\tau}}}, 
\label{eq: condition negligible drift terms}
\end{equation}
under which the $y'$ derivative terms in the Boltzmann equation can be neglected.

We further assume that the phonon distribution is nearly uniform along the $x'$ direction, parallel to $\bm{k}_{\parallel}$. 
Then its $x'$ average, expressed in Eqs.~\eqref{181855_25Jan25} and \eqref{181858_25Jan25}, reduces to
\begin{subequations}
 \begin{align}
 \breve{F}^{(1)}_{s,n}(\omega, \bm{k}_{\parallel}, L^z, z)
 &\sim F^{(1)}_{s,n}(\omega, \bm{k}_{\parallel}, \bm{R}_{\parallel}, z), \label{171723_29Jan25}\\
 \breve{f}^{(1)}_{s,n}(\omega, \bm{q}_{\parallel}, L^z, z)
 &\sim f^{(1)}_{s,n}(\omega, \bm{q}_{\parallel}, \bm{r}_{\parallel}, z).\label{171725_29Jan25}
 \end{align}
\end{subequations}
With this assumption, and using the relations
$L^z(\bm{k}_{\parallel}, \bm{R}_{\parallel}) = - \hbar \Abs{\bm{k}_{\parallel}} \left(\bm{R}_{\parallel}\right)_{y'}$ and
$\widetilde{L}^z(\bm{q}_{\parallel}, \bm{r}_{\parallel}) = - \hbar \Abs{\bm{q}_{\parallel}} \left(\bm{r}_{\parallel}\right)_{y'}$, 
we can rewrite the condition~\eqref{eq: condition negligible drift terms} as
\begin{subequations}
 \begin{align}
 \Abs{\hbar v^{y'}_{\bm{k}n} \Abs{\bm{k}_{\parallel}} \frac{\partial \breve{F}^{(1)}}{\partial L^z}}
 &\ll \Abs{\frac{\breve{F}^{(1)}}{\tau}}, \label{eq: condition negligible drift terms' a}\\
 \Abs{\hbar c^{y'}_{\bm{q}n} \Abs{\bm{q}_{\parallel}} \frac{\partial \breve{f}^{(1)}}{\partial \widetilde{L}^z}}
 &\ll \Abs{\frac{\breve{f}^{(1)}}{\widetilde{\tau}}}.\label{eq: condition negligible drift terms' b}
 \end{align}
\end{subequations}

We substitute Eqs.~\eqref{182532_28Jan25}, \eqref{172201_29Jan25}, \eqref{143309_6Nov24}, and \eqref{143312_6Nov24} 
into Eqs.~\eqref{eq: condition negligible drift terms' a} and \eqref{eq: condition negligible drift terms' b}.  
To simplify the condition, we introduce several approximations.  
We focus on the region $z \simeq 0$ and treat $\Abs{B_{s,n}(\omega, \bm{k}_{\parallel})}$ as approximately independent of $s$ and $n$.  
The reflectance $\mathcal{R}_{nm}$ and transmittance $\mathcal{T}_{nm}$ are taken to be of order unity,  
and the SAM per mode, $S^z_{s,n}(\omega, \bm{k}_{\parallel})$ and $\widetilde{S}^z_{s,n}(\omega, \bm{q}_{\parallel})$, to be of order $\hbar$.  
We further treat $\bm{k}_{\parallel}$ and $\bm{q}_{\parallel}$ as scalars $k$ and $q$, 
and the group velocities $v^{y'}_{\bm{k}n}$ and $c^{y'}_{\bm{q}n}$ as $v$ and $c$, respectively.  
Under these approximations, the leading-order behavior is estimated as
\begin{align}
\hbar v k\Abs{\frac{\partial}{\partial L}\left[\frac{W(L + \hbar)}{W(L)}\right]} &\ll \frac{1}{\tau}, \label{174520_4Oct25}\\
\hbar c q\Abs{\frac{\partial}{\partial L}\left[\frac{W(L + \hbar)}{W(L)}\right]} &\ll \frac{1}{\widetilde{\tau}}.
\label{173200_29Jan25} 
\end{align}
For notational simplicity, we write the OAM density function $W_{\bm{k}_{\parallel}}$ as $W$ and 
abbreviate $L = L^z$ and $\widetilde{L} = \widetilde{L}^z$.  
In the bulk, $W$ takes a constant value $W_0$, as shown in Eqs.~\eqref{125112_25Jan25} and \eqref{125118_25Jan25}, 
and the conditions~\eqref{174520_4Oct25} and \eqref{173200_29Jan25} are satisfied.

Near the nominal boundary, where $0 < W < W_0$, 
the OAM $L$ of a wave packet is of the order of the system size $\Lambda$, and $L \gg \hbar$ holds.  
It then follows that
\begin{equation}
\frac{W(L + \hbar)}{W(L)} \sim 1 + \hbar \, \frac{\partial \ln W(L)}{\partial L}. \label{175005_4Oct25}
\end{equation}
Introducing the phonon mean free paths $l = \tau v$ and $\widetilde{l} = \widetilde{\tau} c$ and assuming $kl \simeq q\widetilde{l}$,  
Eqs.~\eqref{174520_4Oct25} and \eqref{175005_4Oct25} yield
\begin{equation}
\Abs{\hbar^2\, \frac{\partial^2}{\partial L^2}\ln W(L)} \ll \frac{1}{kl}. 
\label{173705_29Jan25}
\end{equation}
For the mesoscopic transport described by the Boltzmann equation, $kl \gg 1$ generally holds.

The condition in Eq.~\eqref{173705_29Jan25} sets a lower bound on the spatial confinement of a wave packet.  
As an illustrative example, we consider an OAM density function of the form
\begin{equation}
 W(L, b)
= \frac{W_0}{\displaystyle\exp\left[b\, \left(\frac{\Abs{L}}{\hbar k\Lambda} -1\right)\right] + 1}, 
\end{equation}
similar to Eq.~\eqref{131000_25Jan25}. 
Here $b$ is a dimensionless parameter characterizing the decay of $W$ near the edge.  
Introducing $L_0 = \hbar k \Lambda$, we obtain
\begin{equation}
\frac{\partial^2}{\partial L^2} \ln W(L, b)
= - \left\{\frac{b}{2L_0 \cosh\left[b(L/L_0 - 1)/2\right]}\right\}^2. 
\end{equation}
To satisfy Eq.~\eqref{173705_29Jan25} for all $L$, the following condition must hold:
\begin{equation}
\left(\frac{\hbar b}{L_0}\right)^2 \ll (kl)^{-1}. 
\end{equation}
This condition can be expressed in terms of the confinement length scale $\xi_{\text{con}}$, 
which quantifies the spatial extent of the wave packet beyond the nominal confining region, as defined in Eq.~\eqref{174048_6Feb25}:
\begin{equation}
 \xi_{\text{con}}\equiv \frac{\Lambda}{b}
\gg \frac{\Lambda\, \sqrt{kl}}{L_0/\hbar}
= \sqrt{\frac{\lambda l}{2\pi}}. \label{174709_29Jan25}
\end{equation}
Here $\lambda = 2\pi/k$ denotes the phonon wavelength.

In summary, neglecting the in-plane spatial variations of the phonon distribution~\eqref{eq: 4 drift terms} 
is justified when the confinement is sufficiently smooth to satisfy Eq.~\eqref{174709_29Jan25}.
Note that the confinement width $\xi_{\mathrm{con}}$ is larger than the transverse shift $\Delta r_{\mathrm{IF}}$ of a wave packet:
$\Delta r_{\mathrm{IF}} \sim \lambda \ll \sqrt{\lambda l} \ll \xi_{\mathrm{con}}$.  
This relation is the same as that stated at the end of the caption of 
Fig.~\ref{165750_6Nov24}.

\bibliography{reference}

\clearpage
\renewcommand{\theequation}{S\arabic{equation}}
\setcounter{equation}{0}
\renewcommand{\thefigure}{S\arabic{figure}}
\setcounter{figure}{0}
\renewcommand{\thetable}{S\arabic{table}}
\setcounter{table}{0}
\makeatletter
\c@secnumdepth = 2
\makeatother
\onecolumngrid
\begin{center}
 {\large \textmd{Supplemental Materials for:} \\[0.3em] {\bfseries 
Boundary condition for phonon distribution functions
at a smooth crystal interface and interfacial angular momentum transfer
}}
\end{center}
\setcounter{page}{1}
\setcounter{section}{0}
\renewcommand{\thesection}{S\arabic{section}}
\renewcommand{\appendixname}{}


\section{Angle dependence and symmetry of power reflectance and transmittance}

\begin{figure}[bp]
 \centering
\includegraphics[pagebox=artbox,width=0.99\textwidth]{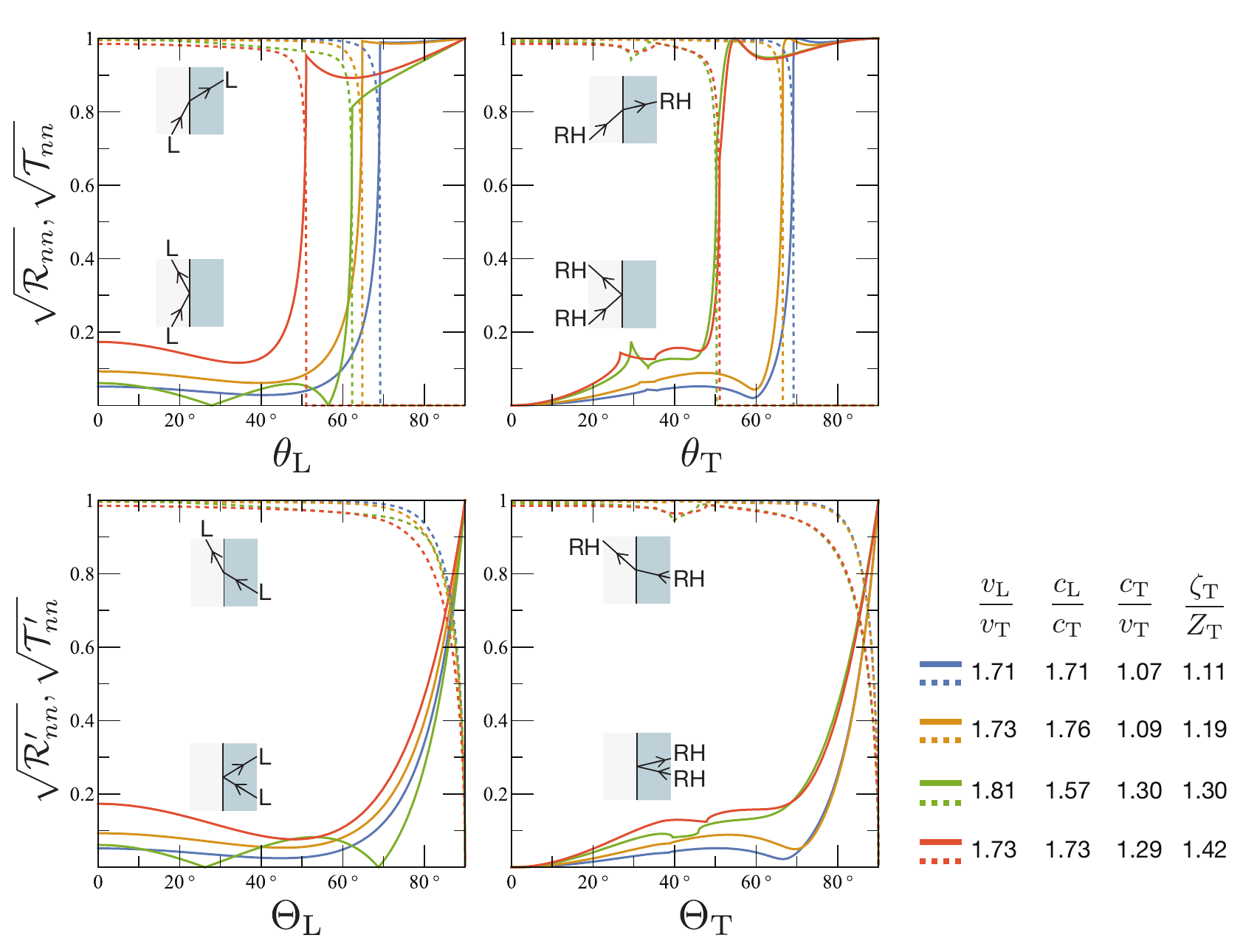}
\caption{Square roots of reflected/transmitted energy flow divided by incident energy flow if no change in elastic wave modes occurs.
The wave modes here consist of longitudinal~(L) and right-handed~(RH)/left-handed~(LH) circularly polarized transverse modes.
}
\label{fig: R and T between the same modes}
\end{figure}

\begin{figure}[tbp]
 \centering
\includegraphics[pagebox=artbox,width=0.99\textwidth]{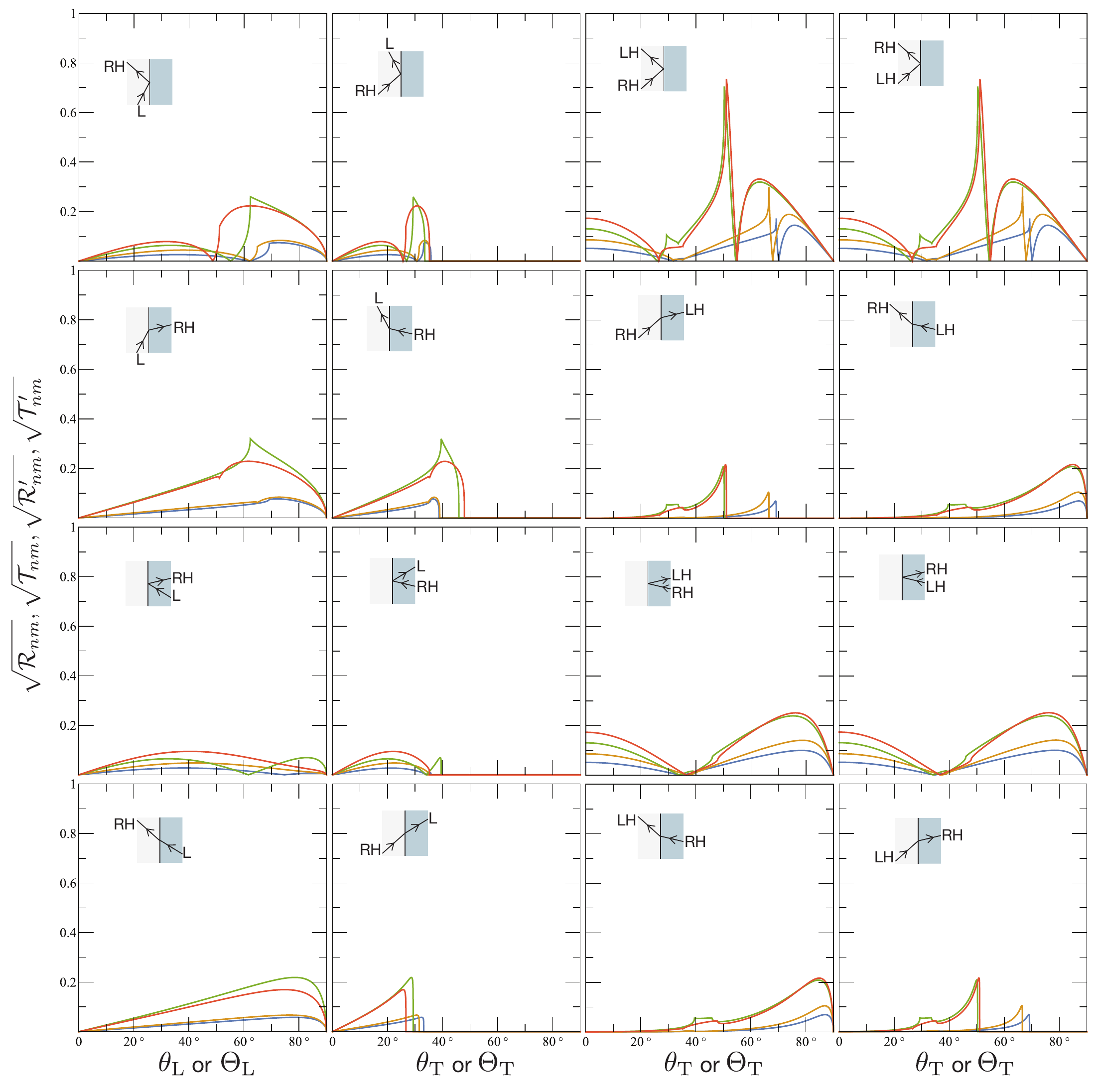}
\caption{Square roots of reflected/transmitted energy flow divided by incident energy flow 
if incident and reflected/transmitted waves are of different mode.
The wave modes here consist of longitudinal~(L) and right-handed~(RH)/left-handed~(LH) circularly polarized transverse modes.
Parameters for each colored curve are the same as Fig.~\ref{fig: R and T between the same modes}.
}
\label{fig: R and T between different modes}
\end{figure}

In the main text, we defined the power reflectance and transmittance for elastic waves as 
$\mathcal{R}_{nm}$, $\mathcal{R}'_{nm}$, $\mathcal{T}_{nm}$, and $\mathcal{T}'_{nm}$.
For representative sets of velocities and acoustic impedances 
$Z_n = \rho v_n$ and $\zeta_n = \widetilde{\rho} c_n$ ($n = \text{L}, \text{T}$),
we show their dependences on the angle of incidence in 
Figs.~\ref{fig: R and T between the same modes} and \ref{fig: R and T between different modes}.
Here the reflectance and transmittance are expressed as functions of the angle of incidence,  
$\mathcal{R}_{nm}(\omega, \bm{k}_{\parallel}) = \mathcal{R}_{nm}(\theta_m)$ and  
$\mathcal{T}_{nm}(\omega, \bm{k}_{\parallel}) = \mathcal{T}_{nm}(\theta_m)$,  
where $\sin \theta_m = v_m \Abs{\bm{k}_{\parallel}} / \omega$.  
Similarly, for the primed quantities we use  
$\mathcal{R}'_{nm}(\omega, \bm{q}_{\parallel}) = \mathcal{R}'_{nm}(\Theta_m)$ and  
$\mathcal{T}'_{nm}(\omega, \bm{q}_{\parallel}) = \mathcal{T}'_{nm}(\Theta_m)$,  
where $\sin \Theta_m = c_m \Abs{\bm{q}_{\parallel}} / \omega$.

These quantities satisfy two symmetry relations: 
one is the Helmholtz reciprocity, 
$\mathcal{R}_{nm} = \mathcal{R}_{mn}$, $\mathcal{R}'_{nm} = \mathcal{R}'_{mn}$, and $\mathcal{T}_{nm} = \mathcal{T}'_{mn}$; 
the other is a mode-exchange invariance, under which 
$\mathcal{R}_{nm}$, $\mathcal{R}'_{nm}$, $\mathcal{T}_{nm}$, and $\mathcal{T}'_{nm}$ 
remain unchanged when the subscripts $(n, m)$ are exchanged from (L, RH, LH) to (L, LH, RH).

To avoid redundancy due to the latter invariance, we omit the cases of reflection and transmission between LH modes in 
Fig.~\ref{fig: R and T between the same modes},  
and between L and LH modes in Fig.~\ref{fig: R and T between different modes}.  
The former symmetry, the Helmholtz reciprocity, can also be verified in 
Fig.~\ref{fig: R and T between different modes} by comparing the curves of the same color 
in the first and second columns, or in the third and fourth columns, 
after converting their incidence angles according to Snell's law.

\section{Interfacial thermal resistance as a test of the boundary conditions}

In this section we examine energy transport across an interface between two crystals held at different temperatures, 
rather than the phonon angular momentum transport discussed in the main text. 
For a smooth interface, the energy flow is described by the acoustic mismatch model established in earlier studies~\cite{Little1959,Khalatnikov1965,Swartz1989}. 
We use this well-understood problem as a benchmark to test the validity of the coarse-grained boundary conditions given in Eq.~(4) of the main text.

We consider two semi-infinite crystals joined at a plane $z=0$. 
The crystal on the left ($z<0$) has a bulk temperature $T_1$, while that on the right ($z>0$) has a lower bulk temperature $T_2$ ($T_2<T_1$). 
We assume that the temperature drop $\Delta T = T_1 - T_2$ is small compared with the average temperature $T = (T_1 + T_2)/2$.

The phonon distribution functions in the left and right crystals, denoted by $F_{\bm{k}n}(z)$ and $f_{\bm{q}n}(z)$, respectively, 
satisfy the Boltzmann equation with a constant relaxation time and without external fields:
\begin{subequations}
 \begin{gather}
 v^z_{\bm{k}n}\frac{\partial F_{\bm{k}n}}{\partial z}  = - \frac{F_{\bm{k}n} - f^{\text{eq}}(\Omega_{\bm{k}n}, T_1)}{\tau} \qquad (z < 0), \\
 c^z_{\bm{q}n}\frac{\partial f_{\bm{q}n}}{\partial z} = - \frac{f_{\bm{q}n} -  f^{\text{eq}}(\omega_{\bm{q}n}, T_2)}{\widetilde{\tau}} \qquad (z > 0).
 \end{gather}
\end{subequations}
Here $f^{\text{eq}}(\omega, T) = [\exp(\hbar\omega/k_{\text{B}}T) - 1]^{-1}$ is the Bose distribution function. 
The stationary solutions can then be written as
\begin{align}
F_{\bm{k}n} &= f^{\text{eq}}(\Omega_{\bm{k}n}, T_1) + 
\begin{cases}
C_{\bm{k}n} \exp\left[z\Big/(-\tau v^z_{\bm{k}n})\right] & v^z_{\bm{k}n} < 0\\
0 & v^z_{\bm{k}n} > 0
\end{cases}
, \\
f_{\bm{q}n} &= f^{\text{eq}}(\omega_{\bm{q}n}, T_2) + 
\begin{cases}
0 & c^z_{\bm{q}n} < 0 \\
D_{\bm{q}n} \exp\left[- z\Big/\widetilde{\tau} c^z_{\bm{q}n}\right] & c^z_{\bm{q}n} >0
\end{cases}, 
\end{align}
where $C_{\bm{k}n}$ and $D_{\bm{q}n}$ are amplitudes representing deviations localized near the interface.

For the distribution functions $F_{\bm{k}n}(z)$ and $f_{\bm{q}n}(z)$ satisfying $\Omega_{\bm{k}n} = \omega_{\bm{q}n}$ and $\bm{k}_{\parallel} = \bm{q}_{\parallel}$, 
the coarse-grained boundary conditions given in Eq.~(4) of the main text lead to the following linear relations:
\begin{equation}
 \left\{
\begin{aligned}
F_{-,n} &= \sum_m \left[\mathcal{R}_{nm} F_{+,m} + \mathcal{T}'_{nm} f_{-,m}\right],\\
f_{+,n} &= \sum_m \left[\mathcal{T}_{nm} F_{+,m} + \mathcal{R}'_{nm} f_{-,m}\right].
\end{aligned}
\right.
\end{equation}
Here $F_{\bm{k}n}(z=-0)$ and $f_{\bm{q}n}(z=+0)$ are written as
\begin{align}
F_{\bm{k}n} &=
\begin{cases}
F_{+,n}(\Omega_{\bm{k}n}, \bm{k}_{\parallel}), & v^z_{\bm{k}n} > 0,\\[3pt]
F_{-,n}(\Omega_{\bm{k}n}, \bm{k}_{\parallel}), & v^z_{\bm{k}n} < 0,
\end{cases}
&
f_{\bm{q}n} &=
\begin{cases}
f_{+,n}(\omega_{\bm{q}n}, \bm{q}_{\parallel}), & c^z_{\bm{q}n} > 0,\\[3pt]
f_{-,n}(\omega_{\bm{q}n}, \bm{q}_{\parallel}), & c^z_{\bm{q}n} < 0,
\end{cases}
\end{align}
where $\bm{k}_{\parallel}$ and $\bm{q}_{\parallel}$ are the wavevector components parallel to the interface.

Since the sum of the reflectance and transmittance coefficients equals unity, 
\begin{align}
 \sum_m \left(\mathcal{R}_{nm} + \mathcal{T}'_{nm}\right) &= 1, 
&\sum_m \left(\mathcal{T}_{nm} + \mathcal{R}'_{nm}\right) &= 1, 
\end{align}
the amplitudes of the localized phonon distributions near the interface are obtained as
\begin{equation}
 \left\{
\begin{aligned}
C_{-,n}(\omega, \bm{k}_{\parallel}) &= 
\left[\sum_m \mathcal{T}'_{nm}(\omega, \bm{k}_{\parallel})\right]
\!\left[f^{\text{eq}}(\omega, T_2) - f^{\text{eq}}(\omega, T_1)\right],\\[4pt]
D_{+,n}(\omega, \bm{q}_{\parallel}) &= 
\left[\sum_m \mathcal{T}_{nm}(\omega, \bm{q}_{\parallel})\right]
\!\left[f^{\text{eq}}(\omega, T_1) - f^{\text{eq}}(\omega, T_2)\right].
\end{aligned}
\right.
\end{equation}

Using the Helmholtz reciprocity relation $\mathcal{T}_{nm} = \mathcal{T}'_{mn}$, 
we obtain
\begin{equation}
 \left\{
\begin{aligned}
 C_{\bm{k}n} &= - \Delta T\, \frac{\partial f^{\text{eq}}(\hbar v_n k/k_{\text{B}}T)}{\partial T}
\left[\sum_m \mathcal{T}_{mn}(\Omega_{\bm{k}n}, \bm{k}_{\parallel})\right]\qquad (k_z < 0)\\[6pt]
 D_{\bm{q}n}&= \Delta T\, \frac{\partial f^{\text{eq}}(\hbar c_n q/k_{\text{B}}T)}{\partial T}
\left[\sum_m \mathcal{T}'_{mn}(\omega_{\bm{q}n}, \bm{q}_{\parallel}) \right] \qquad (q_z > 0)
\end{aligned}
\right. . 
\end{equation}

The energy flux normal to the interface is defined as
\begin{equation}
 j_{\mathfrak{u}}(z) =
\begin{cases}
\displaystyle \sum_{n}\int \frac{\rmd^3 \bm{k}}{(2\pi)^3}\, \hbar\Omega_{\bm{k}n} v^z_{\bm{k}n} F_{\bm{k}n} (z), & z < 0,\\[6pt]
\displaystyle  \sum_{n}\int \frac{\rmd^3 \bm{q}}{(2\pi)^3}\,\hbar\omega_{\bm{q}n} c^z_{\bm{q}n} f_{\bm{q}n} (z), & z > 0,
\end{cases}
\end{equation}
and can be evaluated as
\begin{align}
 j_{\mathfrak{u}}(z<0) &= \sum_n\int\limits_{k_z <0 } \frac{\rmd^3\bm{k}}{(2\pi)^3} 
 \hbar \Omega_{\bm{k}n} v^z_{\bm{k}n} C_{\bm{k}n}\exp
\left[- \Abs{\dfrac{z}{\tau v^z_{\bm{k}n}}}\right] \nonumber\\
 &= \frac{2\pi^2}{15} \frac{(k_{\text{B}}T)^3 k_{\text{B}}\Delta T}{\hbar^3}
\sum_n \frac{1}{v_n^2}
\int^1_0 \frac{\rmd (\cos\theta)}{2}\cos \theta
\exp\left[-\dfrac{\Abs{z/\tau v_n}}{\cos\theta}\right] \sum_m \mathcal{T}_{mn}(\theta)
\end{align}
for the left-hand side, while for the right-hand side we have
\begin{align}
j_{\mathfrak{u}}(z>0) &= \sum_n\int\limits_{q_z > 0 } \frac{\rmd^3\bm{q}}{(2\pi)^3} 
 \hbar \omega_{\bm{q}n} c^z_{\bm{q}n} D_{\bm{q}n}\exp
\left[- \Abs{\dfrac{z}{\widetilde{\tau} c^z_{\bm{q}n}}}\right]\nonumber\\
 &= \frac{2\pi^2}{15} \frac{(k_{\text{B}}T)^3 k_{\text{B}}\Delta T}{\hbar^3}
 \sum_n \frac{1}{c_n^2}
\int^1_0 \frac{\rmd (\cos\Theta)}{2}\cos \Theta
\exp\left[-\dfrac{\Abs{z/\widetilde{\tau} c_n}}{\cos\Theta}\right] \sum_m \mathcal{T}'_{mn}(\Theta).
\end{align}
At the interface, the incoming energy fluxes from both sides are given by
\begin{subequations}
 \begin{align}
 j_{\mathfrak{u}}(z=-0) 
 &= \frac{\pi^2}{15} \frac{(k_{\text{B}}T)^3 k_{\text{B}}\Delta T}{\hbar^3} \sum_{n, m} \int^1_0 \frac{\cos\theta_n \rmd (\cos\theta_n)}{v_n^2}\mathcal{T}_{mn}(\theta_n),
\label{eq: interfacial thermal flux a}\\
 j_{\mathfrak{u}}(z=+0) 
 &= \frac{\pi^2}{15} \frac{(k_{\text{B}}T)^3 k_{\text{B}}\Delta T}{\hbar^3}
\sum_{n, m} \int^1_0 \frac{\cos\Theta_m \rmd (\cos\Theta_m)}{c_m^2}\mathcal{T}'_{nm}(\Theta_m). \label{eq: interfacial thermal flux b}
 \end{align}
\end{subequations}

The continuity of the energy flux across the interface can now be verified explicitly.
First, the Helmholtz reciprocity implies $\mathcal{T}_{mn}(\theta_n) = \mathcal{T}'_{nm}(\Theta_m)$.
Second, Snell's law gives ${\sin\theta_n}/{v_n} = {\sin\Theta_m}/{c_m}$.
Finally, taking a differential of Snell's law for small angular differences yields 
${\cos\theta_n\,\rmd\theta_n}/{v_n} = {\cos\Theta_m\,\rmd\Theta_m}/{c_m}$.
Combining these three relations, we find
\begin{equation}
\frac{\cos\theta_n \rmd (\cos\theta_n)}{v_n^2}\mathcal{T}_{mn}(\theta_n)
= \frac{\cos\Theta_m \rmd (\cos\Theta_m)}{c_m^2}\mathcal{T}'_{nm}(\Theta_m), 
\end{equation}
which shows that the two expressions, 
Eqs.~\eqref{eq: interfacial thermal flux a} and \eqref{eq: interfacial thermal flux b},
are equivalent, i.e., $j_{\mathfrak{u}}(z=-0)=j_{\mathfrak{u}}(z=+0)$.

We thus define a characteristic velocity $\mathscr{V}$ through
\begin{equation}
 j_{\mathfrak{u}}(z=-0) = j_{\mathfrak{u}}(z=+0) 
=: \frac{\pi^2}{15} \frac{(k_{\text{B}}T)^3 k_{\text{B}}\Delta T}{\hbar^3}\,\frac{1}{\mathscr{V}^2}, 
\end{equation}
from which the interfacial thermal resistance is found to scale as $T^{-3}$:
\begin{equation}
k_{\text{B}} \Delta T = 
\frac{15\mathscr{V}^2}{\pi^2}\left(\frac{\hbar}{k_{\text{B}}T}\right)^3 j_{\mathfrak{u}} (z=0),\label{eq: thermal boundary resistance}
\end{equation}
indicating that a finite temperature discontinuity appears at the interface
whenever a heat flux flows between the two crystals.
Equation~\eqref{eq: thermal boundary resistance} is fully consistent with the
interfacial thermal resistance predicted by the
acoustic mismatch model~\cite{Little1959,Khalatnikov1965}.

\end{document}